\documentclass[12pt]{article}        

\usepackage{amsmath}
\usepackage{amssymb}
\usepackage{euscript}

\usepackage{epsfig}

\usepackage{cite}

\usepackage{alltt}

\usepackage{epic}

\def\Order#1{${\cal O}(#1)$}

\def\Oeex#1{${\cal O}(#1)_{\rm EEX}$}
\def\Oceex#1{${\cal O}(#1)_{\rm CEEX}$}
\def\KK{${\cal KK}$}
\def\bbeta{\bar{\beta}}
\def\born{{\rm Born}}
\def\st{\hbox{}} 

\newcommand{\Bmf}{\mathfrak{B}}

\newcommand{\Mmf}{\mathfrak{M}}

\newcommand{\Fmf}{\mathfrak{F}}
\newcommand{\FmI}{\mathfrak{I}}
\newcommand{\sfac}{\mathfrak{s}}

\begin{document}                     

\allowdisplaybreaks

\begin{titlepage}

\begin{flushright}
{\bf  
DESY-99-106  \\
CERN-TH/99-235  \\
UTHEP-99-08-01}
\end{flushright}

\vspace{1mm}
\begin{center}{\bf\Large The Precision Monte Carlo Event Generator ${\cal KK}$  }\end{center}
\begin{center}{\bf\Large For Two-Fermion Final States In $e^+e^-$ Collisions$^{\dag}$ }\end{center}

\vspace{1mm}
\begin{center}
  {\bf   S. Jadach$^{a,b}$,}
  {\bf   B.F.L. Ward$^{c,d}$}
  {\em and}
  {\bf   Z. W\c{a}s$^{b,e}$ }
\\
\vspace{1mm}
{\em $^a$DESY, Theory Division, Notkestrasse 85, D-22603 Hamburg, Germany,}\\
{\em $^b$Institute of Nuclear Physics,
         ul. Kawiory 26a, 30-055 Cracow, Poland,}\\
{\em $^c$Department of Physics and Astronomy,\\
         The University of Tennessee, Knoxville, TN 37996-1200, USA,}\\
{\em $^d$SLAC, Stanford University, Stanford, CA 94309, USA,}\\
{\em $^e$CERN, Theory Division, CH-1211 Geneva 23, Switzerland,}\\
\end{center}

\vspace{1mm}
\begin{abstract}
We present the  Monte Carlo event generator \KK\ version 4.13
for precision predictions of the Electroweak Standard Model
for the process $e^+e^-\to f\bar{f} +n\gamma$, $f=\mu,\tau,d,u,s,c,b$,
at centre-of-mass energies from $\tau$ lepton threshold to 1 TeV,
that is for LEP, SLC, future Linear Colliders, $b,c,\tau$-factories, etc.
Effects due to photon emission from initial beams and outgoing fermions
are calculated in QED up to second order, including all interference effects,
within Coherent Exclusive Exponentiation (CEEX), 
which is based on Yennie--Frautschi--Suura exponentiation.
Electroweak corrections are included in first order, with higher-order 
extensions, using the DIZET 6.21 library.
Final-state quarks hadronize according to the parton shower model using JETSET.
Beams can be polarized longitudinally and transversely.
Decay of the $\tau$ leptons is simulated using the TAUOLA library, taking into
account spin polarization effects as well. In particular the complete
spin correlations density matrix of the initial-state beams and final
state $\tau$'s is incorporated in an exact manner.
Effects due to beamstrahlung are simulated in a realistic way.
The main improvements with respect to KORALZ are: 
(a) inclusion of the initial--final state QED interference,
(b) inclusion of the exact matrix element for two photons,
and (c) inclusion of the transverse spin correlations in $\tau$ decays (as in KORALB).
\end{abstract}
\begin{center}
{\it To appear in Computer Physics Communications}
\end{center}

\vspace{1mm}
\footnoterule
\noindent
{\footnotesize
\begin{itemize}
\item[${\dag}$]
Work supported in part by Polish Government grants 
KBN 2P03B08414, 
KBN 2P03B14715, 
the US DoE contracts DE-FG05-91ER40627 and DE-AC03-76SF00515,
the Maria Sk\l{}odowska-Curie Joint Fund II PAA/DOE-97-316,
and the Polish--French Collaboration within IN2P3 through LAPP Annecy.
\end{itemize}
}
\begin{flushleft}
{\bf 
  DESY-99-106  \\
  CERN-TH/99-235  \\
  UTHEP-99-08-01\\
  July 1999}
\end{flushleft}

\end{titlepage}

\tableofcontents   

\newpage

\noindent{\bf PROGRAM SUMMARY}
\vspace{10pt}

\noindent{\sl Title of the program:} \KK\, version 4.13.

\noindent{\sl Computer:}\\
any computer with the FORTRAN 77 compiler and the UNIX operating system

\noindent{\sl Operating system:}\\
UNIX, program was tested under AIX 4.x, HP-UX 10.x and Linux

\noindent{\sl Programming language used:}
FORTRAN 77 with popular extensions such as long names, etc.

\noindent{\sl High-speed storage required:}  $<$ 10 MB

\noindent{\sl No. of cards in combined program and test deck:}
about 21,800, without JETSET, TAUOLA and PHOTOS.


\noindent{\sl Nature of the physical problem:}
The fermion pair production  is and will be
used as an important data point for precise tests
of the standard electroweak theory at LEP and future linear colliders at higher energies.
QED corrections to fermion pair production (especially $\tau$ leptons)
at $c$-quark and $b$-quark factories has to be known to second order,
including spin polarization effects.
The Standard Model predictions at the per mille precision level, taking
into account multiple emission of photons for realistic experimental
acceptance,  can {\em only} be obtained using a Monte Carlo event generator.

\noindent{\sl Method of solution:}
The Monte Carlo methods are used to simulate most of the two-fermion final-state
processes in $e^+e^-$ collisions in the presence of multiphoton
initial-state radiation. 
The latter is described in the framework of exclusive coherent exponentiation (CEEX)
based on Yennie--Frautschi--Suura exclusive exponentiation (YFS/EEX).
CEEX treats correctly  to infinite order not only infrared cancellations but also
QED interferences and narrow resonances.
The matrix element according to standard YFS exponentiation is also provided for tests.
For quarks and $\tau$ leptons, the appropriate simulation of hadronization or decay is included.
Beam polarization and spin effects, both longitudinal and transverse, 
in tau decays are properly taken into account.

\noindent{\sl Restrictions on the complexity of the problem:}
In the present version, electron (Bhabha), neutrino and top quark final states are not included 
(they will be in a future version).
Additional fermion pair production is not included.
Third-order QED corrections in leading-logarithmic approximation are included
only in the auxiliary YFS/EEX matrix element
(which can be activated with the help of input parameters).
Electroweak corrections should not be trusted above the $t$-quark threshold.
The total cross section for light quarks for $\sqrt{s}<10$ GeV requires an 
improvement
using experimental data.

\noindent{\sl Typical running time:}
On the IBM PowerPC M43P240 installation (266 MHz, 65 CERN units) 
     4 sec per constant-weight event are needed.
This result is for a {\em default/recommended} 
setting of the input parameters, with {\em all} hadronization/decay libraries switched ON.

\newpage

\section{Introduction}
Monte Carlo (MC) event generators have the double purpose to compensate for detector
inefficiencies and to provide theoretical predictions for distributions
and integrated cross sections.
The second task is more important and difficult.
Precision predictions of the Standard Model,
with a total error below 0.5\%, for the process  of production
of the fermion pair in electron--positron scattering ($e^-e^+\to \mu^-\mu^+$) were
first obtained for energies close to the $Z$ resonance with the MC event generator 
KORALZ~\cite{koralz4:1994}  (see also the latest version~\cite{koralz4:1999}).
The prototype of the modern MC event generator for this process was constructed 
earlier~\cite{mustraal-np:1983}, but it could not deliver sub-per cent precision, 
because
it did not include electroweak corrections and second order QED corrections
(it later became a part of the KORALZ package).
KORALZ was originally developed for a simulation of 
the $\tau$-pair production and decay, and later on was extended
to muon, quark and neutrino pairs, 
$e^-e^+\to f\bar{f}, with f=\mu,\tau,d,u,s,c,b,\nu$.
The Bhabha scattering $e^-e^+\to e^-e^+$ was never included in KORALZ 
and a dedicated precision MC event generator BHLUMI~\cite{bhlumi2:1992,bhlumi4:1996}
was developed for this process.
BHLUMI, at the expense of specializing to small scattering angles, 
could deliver at LEP1 energies
the integrated cross section with the record total precision of 0.06\%~\cite{bhlumi-precision:1998}.
It should be stressed, however,  that the high-precision level of KORALZ and BHLUMI
was achieved thanks to exclusive exponentiation~\cite{yfs1:1988} (EEX) 
based on the classical work of Yennie--Frautschi--Suura (YFS)~\cite{yfs:1961}, 
in which the multiple soft and hard
real photons are treated in a completely realistic way,  i.e. four-momenta are generated, 
and the infrared (IR) cancellations between real and virtual
soft photons occur exactly to infinite order.

At the end of LEP2 operation the total cross section for  the process $e^-e^+ \to f\bar{f}$
will have to be calculated with a precision of $0.2\%$--$1\%$, 
depending on the event selection~\cite{2f-proposal}.
The arbitrary differential distributions also have to be calculated
with the corresponding precision.
In future linear colliders (LCs) the precision requirement can be substantially stronger,
especially for the high luminosity option, as in the TESLA case.
The above new requirements necessitate the development of a new calculational framework 
for the QED corrections and the construction of new dedicated MC programs.
The present work is an important step in this direction.

The main limiting factor that prevents us from getting
more precise theoretical predictions for the $e^-e^+\to f\bar{f}$
process is higher-order QED radiative corrections
(the QED part of electroweak Standard Model).
In order to achieve the 0.2\% precision tag,
the virtual corrections have to be calculated up to 2--3 loops and the multiple bremsstrahlung
up to 2--3 hard photons, 
integrating exactly the multiphoton phase space for the arbitrary event selection
(phase-space limits).

The MC event generators KORALZ and BHLUMI, 
although representing the state of the art of MC evaluation
of QED radiative corrections for $e^-e^+\to f\bar{f}$ at
the beginning of the LEP2 run, are strongly limited in their development 
towards higher precision.
The main limitation is rooted in the use of the spin-summed differential 
cross sections in the otherwise so successful YFS/EEX --
instead of using spin amplitudes directly%
\footnote{
  This was a sensible choice; for instance, it has saved precious CPU time,
  which was a big problem in the MC calculations, a decade ago.}.
For this reason, certain interferences such as the one
between initial-state radiation (ISR) and final-state radiation (FSR) had to be neglected,
especially in the calculations beyond first order.
Also, in any processes with more Feynman diagrams, such as the Bhabha process,
EEX suffers from the proliferation of the interference terms, especially beyond first order.
EEX neglects ISR--FSR interferences. However, they were often unimportant --
at the Z peak much below 1\% of the integrated cross section.
The analogous QED interferences among photons emitted from electron and positron in
small-angle Bhabha were also small, of order $0.01\%$,
at the small-angle range of the luminosity measurement.
Further improvement on the precision for off-$Z$-peak (LEP2) and for large-angle Bhabha
scattering definitely requires reintroduction of these interferences.
This is achieved in a natural way by
reformulating the exponentiation entirely in terms of spin amplitudes.
This  turns out to be quite a non-trivial task, and
it was done only recently, see refs.~\cite{ceex1:1999,gps:1998}.
The resulting new ``reincarnation'' of the YFS exponentiation,
called the {\em coherent exclusive exponentiation} (CEEX), was born.
The CEEX scheme is implemented in  the present \KK\ MC program for the first time.
In fact  we have described in ref.~\cite{ceex1:1999} only the first \Oceex{\alpha^1} version, 
with the pure QED matrix element, while in the present program we have already implemented
the bulk of the QED \Oceex{\alpha^2} matrix element, 
and also \Order{\alpha} electroweak (EW) corrections.
This new, \Oceex{\alpha^2}, important development of CEEX, 
together with the wealth of numerical results,
will be published separately~\cite{ceex2:1999}.

\newpage
\subsection{Ultimate MC event generator for two-fermion final states}
Having briefly introduced the reader to the history and the main characteristics
of the subject of the MC event generators for the $e^-e^+\to f\bar{f}$ process,
let us come to the important question:
What is the most complete list of requirements that the precision MC event generator
for the $e^-e^+\to f\bar{f}$ process should fulfil in order to satisfy the needs
of the experiments in the present and future $e^-e^+$ colliders, 
that is for the entire centre-of-mass energy range from 
the $\tau$ production threshold up to 1 TeV?
This would cover experiments at LEP, SLC, $b$-factories, $c$-factories, $\tau$-factories,
future linear colliders such as TESLA, JLC and NLC, and also high-luminosity
experiments at the $Z$ resonance, the so-called $Z$-factories.

In the following we try to answer the above question,
and complete the list of the desired features of the ultimate MC event generator 
for two-fermion final states:
\begin{itemize}
\item
  The total {\em precision} of the integrated cross section has to be at least 0.2\%.
  From the above, it automatically follows that we need for the QED part and electroweak corrections
  the entire first-order \Order{\alpha} and the QED second-order \Order{\alpha^2}, at least
  in the leading-log (LL) approximation with the exponentiation, 
  e.g. with the \Order{\alpha^2L^2} contributions, especially for ISR.
  In fact the QED second-order subleading \Order{\alpha^2L^1} 
  and third-order leading \Order{\alpha^3L^3} are also mandatory, at least for the discussion
  of the theoretical error, but preferably present in the actual MC matrix element.
  The inclusion of $f\bar{f}f'\bar{f'}$ four-fermion final states will often be necessary,
  especially the production of the additional soft fermion pair of light fermions.
  The $Z$-factory option with $10^9$ statistics would be the most demanding experiment,
  asking for precision better than $0.02\%$!
  Note that here we do not attempt to define individual precision requirements for 
  all kinds of distributions and averages, such as charge and spin asymmetries.
  We assume that their precision should correspond to a precision of $0.2\%$-$0.02\%$ 
  for the integrated cross sections.
\item
  The other role of MC event generators is to provide a {\em realistic picture of the process}
  for the detector studies.
  It means that it is highly {\em undesirable} to integrate out certain final-state topologies.
  For instance we know that contributions from collinear real photons and from virtual photons
  combine in such a way that the net effect in the integrated cross section is zero or negligible.
  Nevertheless, for experiments it is of vital importance to have all real photons manifestly
  in the MC event, even very soft and very collinear ones: it helps enormously
  to understand correctly the operation of a detector, 
  and hence to reduce the experimental systematic errors.
  The same is true for the emission of additional light-fermion pairs.
  The above {\em maximal exclusivity} requirement leads directly to {\em exponentiation} in the 
  MC event generator,
  that is to a procedure in which the perturbative expansion is reorganized in such a way that
  contributions from IR real and virtual singularities cancel to infinite order.
  The remaining non-IR corrections are calculated order by order --
  for example in the present \KK\ MC the non-IR corrections are included in \Order{\alpha^2}.
\item
  For the initial beams and outgoing unstable $\tau$ leptons and $t$-quarks  
  {\em spin polarizations}, including {\em all spin correlations}, have to be fully taken  into account, 
  also in the presence of the real bremsstrahlung photons.
  For beams the longitudinal polarization of both beams is the minimum requirement.
  For decaying final fermions, longitudinal polarizations are often not sufficient --
  transverse polarizations are necessary, both for $\tau$ leptons and for $t$-quarks.
  In fact not only polarizations are required, but also the effects of exact spin correlation  
  between spin polarization vectors of two outgoing fermions.
  This again makes it mandatory to the use spin amplitudes and/or spin density matrices for the fermion production
  and decay processes.
\item
  Non-QED {\em electroweak corrections} to the $e^-e^+\to f\bar{f}$ process are usually calculated separately
  and combined later on with QED corrections in an manner
  that is, beyond \Order{\alpha},  an ``ad hoc'' recipe.
  It is not done at the level of the spin amplitudes, but rather in terms of the inclusive distributions,
  having integrated some photon phase space e.g. transverse momenta beforehand.
  Typically such an ad hoc procedure is based on the second- or
  third-order ``structure functions'' for the incoming electron, adding
  subleading first- and second-order corrections.
  It can be questioned whether such an approach is really justified at the precision level of $0.2\%$,
  because it is too far from the solid environment of 
  the Lagrangian, Feynman diagrams and the exact phase space.
  The Monte Carlo based on the spin amplitudes offers a natural realization of such an environment.
  It is therefore the only viable solution for the problem of combining QED and non-QED corrections
  beyond first order, without any unnecessary ambiguities.
  Such a Monte Carlo can then be used to cross-check (calibrate) the popular semi-analytical 
  approaches, which employ all kinds of ad hoc recipes.
  This important role should not be underestimated, as the semi-analytical programs 
  have many advantages of their own and will always be around.
\item
  Effects due to {\em beamstrahlung} will be present at future linear colliders and thus should be
  implemented in the MC event generator.
  The beamstrahlung structure functions should be a ``user function'', 
  supplied (or replaced) easily by the user, without any loss of the efficiency of the MC program.
  Note that the ``luminosity energy spectrum'' will be known from machine simulation only to a certain
  extent; the true distribution will have to be determined from the inspection 
  of the $e^-e^+\to f\bar{f}$ process, most probably for $f=e$, 
  at small ($\sim 1^\circ$) or intermediate ($\sim 10^\circ$) angles.
  In such a case, reliable MC predictions of the Standard Model (SM) integrated cross sections 
  and distributions in the presence of beamstrahlung,
  with the non-trivial event selection criteria, are of vital importance.
  In this context, the $Z$ radiative return, 
  i.e. the $e^-e^+\to \gamma Z$ subprocess, may also play an important role.
\item
  The MC event generator should be maximally {\em upgradable to other processes}.
  Although, in this specification, we concentrate on the precision SM prediction for the 
  two-fermion final states, the MC program should be constructed in such a way that inclusion
  of the other similar standard processes such as $e^-e^+\to W^+W^-$ 
  or non-standard  processes (supersymmetric) should be relatively easy. 
  Also the change of electron beam to muon beam should be possible.
  This requires in practice that the MC program be constructed right from the beginning
  in a {\em highly modular} way. 
  The Object-Oriented Programming (OOP) approach would help.
\item
  For quark final states, the photon emission should be well combined with the
  {\em quark--gluon parton shower}, properly taking  QED and QCD corrections at the NLL level.
\item
  Last but not least the program should also have the option to run in a {\em maximally inclusive} mode
  in which the type of the final state (fermion) is
  chosen randomly, event per event,
  exactly according to its integrated cross section.
  This requirement is not as trivial as it may seem, since the integrated cross section,
  in the MC, is
  typically known at the end of the run (from the average MC weights).
\end{itemize}
Let us note that the above specification goes far beyond the very ambitious (at the time)
specification of the ``ultimate MC'' formulated at the end of the 1989 LEP1 Workshop.
How far are we with \KK\ MC on the road to this ultimate goal?

\begin{table}[ht]
\centering
\setlength{\unitlength}{1mm}
\begin{picture}(160,110)
\put(-2, 00){\makebox(0,0)[lb]{
\epsfig{file=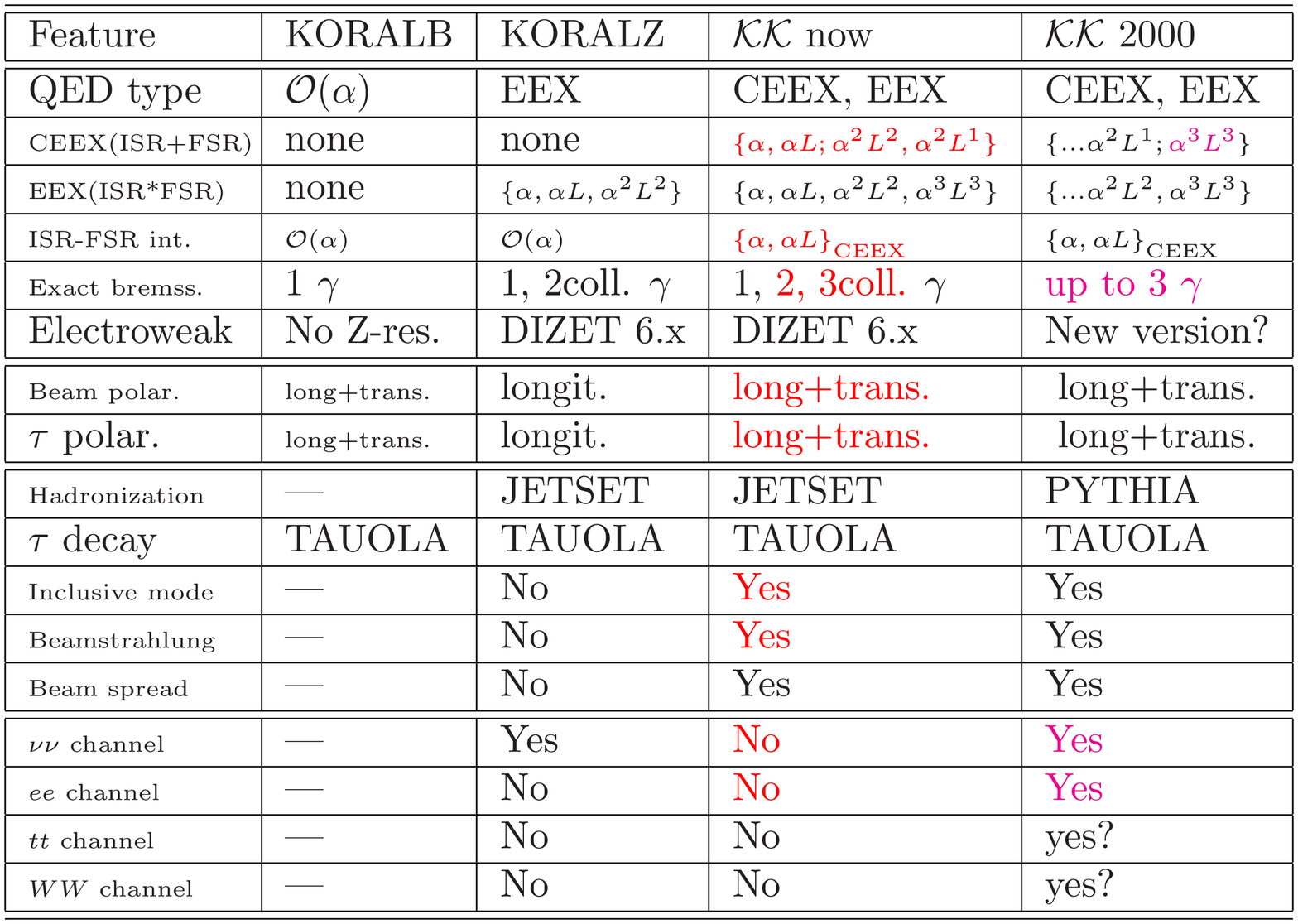,width=160mm,height=110mm}
}}
\end{picture}
\caption{\sf 
  List of features of the present \KK\ MC, compared to older MCs for fermion
  pair production, and future plans.
}
\label{tab:KK-status}
\end{table}

\subsection{How far are we on the road to the Ultimate MC?}
The present specification of the \KK\ MC is summarized in Table~\ref{tab:KK-status},
where we have also given those of KORALZ and KORALB for comparison.

As we see, the present \KK\ MC has already all the functionality of KORALB.
The \KK\ MC fulfils completely our ultimate specification for the spin treatment.
Up to now, the first and the only MC event generator fulfilling the above specification 
was KORALB\cite{koralb2:1995,koralb:1985}.
However, KORALB is limited to $\sqrt{s}<30$ GeV, because of the lack of the $Z$-resonance and
it does not include more than one photon emission --  
it is based on the pure first-order QED calculation,
without exponentiation.
In KORALZ, longitudinal polarization effects are implemented at the level of the differential
distribution for initial beams and for outgoing fermions (including longitudinal spin correlations).
Effects due to transverse spin polarizations are omitted.
In the present \KK\ program these limitations are removed -- the complete longitudinal and transverse
polarizations are implemented for beams and outgoing fermions, including all spin correlations
exactly, as in KORALB, also in the presence of multiple real photons.

As compared to KORALB, there is still one improvement to be done in the \KK\ MC:
the CEEX matrix element does not generally rely on the assumption $m_{\tau} \ll \sqrt{s}$,
but certain parts of the actually implemented virtual corrections may still rely on
this approximation, notably the spin amplitudes for the $\gamma$-$\gamma$ box.
This approximation is not really necessary and will be corrected in the future version.

As compared to KORALZ, the \KK\ MC covers all its functionality, except for the
presence of the neutrino channels.
We plan to implement the neutrino channel in the next version.
The most important new features in the present \KK\ with respect to KORALZ is the inclusion
of ISR--FSR interference, which at LEP2 modifies the total cross section
and charge asymmetry by about 2\%,
the inclusion of the second-order subleading corrections and the inclusion of the exact matrix
element for two hard photons.

The \KK\ MC is the first full-scale event generator to include the bulk of second order NLL
corrections, and it may easily incorporate second-order NNLL corrections.
In fact the exact double bremsstrahlung and exact two-loop virtual corrections are already included,
but for the moment there is no need to complete the missing NNLL corrections
since they are $\sim 10^{-5}$.

\newpage
\section{Physics content}
The present version of \KK\ MC includes first-order \Order{\alpha}
QED and electroweak corrections and almost complete \Order{\alpha^2} QED corrections
due to the emission of photons from the initial- and final-state fermions.
It does not include the emission of an additional fermion pair.

\subsection{Two types of QED matrix element}
The QED part of the calculation is based on the new \Order{\alpha^2}
calculation with coherent exclusive exponentiation (CEEX)
at the amplitude level.
The older QED matrix element based on exclusive exponentiation (EEX) 
at the differential distribution level (amplitudes squared and spin summed)
of the BHLUMI type
is still present and is used as a backup solution, for various tests.
In particular the EEX matrix element includes \Order{\alpha^3L^3} corrections,
which are still absent in our CEEX; it therefore provides a useful estimate
of these corrections.

\subsubsection{CEEX spin amplitudes}

Defining the Lorentz-invariant phase space as
\begin{equation}
\label{eq:lips}
\int d{\rm LIPS}_n(P;p_1,p_2,...,p_n) 
    = \int (2\pi)^4\delta^{(4)}\bigl(P-\sum_{i=1}^n p_i\bigr) \prod_{i=1}^n \frac{ d p^3}{(2\pi)^3 2p^0_i},
\end{equation}
we write the \Order{\alpha^r} CEEX total cross section for the process
\begin{equation}
e^-(p_a) +e^+(p_b) 
  \to f(p_c) +\bar{f}(p_d) +\gamma(k_1) +\gamma(k_2)+...+\gamma(k_n), n=0,1,2,...,n,
\end{equation}
with polarized beams
and decays of unstable final fermions sensitive to fermion spin polarizations,
following refs.~\cite{ceex1:1999,ceex2:1999}, as follows:
\begin{equation}
  \label{eq:sigma-ceex2}
  \sigma^{(r)} =  {1\over {\rm flux}(s)}
  \sum_{n=0}^\infty 
  \int d{\rm LIPS}_{n+2} ( p_a+p_b; p_c,p_d, k_1,\dots,k_n)\;
  \rho^{(r)}_{\rm CEEX}  ( p_a,p_b, p_c,p_d, k_1,\dots,k_n),
\end{equation}
where
\begin{equation}
  \label{eq:rho-ceex2}
  \begin{split}
  \rho^{(r)}_{\rm CEEX} &( p_a,p_b, p_c,p_d, k_1,k_2,\dots,k_n)=
  {1\over n!} e^{Y(\Omega;p_a,...,p_d)}\;\bar{\Theta}(\Omega)\;
    \sum_{\sigma_1,\dots,\sigma_n =\mp 1}\;
    \sum_{\lambda_A,\bar{\lambda}_A=\mp 1}\;
\\&
    \sum_{i,j,l,m=0}^3\;
        \hat{\varepsilon}^i_1                  \hat{\varepsilon}^j_2\;
        \sigma^i_{\lambda_a \bar{\lambda}_a}   \sigma^j_{\lambda_b \bar{\lambda}_b}
    \Mmf^{(r)}_n 
    \left(  \st^{p}_{\lambda}
            \st^{k_1}_{\sigma_1} 
            \st^{k_2}_{\sigma_2}
            \dots
            \st^{k_n}_{\sigma_n}
    \right)
    \left[
    \Mmf^{(r)}_n 
    \left(  \st^{p}_{\bar{\lambda}}
            \st^{k_1}_{\sigma_1} 
            \st^{k_2}_{\sigma_2}
            \dots
            \st^{k_n}_{\sigma_n}
    \right)
    \right]^\star
        \sigma^l_{\bar{\lambda}_c \lambda_c }   \sigma^m_{\bar{\lambda}_d \lambda_d }
        \hat{h}^l_3                             \hat{h}^m_4,
  \end{split}
\end{equation}
where, in order to shorten this and other formulas, we use a compact collective notations
\begin{displaymath}
    \left(  \st^{p}_{\lambda} \right) =
    \left(  \st^{p_a}_{\lambda_a} \st^{p_b}_{\lambda_b}
            \st^{p_c}_{\lambda_c} \st^{p_d}_{\lambda_d} \right)
\end{displaymath}
for fermion four-momenta $p_A$ and helicities $\lambda_A,A=a,b,c,d$,

For $k=1,2,3$,  $\sigma^k$ are Pauli matrices and
$\sigma^0_{\lambda,\mu} = \delta_{\lambda,\mu}$ is the unit matrix.
The components $\hat{\varepsilon}^j_1, \hat{\varepsilon}^k_2, j,k=1,2,3$, 
are the components of the conventional spin polarization vectors 
of $e^-$ and $e^+$ respectively, defined in the so-called GPS fermion rest frames
(see ref.~\cite{gps:1998} for the exact definition of these frames).
We define $\hat{\varepsilon}^0_A=1$
in a non-standard way  (i.e. $p_A\cdot \hat{\varepsilon}_A=m_e, A=a,b$).
The {\em polarimeter} vectors $\hat{h}_C$ are similarly defined
in the appropriate GPS rest frames of the final unstable fermions  ($p_C\cdot \hat{h}_C=m_f, C=c,d$).
Note that, in general, $\hat{h}_C$ may
depend in a non-trivial way on the momenta of all decay products,
see refs.~\cite{jadach-was:1984,tauola2.4:1993,jadach:1985,gps:1998} for details.
We did not introduce polarimeter vectors for bremsstrahlung photons,
i.e. we take advantage of the fact that all high-energy experiments are completely blind
to photon spin polarizations. See next subsection for more on spin effects.

WE define the complete set of spin amplitudes
for emission of $n$ photons in \Oceex{\alpha^r}, $r=0,1,2$ as follows:
\begin{equation}
  \label{M-expon2}
  \begin{align}
   \Mmf^{(0)}_n\left( \st^{p}_{\lambda} \st^{k_1}_{\sigma_1}
                                   \dots \st^{k_n}_{\sigma_n}  \right)
   &=\sum_{\{\wp\}}\;  \prod_{i=1}^n \; \sfac^{\{\wp_i\}}_{[i]}\;
    \beta^{(0)}_0 \left( \st^{p}_{\lambda}; X_\wp \right),
\\
   \Mmf^{(1)}_n\left( \st^{p}_{\lambda} \st^{k_1}_{\sigma_1}
                                   \dots \st^{k_n}_{\sigma_n}  \right)
   &=\sum_{\{\wp\}}\;  \prod_{i=1}^n \; \sfac^{\{\wp_i\}}_{[i]}\;
    \left\{ \beta^{(1)}_0 \left( \st^{p}_{\lambda}; X_\wp \right)
            +\sum_{j=1}^n 
             {\beta^{(1)}_{1\{\wp_j\}} \left( \st^{p}_{\lambda} \st^{k_j}_{\sigma_j} ; X_\wp \right)
                                          \over \sfac^{\{\wp_j\}}_{[j]} }\;
    \right\},
\\ \nonumber
   \Mmf^{(2)}_n\left( \st^{p}_{\lambda} \st^{k_1}_{\sigma_1}
                                   \dots \st^{k_n}_{\sigma_n}  \right)
   &=
\\
    =\sum_{\{\wp\}}\;  \prod_{i=1}^n \; \sfac^{\{\wp_i\}}_{[i]}\;
   &\left\{  \beta^{(2)}_0 \left( \st^{p}_{\lambda}; X_\wp \right)
            +\sum_{j=1}^n 
             {\beta^{(2)}_{1\{\wp_j\}} \left( \st^{p}_{\lambda} \st^{k_j}_{\sigma_j} ; X_\wp \right)
                                          \over \sfac^{\{\wp_j\}}_{[j]} }\;
             +\sum_{1\leq j<l\leq n}\;\!\!\!\!
              {\beta^{(2)}_{2\{\wp_j\wp_l\}}
                   \left( \st^{p}_{\lambda} \st^{k_j}_{\sigma_j}  \st^{k_l}_{\sigma_l} ;X_\wp \right)
                               \over \sfac^{\{\wp_j\}}_{[j]} \sfac^{\{\wp_l\}}_{[l]} }\;
    \right\}.
  \end{align}
\end{equation}
The {\em coherent} sum is taken over set $\{ \wp \}$ of all $2^n$ partitions --
the partition $\wp$ is defined as a vector $(\wp_1,\wp_2,\dots, \wp_n)$;
$\wp_i=1$ for an ISR and  $\wp_i=0$ for an FSR photon;
see the analogous construction in refs.~\cite{greco:1975,greco:1980}.
The set of all partitions is explicitly the following:
\begin{displaymath}
\{ \wp \} = \{
(0,0,0,\dots,0),\; (1,0,0,\dots,0),\;
(0,1,0,\dots,0),\; (1,1,0,\dots,0),\; \dots
(1,1,1,\dots,1) \}.
\end{displaymath}
The $s$-channel four-momentum in the (possibly) resonant $s$-channel  propagator is
$X_\wp = p_a+p_b -\sum_{i=1}^n   \wp_i\; k_i.$

At \Order{\alpha^r} we have to provide
functions $\beta^{(r)}_{k}, k=0,1,...,r$, from Feynman diagrams, which are
infrared-finite by construction~\cite{yfs:1961}.
Their actual precise definitions can be found in refs.~\cite{ceex1:1999,ceex2:1999}.
Here we shall define only the most essential ingredients.
The lowest-order $\beta^{(0)}_{0}$ are just Born spin amplitudes times a certain kinematical factor
\begin{equation}
\beta^{(0)}_{0}\left(\st^{p}_{\lambda} ; X \right)
       = \Bmf \left( \st^{p}_{\lambda}; X \right)\;\;
           { X^2  \over (p_c+p_d)^2 }.
\end{equation}
The Born spin amplitudes $\Bmf \left( \st^{p}_{\lambda}; X \right)$
and other spin amplitudes are calculated using the spinor technique of Kleiss and Stirling (KS)
\cite{kleiss-stirling:1985,kleiss-stirling:1986} reformulated in ref.~\cite{gps:1998} (GPS).
Soft factors $\sfac^{(\omega)}_{[i]},\omega=0,1$, are complex numbers, 
see ref.~\cite{gps:1998} for exact definitions;
here we only need to know their absolute values
\begin{equation}
\left|\sfac^{(1)}_{[i]}\right|^2 = 
            -\frac{e^2Q_e^2}{2} \Bigg( {p_a\over k_i p_a} - {p_b\over k_i p_b}  \Bigg)^2,\quad
\left|\sfac^{(0)}_{[i]}\right|^2 = 
            -\frac{e^2Q_f^2}{2} \Bigg( {p_c\over k_i p_c} - {p_d\over k_i p_d}  \Bigg)^2.
\end{equation}

The factor $\bar{\Theta}(\Omega)$ defines the infrared (IR) integration limits for real photons.
More precisely for a single photon, complementary  domain $\Omega$ includes 
the IR divergence point $k=0$,
which is {\em excluded} from the MC phase space, 
we define a characteristic function $\Theta(\Omega,k)=1$ for $k\in\Omega$ 
and $\Theta(\Omega,k)=0$ for $k\not\in \Omega$.
The characteristic function for the phase space included in the integration is
$\bar{\Theta}(\Omega,k) = 1-\Theta(\Omega,k)$.
The characteristic function for {\em all} photons in the MC phase space is
\begin{equation}
\bar{\Theta}(\Omega) = \prod_{i=1}^n \bar{\Theta}(\Omega,k_i).
\end{equation}
In the present program we opt for an $\Omega$ traditionally defined by the photon energy cut
condition $k^0<E_{\min}$.
Consequently, the YFS form factor~\cite{yfs:1961} reads
\begin{equation}
  \label{eq:Y-function}
  \begin{split}
   &Y(\Omega;p_a,...,p_d)
  =   Q_e^2   Y_\Omega(p_a,p_b)  +Q_f^2   Y_\Omega(p_c,p_d)\\
&\qquad\qquad
     +Q_e Q_f Y_\Omega(p_a,p_c)  +Q_e Q_f Y_\Omega(p_b,p_d) 
     -Q_e Q_f Y_\Omega(p_a,p_d)  -Q_e Q_f Y_\Omega(p_b,p_c),
  \end{split}
\end{equation}
where
\begin{equation}
\label{form-factor}
\begin{split}
Y_\Omega(p_1,p_2) 
     \equiv &\;  2 \alpha \tilde{B}(\Omega,p_1,p_2)   +2 \alpha \Re B(p_1,p_2) \\
     \equiv &   -2 \alpha\;{ 1 \over 8\pi^2} \int {d^3k\over k^0} \Theta(\Omega;k)
                       \bigg({p_1\over kp_1} - {p_2\over kp_2} \bigg)^2 \\
            &   +2 \alpha \Re \int {d^4k\over k^2} {i\over (2\pi)^3} 
                       \bigg( {2p_1+k \over 2kp_1+k^2} -{2p_2-k \over 2kp_2-k^2} \bigg)^2
\end{split}
\end{equation}
is given analytically in terms of logarithms and Spence functions~\cite{ceex2:1999}.
As we see, the above YFS form factor includes terms due to the initial--final state interference (IFI).
In the MC with exponentiation it would be possible to do without $\Omega$ (declare it empty)
and rely uniquely on the IR regularization with small photon mass $m_\gamma$ only~\cite{ceex1:1999}.
In that case formulas~(\ref{form-factor}) for the YFS form factor
would include only the second virtual photon integral.

\subsubsection{EEX differential distributions}
In the case of exclusive exponentiation (EEX), we deal with a spin-summed differential distribution.
We neglect the IFI and each photon is therefore attributed either
to the initial or to the final state fermion pair.
The initial state photon momenta are denoted by $k_i,i=1,2,...,n$, 
and final state momenta as $k'_i,i=1,2,...,n'$.
We shall consider the process
$e^-(p_1)+e^+(p_2) \rightarrow f(q_1)+\bar{f}(q_2)+n(\gamma(k_j)) +n'(\gamma(k_l))$.

In the present description of the EEX matrix element and in the following sections on
the Monte Carlo algorithm, we use an {\em alternative notation for the fermion momenta}:
\begin{equation}
  \begin{split}
    &p_1 \equiv p_a,\qquad    q_1 \equiv p_c\\
    &p_2 \equiv p_b,\qquad    q_2 \equiv p_d.
  \end{split}
\end{equation}
This is because such a notation has already been used for a long time in low-level MC programs
and in the EEX distributions;
we have therefore decided to keep it also in the relevant parts of the paper.
We hope that it will help the user to understand the code more easily.
On the other hand the notation with $p_A,A=a,b,c,d$, and $k_i,i=1,...,n$, is very handy
in building up a very compact tensor notation,
with letters for fermion indices and numerals for photon indices,
in the tensor notation for the multiple photon spin amplitudes;
see ref.~\cite{ceex2:1999}, the previous section, and the relevant parts of the program.

Denoting the Lorentz invariant phase space%
\footnote{This phase space has a slightly different normalization from that of the previous section.},
\begin{equation}
\label{eq:dtau}
 d\tau_n(P;p_1,p_2,...,p_n)=
   \prod_{j=1}^n  {d^3 p_j\over 2p^0_j} \;
   \delta^{(4)}\bigg( P -\sum_{j=1}^n p_j \bigg),
\end{equation}
the \Order{\alpha^r} EEX total cross-section
\begin{equation}
\label{sigma-eex3}
\sigma^{(r)}_{EEX} =
  \sum_{n=0}^\infty \sum_{n'=0}^\infty {1\over n!}{1\over n'!}
  \int d\tau_{n+n'+2}( p_1+p_2; q_1,q_2, k_1,...,k_n, k'_1,...,k'_n)\;
  \rho^{(r)}_{EEX},\quad r=0,1,2,3,
\end{equation}
is expressed in terms of the fully differential multiphoton distribution
\begin{equation}
\label{eq:rho-eex3}
\begin{split}
 \rho&^{(r)}_{EEX}(p_1,p_2, q_1,q_2, k_1,...,k_n, k'_1,...,k'_n) =
  e^{  Q_e^2 Y_{\Omega_I}(p_1,p_2) +Q_f^2 Y_{\Omega_F}(q_1,q_2) }\\
& \prod_{j=1}^n     2\tilde{S}_I(k_j)\;  \bar{\Theta}(\Omega_I;k_j)
  \prod_{l=1}^{n'}  2\tilde{S}_F(k'_l)\; \bar{\Theta}(\Omega_F;k'_l)
  \Bigg\{
  \bbeta^{(r)}_0(X,p_1,p_2,q_1,q_2) \\
& +\sum_{j=1}^n    \frac{\bbeta^{(2)}_{1I}(X,p_1,p_2,q_1,q_2,k_j) }{ \tilde{S}_I(k_j) }
  +\sum_{l=1}^{n'} \frac{\bbeta^{(2)}_{1F}(X,p_1,p_2,q_1,q_2,k_l) }{ \tilde{S}_F(k_l) }\\
& +\sum_{n \geq j>k \geq 1}\;\;
       \frac{\bbeta^{(2)}_{2II}(X,p_1,p_2,q_1,q_2,k_j,k_k) }{ \tilde{S}_I(k_j)\tilde{S}_I(k_k)} 
  +\sum_{n' \geq l>m \geq 1}\;\;
       \frac{\bbeta^{(2)}_{2FF}(X,p_1,p_2,q_1,q_2,k_l,k_m) }{  \tilde{S}_F(k_l)\tilde{S}_F(k_m)}\\
& +\sum_{j=1}^n \sum_{l=1}^{n'}
       \frac{\bbeta^{(2)}_{2IF}(X,p_1,p_2,q_1,q_2,k_j,k_l) }{ \tilde{S}_I(k_j)\tilde{S}_F(k_l)}
  +\sum_{n \geq j>k >l \geq 1}\;\;
       \frac{\bbeta^{(3)}_{3III}(X,p_1,p_2,q_1,q_2,k_j,k_k,k_l) }
             { \tilde{S}_I(k_j) \tilde{S}_I(k_k) \tilde{S}_I(k_l) } 
  \Bigg\}.
\end{split}
\end{equation}
The YFS soft factors for real photons emitted from the initial- and final-state fermions read
\begin{equation}
  \tilde{S}_I(k_j) = -Q_e^2 {\alpha \over 4\pi^2}
                     \Bigg( {p_1\over k_jp_1} -{p_2\over k_jp_2} \Bigg)^2,\qquad
  \tilde{S}_F(k_l)  = - Q_f^2 {\alpha \over 4\pi^2}
                     \Bigg( {q_1\over k_lq_1} -{q_2\over k_lq_2} \Bigg)^2.
\end{equation}
The IR domains $\Omega_I$ and $\Omega_F$ 
are different for ISR and for FSR -- this is easier to implement in the MC.
We define $\Omega_I$ with the condition $k^0<E_{\min}\ll\sqrt{s}$ 
in the rest frame of $P=p_1+p_2$ called the CMS frame,
and $\Omega_F$ with ${k'}^0<E'_{\min}\ll 2m_f$ in the rest frame of $Q=q_1+q_2$
referrred to as QMS.
The rest frame of $X=P -\sum_{j=1,n} k_j$ will be referred to as the XMS frame.
In EEX we generally use small mass approximation $m_e,m_f,\ll \sqrt{s}$
(in the following sections we shall however use finite $m_f$ wherever necessary).
In this case the YFS form factors of eqs.~(\ref{form-factor}) for the
above $\Omega$'s are very simple:
\begin{equation}
\begin{split}
 &Y_e(\Omega_I;p_1,p_2)
  =   \gamma_e \ln {2E_{min}\over \sqrt{(p_1+p_2)^2}}     +{1\over 4}\gamma_e
      +Q_e^2 {\alpha\over\pi} \bigg( -{1\over 2} +{\pi^2\over 3}\bigg),\\
 &Y_f(\Omega_F;q_1,q_2)
  =   \gamma_f \ln {2E_{min}\over \sqrt{(q_1+q_2)^2}}     +{1\over 4}\gamma_f
      +Q_f^2 {\alpha\over\pi} \bigg( -{1\over 2} +{\pi^2\over 3}\bigg),\\
 & \gamma =\gamma_e  = 2 Q_e^2 {\alpha\over \pi} \bigg( \ln {2p_1p_2\over m_e^2} -1 \bigg),\quad
   \gamma_f          = 2 Q_f^2 {\alpha\over \pi} \bigg( \ln {2q_1q_2\over m_f^2} -1 \bigg).
\end{split}
\end{equation}
Again the IR-finite functions $\bbeta^{(r)}_k,k=0,1,...,r$
(defined at the level of squared spin-summed spin amplitudes)
are calculated, in principle, from the Feynman diagrams and they are defined 
completely in ref.~\cite{ceex2:1999}%
\footnote{
  In reality they are constructed there by combining in a clever way
  first-order results and results of the triple convolution of the Altarelli--Parisi kernels.
  This solution has several practical advantages.}.

Here, let us only give the explicit definition of lowest order
\begin{equation}
    \label{beta00}
    \bbeta_0^{(0)}(X,p_1,p_2,q_1,q_2)=
        \frac{1}{4}\sum_{k,l=1,2} {d\sigma^\born \over d\Omega}(X^2,\vartheta_{kl})
\end{equation}
where
\begin{equation}
  \vartheta_{11}= \angle( \vec{p}_1, \vec{q}_1),\;\;
  \vartheta_{12}= \angle( \vec{p}_1,-\vec{q}_2),\;\;
  \vartheta_{21}= \angle(-\vec{p}_2, \vec{q}_1),\;\;
  \vartheta_{22}= \angle(-\vec{p}_2,-\vec{q}_2),\;
\end{equation}
and where all three-vectors are taken
in the rest frame of the four-momentum $X = p_1+p_2-\sum_{j=1}^n k_j$, called the XMS.
The above $\bbeta_0^{(0)}$ function is proportional to the Born
differential cross section $d\sigma^\born(s,\vartheta)/d\Omega$,
for $e^+e^-\rightarrow f\bar{f}$.

\subsection{Spin effects}
Spin effects are treated in the most general way.
For initial beams we use as an input the full spin polarization vectors and for
the outgoing $\tau$ we contract the $\tau$-production spin amplitudes
with the polarimeter vectors of the two decaying $\tau$'s 
defined in exactly the same rest frames of the $\tau$'s where the spin quantization
axes of the  $\tau$'s were defined in the first place.
To find out exactly these frames was a little bit of an exercise, because for the 
calculation of the  spin amplitudes~\cite{ceex1:1999} for the $e^-e^+\to f\bar{f} n(\gamma)$
process we use the Kleiss--Stirling technique, for which these spin quantization frames
have been found in ref.~\cite{gps:1998}.
We call them the GPS frames of the fermions (beam $e^\pm$ and $\tau^\pm$).
At the practical level, we have written a subprogram that performs the Lorentz transformations
from the four GPS frames to the CMS frame.
This subprogram is used to transform all $\tau$ decay products from the GPS $\tau$ rest frames
to CMS frame.
The polarization vectors of the incoming $e^\pm$ have also to be provided in the GPS frames of  $e^\pm$. It  is simple to obtain by consecutive  Lorentz transformation from the experimentally defined 
frames  of  $e^\pm$ to CMS frame and later to GPS frames. 
It amounts to making the three-dimensional Wigner--Wick rotation of the beam polarization vectors from
the experimentally defined frame to the GPS frame 
(in practice it is a rotation around the beams, due to the smallness of the electron mass).
All of the above technique is an extension of the methodology used in KORALB~\cite{koralb2:1995}.
For more details, see refs.~\cite{ceex1:1999,gps:1998}.

Let us finally touch briefly upon another very serious problem 
relevant to implementation of spin effects and its solution.
In eq.~(\ref{eq:sigma-ceex2})  the single spin amplitude $\Mmf^{(1)}_n$ already contains
$2^n(n+1)$ terms (due to $2^n$ ISR--FSR partitions).
The grand sum over spins in eq.~(\ref{eq:sigma-ceex2}) counts
$2^n 4^4 4^4 = 2^{n+16}$ terms! 
Altogether we expect up to $N\sim n2^{2n+16}$ operations
in the CPU time expensive complex (16 bytes) arithmetics.
Typically in $e^-e^+\to \mu^-\mu^+$ the average photon multiplicity with $k^0>1$ MeV
is about 3, corresponding to $N\sim 10^7$ terms.
In a sample of $10^4$ MC events there will be a couple of events with 
$n=10$ and $N=10^{12}$ terms, clearly something that would ``choke''
completely any modern, fast workstation.
There are several simple tricks that help to ease the problem;
for instance, objects such as
$\sum_a \hat{\varepsilon}^a_i \sigma^a_{\lambda\bar{\lambda}}$ and the $\sfac$-factors
are evaluated only once and stored for multiple use.
However, this is not sufficient. 
What really helps to substantially speed up the numerical calculation
in the Monte Carlo program is the following trick of {\em photon spin randomization}.
Instead of evaluating
the sum over photon spins $\sigma_i,\;i=1,...,n$,  in eq.~(\ref{eq:sigma-ceex2}),
we generate randomly one spin sequence of $(\sigma_1,...,\sigma_n)$ per MC event
and the MC weight is calculated only for this particular spin sequence!
In this way we save one hefty $2^n$ factor in the calculation time%
\footnote{The other $2^n$ factor due to coherent summation over partitions cannot
  be eliminated, unless we give up on narrow resonances.}.
Mathematically this method is correct, i.e.
the resulting cross section and all MC distributions
will be the same as if we had used in the MC weight 
the original  eq.~(\ref{eq:sigma-ceex2})
(see a formal proof of the above statement in Section~4 of ref.~\cite{mcguide:1999}).
Let us stress again that it is possible to apply this photon spin randomization
trick because
(a) the typical high-energy experiment is blind to photon spin polarization, so that
we did not need to introduce  
the polarimeter vectors for photons in eq.~(\ref{eq:sigma-ceex2}), and
(b) for our (circular) choice of photon spin polarizations 
the cross section is rather weakly sensitive to them,
so the method does not lead to any significant loss in the MC efficiency.

\subsection{Electroweak corrections}
Electroweak (EW) corrections are implemented using the DIZET library~\cite{dizet:1989},
more precisely using its most recent version, 6.21,
the same as is used in the ZFITTER~\cite{zfitter6:1999} semi-analytical program%
\footnote{We would like to thank the authors of DIZET for help in implementing DIZET in 
the \KK\ MC.}.
The method we use to implement EW corrections is rather simillar to that 
in KORALZ~\cite{koralz4:1994}:
they enter through modified coupling constants in the Born cross section, and in order
to save CPU time, correction factors are stored in the look-up tables; 
see the relevant section later.

In the absence of EW corrections,
coupling of two neutral bosons $\gamma$ and $Z$ are defined in a conventional way:
\begin{equation}
  \begin{split}
    &G^{Z,f}_{\lambda}     = g^{Z,f}_{V} -\lambda g^{Z,f}_{A},\;\;\;
     G^{\gamma,f}_{\lambda}= g^{Z,f}_{V},\; \lambda=+,-=R,L,
\\
    &g^{\gamma,e}_{V} =Q_e=1,\; g^{\gamma,f}_{V} =Q_f,\; 
     g^{\gamma,e}_{A} =0,\;     g^{\gamma,f}_{A} =0,\;
\\
    &g^{Z,e}_{V}  = {2 T^3_e -4 Q_e \sin^2\theta_W \over 4 \sin\theta_W \cos\theta_W},\;
     g^{Z,f}_{V}  = {2 T^3_f -4 Q_f \sin^2\theta_W \over 4 \sin\theta_W \cos\theta_W},\;
\\
    &g^{Z,e}_{A}  = {2 T^3_e                        \over 4 \sin\theta_W \cos\theta_W},\;
     g^{Z,f}_{A}  = {2 T^3_f                        \over 4 \sin\theta_W \cos\theta_W},\;
  \end{split}
\end{equation}
where $T^3_f$ is the isospin of the left-handed component of the fermion
($T^3_d=-1/2,\; T^3_e=-1/2$).

As in KORALZ~\cite{koralz4:1994}, we implement the EW corrections
as follows:
the $\gamma$ and $Z$ propagators are multiplied
by the corresponding two functions (scalar formfactors due to vacuum polarizations):
\begin{equation}
         H_{\gamma} = {1   \over 2-\Pi_\gamma},\;\;
         H_{Z} =
                16\sin^2\theta_W \cos^2\theta_W \;
                {G_{\mu} M_Z^2 \over \alpha_{_{\rm QED}}  8\pi \sqrt{2}} \;
                \rho_{\rm EW}.
\end{equation}
In addition the vector couplings of $Z$ get multiplied by extra form factors.
First of all we replace
\begin{equation}
  \begin{split}
    &g^{Z,e}_{V}  = {2 T^3_e -4 Q_e \sin^2\theta_W   \over 4 \sin\theta_W \cos\theta_W}
     \;\;{\rm by}\;\; {2 T^3_e -4 Q_e \sin^2\theta_W F^{e}_{EW}(s) \over 4 \sin\theta_W \cos\theta_W}
\\
    &g^{Z,f}_{V}  = {2 T^3_f -4 Q_f \sin^2\theta_W   \over 4 \sin\theta_W \cos\theta_W}
     \;\;{\rm by}\;\; {2 T^3_f -4 Q_f \sin^2\theta_W F^{f}_{EW}(s) \over 4 \sin\theta_W \cos\theta_W}
  \end{split}
\end{equation}
where  $F^{e}_{EW}(s)$ and $F^{f}_{EW}(s)$ are electroweak form factors provided
by the DIZET program and they correspond to electroweak vertex corrections.

The electroweak box diagrams require a more complicated treatment.
In the Born spin amplitudes we have essentially two products of the coupling constants:
\begin{equation}
  \begin{split}
    &G^{Z,e}_{\lambda} G^{Z,f}_{-\lambda}
      =(g^{Z,e}_{V} -\lambda g^{Z,e}_{A})(g^{Z,f}_{V} +\lambda g^{Z,f}_{A})
      =         g^{Z,e}_{V} g^{Z,f}_{V}   -\lambda g^{Z,e}_{A} g^{Z,f}_{V} 
       +\lambda g^{Z,e}_{V} g^{Z,f}_{A}           -g^{Z,e}_{A} g^{Z,f}_{A},
\\
    &G^{Z,e}_{\lambda} G^{Z,f}_{\lambda}
      =(g^{Z,e}_{V} -\lambda g^{Z,e}_{A})(g^{Z,f}_{V} -\lambda g^{Z,f}_{A})
      =         g^{Z,e}_{V} g^{Z,f}_{V}   -\lambda g^{Z,e}_{A} g^{Z,f}_{V} 
       -\lambda g^{Z,e}_{V} g^{Z,f}_{A}           +g^{Z,e}_{A} g^{Z,f}_{A}.
  \end{split}
\end{equation}
In the above, the following replacement is done for the doubly-vector component
\begin{equation}
g^{Z,e}_{V} g^{Z,f}_{V} \Longrightarrow
    {  4 T^3_e T^3_f  
      -8 T^3_e Q_f  F^{f}_{EW}(s)   
      -8 T^3_f Q_e  F^{e}_{EW}(s)  
      +16  Q_f Q_f  F^{ef}_{EW}(s,t)
      \over 16\sin^2\theta_W \cos^2\theta_W },
\end{equation}
where the new form factor $F^{ef}_{EW}(s,t)$ corresponds to electroweak
boxes and is angle dependent.

In CEEX the Born spin amplitudes modified in the above way enter everywhere as building block, 
also in the spin amplitudes for the single and multiple real photons.
In EEX the Born spin amplitudes with the modified couplings enter into the Born differential
cross section, which is the basic building block in all EEX differential distributions. 
In both cases it thus constitute an improvement
with respect to KORALZ, electroweak corrections are calculated
for every occurrence of Born-like building block of the amplitude
and not only once per event. Note also that we do not need
to discuss final-state mass terms at this point.

\subsection{Beamstrahlung}
Beamstrahlung is the effect of loss of the beam energy due to the electron--bunch interaction.
It is essentially the emission of 
two effective photons strictly collinear with the beams,
as an additional strictly collinear ISR.

{\em 
In the MC program, we require that
the 2-dimensional beamstrahlung structure function be the ``user function'', 
supplied or easily replaced by the user, without any loss of efficiency of the MC program.}

The most general form of the modification of the differential distributions 
due to beamstrahlung in $e^-(p_a)+e^+(p_b)\to X$ reads as follows:
\begin{equation}
  \label{eq:beamstrahlung}
  d\sigma^{bst}(p_a,p_b;X) = \int dz_1 dz_2 {\cal D}(z_1,z_2,\sqrt{s}) d\sigma(z_1p_a,z_2p_b;X),
\end{equation}
where ${\cal D}(z_1,z_2)$ is the double differential beamstrahlung
``luminosity spectrum'' for simultaneous beamstrahlung from both beams.
The luminosity spectrum is assumed to be in the most general form
\begin{equation}
  {\cal D}(z_1,z_2) = 
   \delta(1-z_1)\delta(1-z_2) \rho_0
  +\delta(1-z_1) \rho_1(z_2)
  +\delta(1-z_2) \rho_1(z_1)
  +\rho_2(z_1,z_2),
\end{equation}
where $\rho_0$ is a constant and $\rho_1(z)$ and $\rho_2(z_1,z_2)$
are analytical functions for $z\in [0,1)$.
More precisely we allow for power-like integrable singularities at $z_i=1$:
\begin{equation}
  \label{eq:general}
  \rho_1(1-\epsilon) \sim \epsilon^{\alpha},\quad 
  \rho_2(1-\epsilon,z_2) \sim \epsilon^{\beta},\quad 
  \rho_2(z_1,1-\epsilon) \sim \epsilon^{\beta}.
\end{equation}
The functions $\rho_i$ are regarded as completely arbitrary and we require
that the \KK\ MC be able to cope
efficiently with any beamstrahlung luminosity spectrum of the above most general type.

In the present version of the program,  we include the
beamstrahlung spectra as implemented in the CIRCE package of T. Ohl \cite{circe:1996}.
The CIRCE package is based on the results from
the machine simulation program  {\tt GuineaPig} for linear colliders \cite{guineapig}.
Generation of the $z_i$ variables
is done at the very beginning of the generation, together with the emission of the QED
ISR total photon energy.
The technical problem to be solved is hence the following: 
How to generate up to three variables, the two variables $z_1, z_2$ for beamstrahlung 
and one $z=1-v$ for ISR, 
according to an arbitrary, highly singular probability density distribution?
Not only are the beamstrahlung spectrum and ISR photon distribution strongly singular, but
in addition the MC algorithm has to deal efficiently with singularities due to
resonances and thresholds in the reduced CMS energy variable $s''=s z_1 z_2 z$.

Furthermore, owing to the presence of the $\delta$'s
in the general beamstrahlung distribution of eq.~(\ref{eq:general}),
four branches in the MC generation are present: one 1-dimensional, two 2-dimensional
and one 3-dimensional, corresponding to each term in  eq.~(\ref{eq:general}).
In each branch the corresponding subset of the $z_1,z_2$ and $z$ variables is generated,
using the special general-purpose MC generation program Foam~\cite{foam:1999},
developed recently for exactly this type of problem.
In the present version, we also include another solution for the above task based
on the classical Vegas program~\cite{lepage:1978},
customized to our needs.
This second solution based on Vegas is, however, much less efficient 
than the principal one (Vegas will probably be removed in the future version).
Vegas is extremely efficient in calculating the value of an integral using the MC method, but
our problem is different: the
variables $z_1,z_2$ and $z$ are generated at the beginning of the whole \KK\ MC algorithm;
consequently they have to be generated with the weight equal 1
(otherwise we would waste CPU time for calculating for instance the complicated QED matrix element
for ``unworthy'' weighted events).
To this task, Vegas is not very well suited (see a more detailed
discussion in ref.~\cite{foam:1999}).
The new MC tool Foam is specialized for exactly this difficult
task of generating {\em unweighted events} efficiently,
according to an arbitrary multidimensional probability distribution.

Even if we use powerful MC tool such as Foam, it is worthwhile to improve the efficiency of the
MC generation by the appropriate change of integration variables (mapping).
Let us take the most complicated case with $\int dz_1 dz_2 dz$
where $z=s'/s$, and $s'$ is ``after ISR''.
By the trial and error method, we have found that the best mapping for $\sqrt{s}<$ 1 TeV is:
\begin{equation}
x_1=1-z_1,\;\; x_2=1-z_2,\;\; v=1-z
 x_1  = r_1^{1/\alpha},\;\; 
 x_2  = r_2^{1/\alpha},\;\;
 v    = r_3^{1/\gamma} v_{\max},
\end{equation}
where $0<r_i<1$ are used by Foam, $\alpha$ is from CIRCE
and, for the ISR, we have $\gamma\sim 2(\alpha_{QED}/\pi)\ln(s/m_e^2)$.
The integral gets transformed into:
\begin{equation}
\begin{split}
\int dz_1 dz_2 dz 
 &F(z_1,z_2, s z_1 z_2 z) \Theta(1-z_1 z_2 z -v_{\max})\\
 &= \int_0^1 dr_1 dr_2 dr_3\; 
   {v v_{\max}  \over r_3 \gamma} \; 
   {x_1 \over r_1 \alpha}\;{x_2 \over r_2 \alpha} \Theta(1-z_1 z_2 z -v_{\max}) 
   F(z_1,z_2, s z_1 z_2 z) \\
 &= \int_0^1 dr_1 dr_2 dr_3\;
   {v_{\max} \Theta(1-z_1 z_2 z -v_{\max})
    \over (\alpha x_1^{\alpha-1}) (\alpha x_2^{\alpha-1}) (\gamma v^{\gamma-1})} 
   F(z_1,z_2, s z_1 z_2 z)
\end{split}
\end{equation}
Note that this mapping only takes  care%
\footnote{
  Surprisingly, a more sophisticated mapping gives worse efficiency
  for both Foam and Vegas!}
of singularities at $z_i=1$;
for the $Z$ resonance peak and the phase-space limit $1-z_1 z_2 z <v_{\max}$,
the importance sampling is done by Foam (or Vegas).

For energies up to 1 TeV the efficiency of the MC is better by a factor of more than 10 
with Foam than with Vegas.

\newpage
\section{Monte Carlo algorithm}
\label{sec:mc-algorith}
In this section we describe in detail the numerical Monte Carlo algorithm used to generate
final-state four-momenta, i.e. points within the Lorentz invariant phase space,
according to eq.~(\ref{eq:sigma-ceex2}) for CEEX and eq.~(\ref{eq:rho-eex3}) for EEX.
We shall not only describe the actual algorithm implemented in the present MC but
also try to explain why we opted for certain solutions, and not for others.

We start with a brief discussion of some important general issues.
First of all, we would like to stress that we treat the MC technique as a means
of integrating {\em exactly} over the phase space, without any approximation%
\footnote{
  It is necessary to stress it because in the MC event generators for QCD this may not be the case.
  In particular, the four-momentum conservation may be imposed at the end of the generation
  by certain ad hoc adjustments, which introduce an uncontrolled correction to phase-space normalization.}.
Generally, our approach is that of the textbooks on quantum mechanics:
the differential cross section is the phase-space times the scattering amplitudes from Feynman
diagrams, squared, spin-summed.
The MC technique is basically the numerical method of integration
over the phase space.
Our MC is in fact more than the phase-space integration, because we require that
 events (lists of four-momenta) be
generated with weight equal 1, 
i.e. we directly {\em simulate} the scattering process.
It is quite a strong restriction on the MC algorithm,
see below, and it means that our MC program
is not merely a {\em phase-space integrator}, but the full-scale MC {\em event generator} (MCEG).

Concerning the Monte Carlo techniques, let us remind the reader
that there are only a handful of elementary techniques 
such as  weighting-rejection, mapping and multibranching%
\footnote{
  The multibranching technique is referred to in other papers\cite{kleiss-pittau:1994} 
  as a multichannel technique.}
-- the whole art being to combine them into one bigger algorithm, see
ref.~\cite{mcguide:1999} for more detailed discussion. 
We shall also generally follow the notation and terminology of ref.~\cite{mcguide:1999}.
There also exists  a special group of
{\em self-adapting} MC techniques/programs, 
such as the very popular MC integrator Vegas~\cite{lepage:1978},
which work, in principle, for arbitrary integrand distributions 
(just like a standard Gaussian integration program integrates any external function).
We use in \KK{}, for this role, the newly developed {Foam}~\cite{foam:1999}
self-adapting program and the old {Vesko1} (part of YFS2~\cite{yfs2:1990})
as building blocks of our MC, along with other elementary techniques.

In general, we tried to minimize the use of the multibranching technique
and rely as much as possible on the method of reweighting, constructing several layers
of weights -- the total weight being their product.
Nevertheless, we have three multibranchings,
 one for the types of the final fermions
$f=e,\mu,\tau,d,u,s,c,b$, 
another one for the photon partitions
(for $n$ photons there are $2^n$ ways of attributing photons to ISR or FSR),
and finally for helicities of the bremsstrahlung photons
(for $n$ photons there are $2^n$ helicity assignments).
Each of these three multibranchings is well justified and unavoidable see below.

In the development of the present MC algorithm there are certain {\em critical issues},
in other words, there are some problems that had to be solved, 
otherwise it would have been practically impossible to realize the entire MC.
Typically, we have seen two possible solutions of an important problem,
and we have opted for one of them.
In the following we list these critical issues, explaining how the solution was found
and/or the critical option chosen.
\begin{itemize}
\item
  {\em Problem of the sum over partitions in the weight due to IFI}.
  It was realized already in ref.~\cite{mustraal-cpc:1983} that the effect of
  the quantum-mechanical interference between photon emissions from initial- and final-state charges
  (IFI) can be introduced by means of the weight
  $w^{\rm IFI} = \rho(q_c,q_d,k_1,...,k_n)/\rho'(q_c,q_d,k_1,...,k_n)$, where 
  $\rho$ includes IFI and $\rho'$ neglects it.
  In MC events are generated according to $\rho'$ in the first place.
  Both $\rho$ and $\rho'$ include the sum over $2^n$ ways of attributing photons 
  to initial and final states, the so-called photon partitions.
  The natural way of generating $\rho'$ is to use multi-branching 
  and generate an event for a single partition at a time. 
  Generation of $\rho'$ involves the introduction of the additional ``kinematical'' weight of its own.
  The possible complication to be avoided is the sum  over these kinematical weights
  over all $2^n$ partitions. 
  Fortunately, this can be done: it is sufficient to multiply $w^{\rm IFI}$
  by the kinematical weight for a single partition, the one that is actually generated. 
  The formal proof of this can be found in ref.~\cite{mcguide:1999}.
  The acceptance rate for $n$ photons due to $w^{\rm IFI}$ is $2^{-n}$.
  It is a severe problem and only the partial, brute-force solution is applied at present.
  A better solution, taking into account that IFI is destructive
  for backward scattering and constructive for forward scattering, is needed in the future.
\item
  {\em Photon helicity generation}.
  The sum over partitions in $w^{\rm IFI}$ already costs a lot of CPU time.
  Another sum over $2^n$ photon helicities would render the project impractical,
  even with the present CPU processors. 
  The solution is to introduce multibranching for $2^n$ photon helicities and generate photon helicities
  event by event%
  \footnote{
    Such a ``randomization'' of photon helicities was already used 
    in ref.~\cite{kleiss-stirling:1986}.}. 
  It is justified, because high-energy experiment
  detectors are insensitive to photon polarizations; consequently
  the various photon helicity configurations (branches) contribute equally.
  Invoking the principles of ref.~\cite{mcguide:1999},
  we can calculate weights like $w^{\rm IFI}$ for just one helicity configuration, 
  the actual one. 
  The sum over $2^n$ photon helicities is avoided.
\item
  {\em Fermion-type generation versus photon-energy generation}.
  In the MC we generate flavours according to cross sections 
  that take ISR into account in an approximate way%
  \footnote{ 
    Fermion-type (flavour) could be generated more simply, according to a crude cross-section
    equal to the Born cross-section, and the MC would correct the flavour rate in a way depending
    on the radiative correction, differently for each type of a fermion.
    Such a solution is unacceptable because, in the presence of the $Z$, ISR corrections are huge, 
    a factor $\sim3$, and the method would be rather inefficient.},
  and the MC weight corrects it to the true value later.
  The approximate cross section involves a numerical integral over the ISR total energy,
  which has to be done before the MC generation starts.
  The integral can be done 
  (i)  for each flavour separately, 
  (ii) or just once, for the flavour-summed cross section.
  In the MC generation the above corresponds to two options: 
  (i)  the fermion type is generated first, and the total energy loss due to ISR,
       just for this fermion, is generated next;
  (ii) the total energy loss due to ISR is generated first, for all fermions, 
       and the type of the fermion using the Born cross section for 
       the reduced centre-of-mass energy is generated next.
  We have chosen the second option.
\item
  {\em Comoving frame for the generation of FSR photons.}
  Generating FSR photons is much more difficult than  ISR
  photons because momenta of the final fermions ``move'', 
  i.e. they are themselves variables in the phase-space integral, 
  contrary to beams in ISR, which are fixed (except for beamstrahlung). 
  Moreover, FSR emission distributions are also the simplest in the rest frame of the final fermions.
  For the construction of the FSR algorithm, it was critical to reparametrize
  the phase space in such a way that photon integrals are formulated in the reference frame attached
  to outgoing fermions. 
  This comoving frame was used for FSR generation in the earliest version
  of the YFS3 MC which was incorporated in the KORALZ program since the version
  distributed at the time of the 1989 LEP1 workshop.
  The detailed derivation of the above phase space re-parametrization introducing the comoving
  frame can be found in ref.~\cite{kinematicon:1999}.
\item
  {\em Common IR boundary for ISR and FSR.}
  Since ISR photons are generated in CMS, and FSR photons in the rest frame of the final fermions (QMS),
  the easiest is therefore to introduce the IR cut for real photons in terms of minimum
  energy in these two frames. 
  This defines the IR boundary, and IR domains inside them, 
  which are different for ISR and FSR real photons 
  (the Lorentz-boosted sphere is ellipsoidal).
  This is not a problem, as long as the IFI is neglected.
  For the CEEX, however, the IFI is present and the IR boundary has to be common 
  (for the basic MC in which IFI is neglected).
  The problem may look very difficult; however, an inspection of the proof
  of the independence of the total cross section on the IR boundary for the non-IFI case
  tells us that we may simply take the common IR domain, which contains both ISR and FSR domains,
  and, for each event, we may ``remove from the record'' all photons that are 
  inside the new common IR domain.
  The above is true, however, only for events weight 1, not for the weighted
  events, that are present internally in the MC.
  For weighted events, such a procedure of ``photon-removal'' can still be done, 
  provided it is accompanied by the additional weight, calculable analytically.
  The above procedure of the ``photon-removal with the weight'' was already implemented in the earliest
  version of YFS3 in KORALZ
  (although it was not really necessary for the EEX matrix element there);
  however,  it was never documented%
  \footnote{
    The analogous procedure was introduced in BHLUMI, 
    where it was documented~\cite{bhlumi2:1992}, and later on exploited
    for the construction of BHWIDE~\cite{bhwide:1997}).}.
\item
  {\em Small photon mass as IR regulator in the MC.} There exists all the time an interesting option
  in the MC implementation of the YFS exponentiation, which was never exploited --
  that is to use the photon mass as an IR regulator in the MC.
  The back-of-the-envelope estimate shows that its disadvantages are a slightly more complicated
  algorithm for the generation of the photon momenta and higher photon multiplicity.
  A clear advantage is that such a cut-off is Lorentz-invariant: this would therefore make it
  automatically the same for ISR and FSR, so that we would not need to do any gymnastics 
  with the IR boundaries, as described in the previous point.
\end{itemize}

The user who is familiar with our programs KORALZ and BHLUMI may be puzzled by the fact that we often
we point out to some parts of the basic MC algorithm for QED bremsstrahlung,
which have existed for a decade but are still unpublished or undocumented;
let us comment briefly on this point.
On the one hand, the full-scale ISR and FSR multiphoton generator YFS3 
with the EEX matrix element was never published as a stand-alone program,
for instance because of the lack of EW corrections, and it was absorbed into KORALZ.
Its ISR algorithm was rather well documented in ref.~\cite{yfs2:1990},
but its FSR algorithm was never in fact documented in detail.
In the following sections we do this for the first time.
Another reason for this time-lag is that the role of the MC programs at LEP1 
for fermion pairs, perhaps with the exceptions of Bhabha scattering and $\tau$ pair production, 
was limited to removing detector acceptance,
and was not really a primary source of the SM predictions.
The simple two-particle final state could be described fairly well with  simple
semi-analytical calculations such as ZFITTER.
In LEP2 the role of the MC as a source of SM predictions seems to be more important,
mainly because of the rise of the IFI corrections,
and because of the very strong phenomenon of the $Z$ radiative return%
\footnote{ It is even more true for the $W^-W^+$ channel because of the 4-body final-state kinematics.}.

In the following we shall describe the algorithm of the Monte Carlo generation
of the events according to CEEX and/or EEX differential distributions.
The algorithm is built using elementary methods of weighting
and multibranching
with occasional use of the mapping (change of integration variables) 
wherever it is possible and/or profitable.
The weights are usually products of several component weights ordered in a ``chain''.
Their role is to counter another ``chain'' of simplifications  that were introduced in order
to turn very difficult original differential distributions into simpler ones,
in which we can integrate manually over certain integration variables
(that is to generate them with the help of mapping), one by one.
The remaining small subset of variables for which we are not able to perform
the manual/analytical integration/mapping is treated with the help of the self-adapting
Monte Carlo tool -- in our case it is { Foam} and/or { Vesko1} (optionally { Vegas}).
The above ``bottom-to-top'' procedure of simplifications
countered by weights, multibranchings, and integrations/mappings
will be described in the following subsections,
starting with removing IFI (countered by $w^{\rm IFI}$), then simplifying the
SM/QED matrix element, reorganizing phase space (mapping), 
integrating simplified ISR and FSR emission distributions (mapping),
summing over photon multiplicities, etc., such that in the end we are left with
only the integral over photon energy loss due to ISR and optionally due to beamstrahlung,
which is fed into { Vesko1} or { Foam}.
In the MC program the order of action is ``top-to-bottom'', i.e.
the photon energy loss due to ISR is generated first (together with beamstrahlung, if present),
then the type of the fermion,
the ISR photon multiplicity and momenta,
the FSR photon multiplicity and momenta,
and the series of weights brings us back to the desired EEX or CEEX differential distributions.
The order of the multibranching and weighting elementary
methods can be interchanged, with some care,
and the rules for doing it are given in ref.~\cite{mcguide:1999}.
We shall exploit this possibility, as already indicated.

\subsection{Weights and distributions in general}

Let us describe the whole organization of the weights and distributions in our MC.
We have four principal distributions, pure phase space, model,
 crude and primary.
Their ratios are the principal weights in the MC.
\begin{itemize}
\item
  {\em The pure Lorentz-invariant phase space} (LIPS) distribution of eq.~(\ref{eq:dtau}), 
  which includes four-momentum conservation $\delta$'s.
  It will not be generated directly in our MC%
      \footnote{ It is integrable, but not analytically, except for $n=2,3$.}.
  It is our {\em basic reference differential distribution} in the sense that
  all other differential distributions of interest to us can be expressed
  in its terms:
  \begin{equation}
    d\sigma(r_1,...,r_n) = \rho(r_1,...,r_n)\; d\tau_n(P;r_1,...,r_n)
  \end{equation}
  where the {\em density distributions} (we shall use this terminology)
  \begin{equation}
     \rho(r_1,...,r_n)\; = d\sigma(r_1,...,r_n) / d\tau_n(P;r_1,...,r_n),
  \end{equation}
  are analytical%
  \footnote{We limit ourselves to the fermion pair production process. 
    The $\tau$ decays and quark--gluon parton showers are not included in the discussion.}, 
  except for some simple discontinuities, with no $\delta$'s.
\item
  {\em The model distribution} is a density distribution corresponding to a {\em physical model}
  \begin{equation}
     \rho^{\rm Mod}(r_1,r_2,\dots,r_n)\; = d\sigma^{\rm Mod}(r_1,r_2,\dots,r_n)
                                         / d\tau_{n+2}(    P;r_1,r_2,\dots,r_n),
  \end{equation}
  and it is our ultimate aim to generate MC events according to this differential distribution.
  Usually, we have to deal with several variants (perturbative orders) of the model distribution,
  with similar properties, that is to say, peaks.
\item
  {\em The crude distribution} is a density distribution
  \begin{equation}
     \rho^{\rm Cru}(r_1,r_2,\dots,k_n)\; = d\sigma^{\rm Cru}(r_1,r_2,k_1,...,k_n)
                                            / d\tau_{n}( P;r_1,r_2,k_1,...,k_n),
  \end{equation}
  which is rather close to the {\em physical model distribution},
  in fact it is close to all model distributions of a certain class, 
  and it should be {\em maximally simple}.
  It should be Lorentz-invariant, 
  with no unnecessary traces of the internal technicalities of the MC,
  just a maximally simple function of dot-products of the four-momenta.
  It should have the same pattern of peaks as all model distributions of a certain class.
Here and later, $r_1, r_2$ will correspond to the momenta of the outgoing fermions, 
whereas $k_i$ will denote the momenta of the photons. In such cases we will write  
the phase-space dimension explicitly as $n+2$.
\item
  {\em The primary distribution} is a density distribution, 
  \begin{equation}\
     d\rho^{\rm Pri}(\xi_1,\xi_2,\dots,\xi_n),
  \end{equation}
  defined primarily in the space $\Sigma$ of variables $\xi_i$ with the following properties:
  \begin{itemize}
     \item
       The integral $\int d\rho^{\rm Pri}(\xi_1,\dots,\xi_n)$ is known independently;
       in most cases from analytical integration, or an independent numerical integration of the
       Gauss type, or from a special independent MC run.
     \item
       A well-defined mapping $r \to \xi$ exists. 
       The image of this mapping $\Sigma_{\rm LIPS}$ does not necessarily cover the entire $\Sigma$.
       For $\xi\in\Sigma_{\rm LIPS}$ the inverse maping $\xi \to r$ exists.
  \end{itemize}
  We may therefore define
  \begin{equation}
     \rho^{\rm Pri}(r_1,r_2,\dots,k_n)\; = d\sigma^{\rm Pri}(\xi_1,\xi_2,\dots,\xi_n)
                                         / d\tau_{n+2}(   P;r_1,r_2,\dots,k_n),
  \end{equation}
  restricted to $\xi\in \Sigma_{\rm LIPS}$ and
  which is the distribution actually generated at the lowest level of the Monte Carlo.
  A 0 weight will be assigned to MC points $\xi\not\in \Sigma_{\rm LIPS}$.
  The $\rho^{\rm Pri}$ corresponds to
  events generated according to $d\rho^{\rm Pri}$, with all weights ignored (set to 1);
  it is ``inelegant'', with traces of many technicalities, not necessarily Lorentz-covariant, 
  roughly similar to the physical model distribution.
  Its unique, great property is that it {\em can} actually be generated and integrated.
  Its integral $\int\sigma^{\rm Pri}$ sets the whole normalization of the MC.
\end{itemize}

The main rationale for introducing the intermediate {\em crude distribution},
which obviously stands between the primary and model distributions,
is a very strong practical need of modularity of the MC program.
For instance we want to use the same low-level MC event generator
for both EEX and CEEX models.
(As the example of BHLUMI shows, on top of the low-level with exponentiation one may even impose a model
distribution without exponentiation.)
We definitely want the MC event generator to have  a well defined low-level MC part, which generates
weighted events according to the {\em crude distribution},
that is with the weight
\begin{equation}
  \label{eq:wt-crude}
  \begin{split}
   &W^{\rm Cru}(r_1,r_2,\dots,r_n)\; 
    =  {d\sigma^{\rm Cru}(r_i(\xi_j))   \over d\sigma^{\rm Pri}(r_i)}
    =  {   \rho^{\rm Cru}(r_i)          \over    \rho^{\rm Pri}(r_i)},\quad \xi\in \Sigma_{\rm LIPS},
\\& 
   W^{\rm Cru}(r_1,r_2,\dots,r_n) =0, \quad \xi\not\in \Sigma_{\rm LIPS},
  \end{split}
\end{equation}
with the importance sampling corresponding to the entire group of physical models.
The above weight is provided by the low-level MC
to the outside world {\em numerically},
without any further details on how the event 
$(r_1,r_2,\dots,r_n)$
was actually produced -- {just as a black box}.

The {\em model weight} for the $m$-th model is the ratio
\begin{equation}
  \label{eq:wt-model}
  W^{\rm Mod}_m(r_1,...,r_n)\; 
  =  {d\sigma^{\rm Mod}_m(r_1,...,r_n)   \over d\sigma^{\rm Cru}(r_1,...,r_n)}
  =  {   \rho^{\rm Mod}_m(r_1,...,r_n)   \over    \rho^{\rm Cru}(r_1,...,r_n)},
\end{equation}
calculated in a separate module, and in this module
the crude distribution $\rho^{\rm Cru}$  needs to be known {\em functionally} i.e., 
it is calculated locally,
using a certain analytical expression in terms of the four-momenta of the event,
and without any access to information from the lower-level MC 
(we even assume that such an information is already trashed).
Of course, the total weight is the product of the two:
\begin{equation}
  W^{\rm Tot}_m\; = W^{\rm Cru}\; W^{\rm Mod}_m,
\end{equation}
and the total cross section is given by
\begin{equation}
  \sigma^{\rm Tot}_m\; = \langle W^{\rm Tot}_m\rangle\; \sigma^{\rm Pri}
\end{equation}
This organization is definitely fully modular.

Note also that although we did not require that the crude distribution be analytically integrable,
such a property is strongly welcomed for the purpose of the technical tests.

\subsection{Crude differential distribution for EEX}

We define the crude differential distribution with respect to the standard Lorentz 
invariant phase space as follows:
%
\begin{equation}
\label{eq:dist-crude}
\begin{split}
\rho^{\rm Cru}_{[\dot{n},n']} &(q_1,q_2; \dot{k}_1,\dots,\dot{k}_{\dot{n}}; k'_1,\dots,k'_{n'} )
\equiv 
  {d\sigma_{Cru} \over 
      d\tau_{n+n'+2} ( P; q_1,q_2, \dot{k}_1,\dots,\dot{k}_{\dot{n}}, k'_1,\dots,k'_{n'}) }
\\
&={1\over \dot{n}!}{1\over n'!}\;
  { \sigma_{\born}(s_X) \over 4\pi }\; {s_X\over s_Q}\; {2 \over \beta_f } \;
  \prod_{j=1}^{\dot{n}}  2\tilde{S}_e(\dot{k}_j)  \bar\Theta_{e}(\dot{k}_j)\;
  e^{ \gamma_e \ln\varepsilon_e} 
  \prod_{l=1}^{n'} 2\tilde{S}_f(k'_l) \bar\Theta_{_f}(k'_l)\;
  e^{ \gamma_f \ln\varepsilon_f }\;,
\end{split}
\end{equation}
where $ \varepsilon_e = 2E_{\min} /\sqrt{2p_1p_2} $,
      $ \varepsilon_f = 2E'_{\min} /\sqrt{2q_1q_2}$
      $\beta_f = (1- 4 m_f^2/ s_Q)^{1/2}$,
      $s_Q=2q_1q_2+2m_f^2 $,
and we ``dotted'' the ISR photons in order to avoid a notation clash for CEEX
in the next section.
The above distribution features a maximum resemblance to the actual QED matrix element 
from the point of view of peaks and singularities, and is very simple.
The infrared and collinear singularities are in $\tilde{S}$ soft factors.
The $\sigma_{\rm Born}(s_X)$ has a resonance peak in $s_X$.
The flux-like $s_Q/ s_X$ factor
is already present at \Order{\alpha^1} in the QED matrix element, 
and it can also be obtained from the leading-log approximation at any order.
It is also necessary to include it in the primary distribution,
in order to get reasonable MC weights at \Order{\alpha^1} and higher orders.

How did we get the above distribution?
It was rather simple. We have taken the \Order{\alpha^0}
version of  eq.~(\ref{sigma-eex3}), 
with only the $\bbeta^{(0)}_0$ term of eq.~~(\ref{beta00}) in which we ``averaged''
over the angles in $d\sigma^{\rm Born}$ and
we have taken YFS form factors with the first term only, 
so that the IR finiteness is preserved.
The kinematical factor ${s_X/ s_Q}$ was adjusted to \Order{\alpha^1} and LL.
The factor $2/ \beta_f$ is a pure convention.

The above crude distribution is for EEX only; in CEEX it would only fit one single partition.

\subsection{Crude for CEEX and multibranching over partitions}
\label{ssec:partitions}

In the EEX model we neglect IFI; we can therefore consider each bremsstrahlung photon
to be attributed to the initial or final state; this is also true for the crude distribution for EEX.
In CEEX the crude distribution of eq.~(\ref{eq:dist-crude}) is no longer usable,
because it does not include the sum over partitions.
To see it better, think about constructing the model weight as in eq.~(\ref{eq:wt-model}),
with the CEEX distribution in the numerator and the EEX crude distribution of eq.~(\ref{eq:dist-crude})
in the denominator. The resulting weight would be wildly fluctuating.
Let us therefore approach the problem in a more
systematic way and construct a realistic crude distribution for CEEX 
by simplifying the density distribution in eq.~(\ref{eq:sigma-ceex2}),
much as we did for EEX.
Neglecting all spin effects (unpolarized case)
and taking the \Order{\alpha^0} version of eq.~(\ref{eq:sigma-ceex2}) we get
\begin{equation}
  \begin{split}
  \rho&(p_c,p_d,\; k_1,\dots,k_n)=
  {d\sigma^{(0)} \over d{\rm LIPS}_{n+2} ( P ;\; p_c,p_d,\; k_1,\dots,k_n)}\\
&= {1\over n!} { e^{Y(\Omega;p_a,...,p_d)}\; \bar{\Theta}(\Omega)\; \over {\rm flux}(s)}
  {1\over 4}
  \sum_{\sigma_i=\mp 1}\;
  \sum_{\lambda_i=\mp 1}\;
  \Mmf^{(0)}_n 
  \left(  \st^{p}_{\lambda} \st^{k_1}_{\sigma_1} \st^{k_2}_{\sigma_2}
          \dots \st^{k_n}_{\sigma_n} \right)
  \left[
  \Mmf^{(0)}_n 
  \left(  \st^{p}_{\lambda} \st^{k_1}_{\sigma_1}  \st^{k_2}_{\sigma_2}
          \dots \st^{k_n}_{\sigma_n} \right)
  \right]^\star\\
&={ e^{Y} \bar{\Theta} \over 4s} {1\over n!}
  \sum_{\sigma_i,\lambda_i} \sum_{\{\wp\}} \sum_{\{\wp'\}} 
  \left(
    \prod_{i=1}^n \sfac_{[i]}^{\wp_i}
    \Bmf \left( \st^{p}_{\lambda}; X_{\wp_i} \right)\; { X^2_{\wp_i}  \over s''}
  \right)\!\!
  \left(  
    \prod_{j=1}^n \sfac_{[j]}^{\wp'_j}
    \Bmf \left( \st^{p}_{\lambda}; X_{\wp'_j} \right)\; { X^2_{\wp_j}  \over s''}
  \right)^\star,
  \end{split}
\end{equation}
where $s''=(p_c+p_d)^2$.
In the crude distribution we, of course, want to neglect the IFI.
Neglecting it in the above formula means dropping non-diagonal terms $\wp' \neq \wp$.
In addition, in order to preserve IR cancellation we simplify the YFS form factor as well:
\begin{displaymath}
  Y(\Omega;p_a,...,p_d) \to \gamma_e\ln\varepsilon_e +\gamma_f\ln\varepsilon_f.
\end{displaymath}
As a result we obtain
\begin{equation}
  \rho(p_c,p_d,\; k_1,\dots,k_n)
 = {1\over n!} { e^{\gamma_e\ln\varepsilon_e +\gamma_f\ln\varepsilon_f} \over 4s}
  \sum_{\{\wp\}}
    \prod_{i=1}^n  \bar{\Theta}(k_i) 
    \sum_{\sigma_i}  \left|\sfac_{[i]}^{\wp_i}\right|^2
    \sum_{\lambda_i} \left|\Bmf \left( \st^{p}_{\lambda}; X_{\wp_i} \right)\right|^2\; 
    { X^4_{\wp_i}  \over (s'')^2}.
\end{equation}
We can easily identify
\begin{displaymath}
  \sum_{\sigma_i} \left|\sfac_{[i]}^{\omega_i}\right|^2 
  = -8\pi^3 \tilde{S}_\omega(k_i),\quad
  \tilde{S}_1(k_i) \equiv \tilde{S}_I(k_i),\;
  \tilde{S}_0(k_i) \equiv \tilde{S}_F(k_i),\;
\end{displaymath}
and also the Born-like expression
\begin{displaymath}
    \sum_{\lambda_i} |\Bmf \left( \st^{p}_{\lambda}; X_{\wp_i} \right)|^2\;
    { X^2_{\wp_i}  \over s''}\;
    \sim {d\sigma^{\rm Born} \over d\Omega} (s,s'',t,u,t',u',X^2_{\wp_i}),
\end{displaymath}
which is weakly dependent on the dot-products
$s=2p_a p_b,$ $ s''=2p_c p_c,$ $ t=-2p_a p_c,$ $ t'=-2p_b p_d,$ $ u=-2p_a p_d,$ $ u'=-2p_b p_c$;
it strongly depends on the $X^2_{\wp_i}$ in the $Z$ resonance propagator.
We do not enter into the details of defining the above because
we replace it by an ``angular average'' anyway:
\begin{displaymath}
    \sum_{\lambda_i} |\Bmf \left( \st^{p}_{\lambda}; X_{\wp_i} \right)|^2\;
    { X^2_{\wp_i}  \over s''}
    \to {\sigma^{\rm Born}(X^2_{\wp_i}) \over 4\pi}.
\end{displaymath}
Finally, the crude distribution for CEEX we define as follows
\begin{equation}
  \label{eq:dist-crude-ceex}
  \begin{split}
  \rho^{\rm Cru}_{[n]} (p_c,p_d,\;& k_1,\dots,k_n)
 ={d\sigma^{\rm Cru}_{\rm CEEX} \over d\tau_{n+2} ( P ;\; p_c,p_d,\; k_1,\dots,k_n)}\\
&= {1\over n!} \sum_{\{\wp\}}
   {1\over s}  e^{\gamma_e\ln\varepsilon_e +\gamma_f\ln\varepsilon_f}\;
    {\sigma^{\rm Born}(X^2_{\wp_i}) \over 4\pi}
    { X^2_{\wp_i}  \over s''}
    {2\over \beta_f}
    \prod_{i=1}^n  \bar{\Theta}(k_i) 
    \tilde{S}_{\wp_i}(k_i).
  \end{split}
\end{equation}
(The overall factor like $2/\beta_f$, is just a convention.)
It is almost obvious that the above distribution is the crude distribution
for EEX of eq.~(\ref{eq:dist-crude}) summed over partitions --
it is important to check it, at least to get statistical factors right.
(We shall come to the question of the common IR boundary $\Omega$ later on.)
For example for $n=2$, omitting final fermion momenta, we have
\begin{equation}
  \rho^{\rm Cru}_{[2]} (k_1,k_2)=
  {1\over 2!} \left(
  2!   \rho^{Cru}_{[2,0]} (  k_1,k_2  ;\;          )
 +1!1! \rho^{Cru}_{[1,1]} (  k_1      ;   k_2  )
 +1!1! \rho^{Cru}_{[1,1]} (  k_2      ;   k_1  )
 +2!   \rho^{Cru}_{[0,2]} (\;         ;   k_1,k_2  )
  \right)
\end{equation}
and putting together $[1,0]$ and $[0,1]$  we have
\begin{equation}
  \rho^{\rm Cru}_{[2]} (k_1,k_2)=
  \rho^{Cru}_{[2,0]} (  k_1,k_2  ;\;          )
 +\rho^{Cru}_{[1,1]} (  k_1      ;   k_2  )
 +\rho^{Cru}_{[0,2]} (\;         ;   k_1,k_2  ).
\end{equation}
As we now see, for arbitrary photon multiplicity we have the following relations
between crude distributions for CEEX and EEX
\begin{equation}
  \label{eq:relation-of-two-crudes}
  \rho^{\rm Cru}_{[n]} (k_1,...,k_n)=
  \sum_{\dot{n}+n'=n}
  \rho^{Cru}_{[\dot{n},n']} (  k_1,...,k_{\dot{n}}  ;   k_1,...,k_{n'} )
\end{equation}
where again we have ``folded in'' together all
\begin{displaymath}
  \left( {n \atop n'} \right) = {n! \over \dot{n}!\; n'!}
\end{displaymath}
contributions with the same ISR and/or FSR multiplicity.

In the Monte Carlo we generate for a given $n$ one of the 
$\rho^{Cru}_{[\dot{n},n']}$ distributions and we implicitly understand
that {\em in order to get all $2^n$ partitions} we have to {\em undo} the above
``folding in'', that is we have to perform the proper Bose symmetrization
of the photon momenta; in other words, we must
permute the momenta of all photons randomly, so that no trace is left of
the primary ISR/FSR origin of a given photon.
The above is always understood when we say that,
we generate  in the MC, all $2^n$ partitions using multibranching methods.

The above rule is important to remember, because the MC code itself may be misleading
for two reasons:
first of all we exploit the equivalence principle of ref.~\cite{mcguide:1999},
in order to reorganize our weights, 
and the model weight for CEEX is multiplied by
the crude weight for one of the partitions only, and not by the sum 
of crude weights over all partitions (branches) --
so one may easily get the wrong impression that the multi-branching over partitions is absent.
Secondly, in the low-level MC, when generating the {\em primary} primitive
distribution, we do not generate first $n$ and later $n',$ with $\dot{n}=n-n'$,
but rather $n'$ and $\dot{n}$ independently of each other,
taking advantage of the particular
properties of the Poisson distributions that govern them and allow us to do it.

The bottom line is the following:
the crude distributions for CEEX and EEX can be, and are provided by the same low-level
Monte Carlo, because of the identity (\ref{eq:relation-of-two-crudes}),
and because the multibranching over partitions 
is practically equivalent to Bose symmetrization.
Understandably, this has great practical importance.

\subsection{Model weight and total weight}

The model weight for the \Order{\alpha^{(r)}} EEX is the following
\begin{equation}
\label{eq:wtmot-eex}
  W^{(r)}_{\rm EEX} (q_1,q_2; \dot{k}_1,\dots,\dot{k}_{\dot{n}}; k'_1,\dots,k'_{n'}) = 
  { \rho^{(r)}_{EEX}(p_1,p_2, q_1,q_2, \dot{k}_1...,\dot{k}_{\dot{n}}, k'_1...,k'_{n'})
    \over 
    \rho^{\rm Cru}_{[\dot{n},n']} (q_1,q_2; \dot{k}_1,\dots,\dot{k}_{\dot{n}}; k'_1,\dots,k'_{n'}) 
    },
\end{equation}
where the model distribution in the numerator is from eq.~(\ref{eq:rho-eex3}),
and the crude distribution in the denominator is from eq.~(\ref{eq:dist-crude}).

The model weight for the \Order{\alpha^{(r)}} CEEX is the following
\begin{equation}
  \label{eq:wtmot-ceex}
  W^{(r)}_{\rm CEEX} ( p_a,p_b, p_c,p_d,\;  k_1,k_2,\dots,k_n)=
  { \rho^{(r)}_{\rm CEEX}( p_a,p_b, p_c,p_d, k_1,k_2,\dots,k_n)
    \over 
    \rho^{\rm Cru}_{[n]} (p_c,p_d,\; k_1,k_2,\dots,k_n) (2\pi)^{3(n+2)-4}
    },
\end{equation}
where the model distribution in the numerator is from eq.~(\ref{eq:rho-ceex2}),
and the crude distribution in the denominator is from eq.~(\ref{eq:dist-crude-ceex}).
The factor $(2\pi)^{3(n+2)-4}$ is due to the difference in the normalization 
of $d$LIPS$_n$ and $d\tau_n$.

As explained in the previous section the corresponding total weight is
\begin{equation}
  \begin{split}
   &W^{(r){\rm Tot}}_{\rm CEEX} ( p_a,p_b, p_c,p_d,\;  k_1,k_2,\dots,k_n)=\\
   &=W^{(r)}_{\rm CEEX} ( p_a,p_b, p_c,p_d,\;  k_1,k_2,\dots,k_n)\;
    W^{\rm Cru}( p_a,p_b, p_c,p_d,\;  k_1,k_2,\dots,k_n)
  \end{split}
\end{equation}
and 
\begin{equation}
  \begin{split}
   &W^{(r){\rm Tot}}_{\rm EEX} (q_1,q_2; \dot{k}_1,\dots,\dot{k}_{\dot{n}}; k'_1,\dots,k'_{n'}) =\\ 
   &=W^{(r)}_{\rm EEX} (q_1,q_2; \dot{k}_1,\dots,\dot{k}_{\dot{n}}; k'_1,\dots,k'_{n'})\;
    W^{\rm Cru}(q_1,q_2; \dot{k}_1,\dots,\dot{k}_{\dot{n}}; k'_1,\dots,k'_{n'}),
  \end{split}
\end{equation}
where $W^{\rm Cru}$ is exactly the same, 
provided we do for CEEX the  proper Bose--Einstein symmetrization.
However, even this is strictly speaking unnecessary because
the CEEX matrix element is Bose-symmetric anyway, by construction.

Only one of many model weights is the {\em principal weight} used for
a rejection of the events.
Of course we choose the best one, that is the \Order{\alpha^{(2)}} CEEX-type.
The other ones are available, and we use them for tests.

\subsection{Phase-space reorganization}

Our starting point is the phase-space integral of eq.~(\ref{eq:dist-crude}) 
for the crude total cross section, which can be rewritten as follows
\begin{equation}
\label{eq:crude-start}
\begin{split}
  \sigma^{\rm Cru} &=
  \int d s_X \;
  \sum_{n=0}^\infty \sum_{n'=0}^\infty
  \int d\tau_{n+1} ( P ;\;     k_1,\dots,k_n,\;   X)\;
  {1\over n!} \prod_{j=1}^n    \tilde{S}_e(k_j)  \bar\Theta_{e}(k_j)\;
\\
 &\int d\tau_{n'+2} ( X;\;,  k'_1,\dots,k'_{n'}\;  q_1,q_2)
  {1\over n'!}\prod_{l=1}^{n'} \tilde{S}_f(k'_l) \bar\Theta_{_f}(k'_l)\;
  { \sigma_{\born}(s_X) \over 4\pi }\; {s_X\over s_Q}\; {2 \over \beta_f } \;
  e^{ \gamma_e \ln\varepsilon_e} 
  e^{ \gamma_f \ln\varepsilon_f }\;
\end{split}
\end{equation}
where $P=p_1+p_2$.

The integral above is Lorentz-invariant and, in principle, can be evaluated in any
reference frame, 
not necessarily in the laboratory frame where $\vec{P}=0$ and $p_1=(p^0,0,0,p^3)$,
which we call PMS.
The implicit assumption is also that all momenta in the phase space are in the same reference frame
 at the time of the evaluation.
The reality of the MC world is more complicated.
Although the final product, the list of all four-momenta (MC event), is given to the user
in one universal frame, the laboratory frame,
in the intermediate stages the integral is split into Lorentz-invariant parts
and each part is worked out separately, in the {\em local } Lorentz frame, which is the most convenient
for generating subgroup of four-momenta in this subintegral.
Then, momenta of this subgroup/subintegral are Lorentz-transformed to the laboratory.
The MC generator is not fully documented if the relevant {\em local } frames
and the Lorentz transformation connecting them to one another and to the laboratory system
are not unambiguously defined.
In this section we shall make an effort to do it.

In the case of the above integral we take advantage of the Lorentz invariance
of $d\tau_{n'+2} ( X;\;,  k'_1,\dots,k'_n\;  q_1,q_2)$ and we transform
all its variables to the frame where $X=\hat{X}=(\sqrt{s_X},0,0,0)$, the XMS frame,
and put bars on top of them to mark this:
\begin{equation}
\begin{split}
  \sigma^{\rm Cru} &=
  \int d s_X \;
  \sum_{n=0}^\infty \sum_{n'=0}^\infty
  \int d\tau_{n+1} ( P ;\;     k_1,\dots,k_n,\;   X)\;
  {1\over n!} \prod_{j=1}^n    \tilde{S}_e(k_j)  \bar\Theta_{e}(k_j)\;\\
 &\int d\tau_{n'+2} ( \bar{X};\;,  \bar{k}'_1,\dots,\bar{k}'_n\;  \bar{q}_1,\bar{q}_2)\;
  {1\over n'!}\prod_{l=1}^{n'} \tilde{S}_f(\bar{k}'_l) \bar\Theta_{_f}(\bar{k}'_l)\;
  { \sigma_{\born}(s_X) \over 4\pi }\; {s_X\over s_Q}\; {2 \over \beta_f } \;
  e^{ \gamma_e \ln\varepsilon_e} 
  e^{ \gamma_f \ln\varepsilon_f },\;
\end{split}
\end{equation}
This operation is still not defined unambiguously, unless we specify the direction
of two space-like axes in the new frame or, equivalently, write down explicitly
the Lorentz transformation $L_X$ from XMS to CMS and back.

Before we do this, let us note that there are at least three main ways of fixing
the $z$-axis in the XMS frame.
Two possibilities are to use as a direction for the $z$-axis three-momenta of beams,
$\vec{p}_1$ or $-\vec{p}_2$ (they are not the same if $s\neq s_X$).
We call them XMS1 and XMS2 correspondingly.
Note that these frames were introduced in refs.~\cite{kleiss-widelki,mustraal-cpc:1983}.
The other choice is to use as $z$-axis in the XMS frame the direction
of the $\vec{P}$, that is the direction of the boost connecting PMS and XMS
(that is the direction opposite to the total four-momentum of the ISR photons).
This choice we shall call simply XMS, as it was used in ref.~\cite{koralb:1985}.
A third choice is to do a so-called parallel boost $B_X$ along the direction
of the $\vec{X}$ in PMS.
The corresponding transformation matrix is
\begin{equation}
\label{eq:parallel-boost}
  B_X =  \begin{bmatrix} 
         {X^0\over M_X},      & {\vec{X}^T\over M_X}  \\
                              & \\
         {\vec{X}\over M_X},  & I + {\vec{X}\otimes \vec{X}\over M_X(M_X+X^0) }
         \end{bmatrix}, \; X^2=M_X^2,
\end{equation}
where $T$ marks the matrix transposition and $\otimes$ marks the tensor product.
This is our choice in the present version of the program (the same as in the YFS3 MC).
Note that, in general, in the XMS frame defined with the above transformation,
the $z$-axis is not parallel to $\vec{p}_1$ or $-\vec{p}_2$.
The transformation from our XMS to CMS is
\begin{equation}
  \begin{split}
    k'_i|_{CMS} =L_{X}  \bar{k}'_i,\;\;\;  
    q_i|_{CMS}  =L_{X}  \bar{q}_i,\;\;\;
    L_{X} = B_X.
  \end{split}
\end{equation}
The transformations for other types of XMS are given explicitly in
ref.~\cite{kinematicon:1999}.

Now comes an important point.
As already indicated, the emission of the FSR photons is done
in the comoving frame attached to the momenta $q_i$ of outgoing fermions,
that is in the frame where $\vec{Q}=\vec{q_1+q_2}=0$
and $q_1=(q_1^0,0,0,|q_1^3|)$. We shall call it QMS.
So why do we not transform immediately from XMS to QMS, 
and generate photons there?
The problem is that $Q=X-\sum k'_i$, and in order to get from XMS to QMS we have
to know $k'_i$ in the first place. We are stuck.
The solution is to reparametrize the FSR integral with the help of the integration
over the Lorentz group; the details are given in  ref.~\cite{kinematicon:1999}.
Here we just apply the result of this work and obtain the new formula:
\begin{equation}
\label{eq:intermediate}
\begin{split}
  \sigma^{\rm Cru} =&
  \int d s_X \;
  \sum_{n=0}^\infty {1\over n!} 
  \prod_{j=1}^n   {d^3 k_j\over 2k^0_j}  
                   2\tilde{S}_e(k_j)  \bar\Theta_{e}(k_j)\;
  \delta\left(s_X - (P- \sum_{j=0}^n k_j)^2 \right)
  e^{ \gamma_e \ln\varepsilon_e}\\
 &\int d\psi d\cos\omega
  { \sigma_{\born}(s_X) \over 4\pi }\; \\
 & \sum_{n'=0}^\infty {1\over n'!} \int d s_Q\;
  \prod_{l=1}^{n'} {d^3 \tilde{k}'_l\over 2 {\tilde{k'}}_l^0}\; 
                    2\tilde{S}_f(\tilde{k}'_l) \bar\Theta_{_f}(\tilde{k}'_l)\;
  \delta\left(s_X - (\hat{Q} + \sum_{j=0}^{n'} \tilde{k}'_j)^2 \right)
  e^{ \gamma_f \ln\varepsilon_f },\\
  \tilde{K} \equiv& \sum \tilde{k}'_j,\;\; 
  \hat{Q}   \equiv (\sqrt{s_Q},0,0,0),\;\; 
  \hat{q}_1 \equiv {\sqrt{s_Q}\over 2} (1,0,0, \beta_f),\;\; 
  \hat{q}_2 \equiv {\sqrt{s_Q}\over 2} (1,0,0,-\beta_f),\;\;
\end{split}
\end{equation}
in which those variables with a tilde are defined in the QMS.
Note that the Jacobian due to the reparametrization of the FSR integral
cancels exactly the factor $(s_X/ s_Q) (2 / \beta_f)$.
The explicit transformation from  QMS to XMS defines the meaning of 
the new $\psi, \omega$ integration variables: 
\begin{equation}
  \label{eq:LA}
  \begin{split}
   &\bar{k}_i = L_A k_i,\;\;  \bar{q}_i = L_A \hat{q}_i,\;\;\\
   &L_A  = R_3(\psi) R_2(\omega) B^{-1}_{\hat{X}},\;\;
    \hat{X}= \hat{Q} - \sum \tilde{k}'_j.
  \end{split}
\end{equation}
The most important fact is that the explicit integration over $q_1$ and $q_2$ has disappeared
completely!
The angles $\psi,\omega$ parametrize the three-dimensional
orientation of the momenta set $(\bar{k}'_1,\dots,\bar{k}'_n, \bar{q}_1,\bar{q}_2)$
as a whole in XMS.

The above treatment of the FSR phase space is the simplest one --
this is why we adopted it.
Note that the angles $\psi,\omega$  have no direct geometric meaning
of polar angles of a certain momentum in a certain frame -- they are
just parameters in the Lorentz transformation.
In particular, momenta $\vec{\bar{q}}_1$ and $-\vec{\bar{q}}_1$ do not coincide
with the $z$-axis in XMS.
Such an arrangement is, however, possible.
It corresponds to a different transformation $L_A$ and the integral (\ref{eq:intermediate})
would look slightly different.
Such an alternative solution is described in detail in ref.~\cite{kinematicon:1999}.

The crude integral of eqs.~(\ref{eq:dist-crude}) and (\ref{eq:intermediate}) can be rewritten as follows:
\begin{equation}
\label{eq:ready-to-go}
\begin{split}
  \sigma^{\rm Cru}
 =& \sum_{f=\mu,\tau,d,u,s,c,b}\;\;
  \sum_{n=0}^\infty \sum_{n'=0}^\infty 
  \int d\tau_{n+n'+2} ( P;  q_1,q_2, k_1, \dots, k_{n}, k'_1, \dots, k'_{n'} )
\\ &\qquad\qquad\qquad\qquad \times
   \rho^{\rm Cru}_{[n,n']}  (q_1,q_2; k_1, \dots, k_{n}; k'_1, \dots, k'_{n'} )
\\
 =&\sum_{f=\mu,\tau,d,u,s,c,b}\;\;
  \int d s_X\;  \sigma^f_{\born}(s_X) \;\;
  \int {d\psi d\cos\omega \over 4\pi } \;
\\
  &\sum_{n=0}^\infty {1\over n!} 
   \prod_{j=1}^n   {d^3 k_j\over k^0_j}  \tilde{S}_e(k_j)  \bar\Theta_{e}(k_j)\;
   \delta\left(s_X - (P- \sum_{j=0}^n k_j)^2 \right)
   e^{ \gamma_e \ln\varepsilon_e}\\
  &\sum_{n'=0}^\infty {1\over n'!} \int d s_Q\;
   \prod_{l=1}^{n'} {d^3 \tilde{k}'_l\over  {\tilde{k'}}_l^0}\; 
                   \tilde{S}_f(\tilde{k}'_l) \bar\Theta_{_f}(\tilde{k}'_l)\;
   \delta\left(s_X - (\hat{Q} + \sum_{j=0}^{n'} \tilde{k}'_j)^2 \right)
   e^{ \gamma_f \ln\varepsilon_f }.
\end{split}
\end{equation}
This clearly factorizes into independent ISR and FSR parts,
with the integration over the effective mass $s_X$ pulled out as a principal
integration variable.
The above integral is ready for the MC generation.

Let us stress again that in all the above reorganization 
of the phase space for the $\sigma^{\rm Cru}$,
from eq.~(\ref{eq:dist-crude}) through eq.~(\ref{eq:crude-start}) to eq.~(\ref{eq:ready-to-go}),
we only changed variables with {\em no approximations},
with full control of the Jacobians of all mappings.
For completeness we write the total transformation from QMS down to CMS:
\begin{equation}
  \label{eq:LA-transform}
  \begin{split}
   &k'_i|_{CMS} =B_{X}  L_A \tilde{k}'_i,\;\;  
    q_i|_{CMS}  =B_{X}  L_A \hat{q}_i,\\
   &L_A =R_3(\psi) R_2(\omega) B^{-1}_{\hat{X}},\;\;
    X=X|_{CMS},\;\; \hat{X} = \hat{Q}-\sum_j \tilde{k}'_j
  \end{split}
\end{equation}
and $\hat{X}$ is defined in the rest frame of the outgoing fermions.

In the following sections we shall introduce variables that are used
directly in the MC generation, separately for ISR and FSR,
and we shall define {\em primary} differential distributions generated
at the lowest level of the MC algorithm.
We start with the FSR -- in this case the corresponding {\em primary} differential distribution
sums up to unity, and next we elaborate on the case of ISR.
The case of FSR will be described in detail, while the case of ISR will only
be summarized, since it was already discussed
in ref.~\cite{yfs2:1990}.

\newpage
\subsection{FSR momenta}
In the following we shall  describe the MC algorithm for the generation of the FSR photon momenta.
Although it was used for almost a decade~\cite{yfs3-pl:1992,koralz4:1994},
it is the first time that it is described in full detail.
Let us consider the FSR part of the crude integral of eq.~(\ref{eq:ready-to-go}):
\begin{equation}
  \label{eq:fsr-integral}
  \begin{split}
  & \Fmf_{n'}=
    {1\over n'!}  \int\limits_{4m_f^2}^{s_X} d s_Q\;
    \prod_{j=1}^{n'} 
        \int {d^3 \tilde{k}'_j\over  {\tilde{k'}}_j^0}\; 
        \tilde{S}_f(\tilde{k}'_j) \Theta( \tilde{k}'_j-E'_{\min} )\;
        \delta\Bigg(s_X - \bigg(\hat{Q} + \sum_{l=0}^{n'} \tilde{k}'_l\bigg)^2 \Bigg)
        e^{ \gamma_f \ln\varepsilon_f },
\\
  & \gamma_f  = Q_f^2\; {\alpha\over \pi}\; {1+\beta_f^2 \over \beta_f}\;
                \bigg( \ln { 1+\beta_f \over 1-\beta_f } -1 \bigg)
              = Q_f^2\; {\alpha\over \pi}\; {1+\beta_f^2 \over \beta_f}\;
                \bigg( \ln { (1+\beta_f)^2 \over \mu_f^2} -1 \bigg),
\\
  & \beta_f = (1-\mu_f^2)^{1/2},\;\; 
    \mu_f^2 = {4m_f^2\over s_Q},\;\;
    \varepsilon_f = { 2E'_{\min} \over \sqrt{s_Q}},\;\;
    \hat{Q}= (\sqrt{s_Q},0,0,0),
\end{split}
\end{equation}
where we restored finite $m_f$,
photon momenta $\tilde{k}'_l$ are defined in the QMS rest frame of the outgoing fermions --
the natural reference frame to describe the emission of the FSR photons --
and $E'_{\min}$ is the real photon minimum energy in this frame.
Let us now express the photon momenta in units of ${1\over 2}\sqrt{s_Q}$,
as well as introduce polar parametrization and other auxiliary notation:
\begin{equation}
  \begin{split}
    &\tilde{k}'_j \equiv {\sqrt{s_Q}\over 2}\; \bar{k}_j
     \equiv {\sqrt{s_Q}\over 2}\; x_j\; (1,\sin\theta_j \cos\phi_j, \sin\theta_j \sin\phi_j, \cos\theta_j ),
\\
    & \tilde{K}'=\sum_{l=0}^{n'} \tilde{k}'_l \equiv {\sqrt{s_Q}\over 2}\; \bar{K}.
\end{split}
\end{equation}
With the help of the above we are able to eliminate the $\delta$-function:
\begin{equation}
  \begin{split}
  \int\limits_{4m_f^2}^{s_X} d s_Q\; 
       \delta\Bigg(s_X -  \bigg(\hat{Q} + \sum_{l=0}^{n'} \tilde{k}'_l\bigg)^2 \Bigg)
 &=\int\limits_{4m_f^2}^{s_X} d s_Q\; 
       \delta\Bigg(s_X - s_Q \bigg( 1+\bar{K}^0 + {1\over 4} \bar{K}^2 \big)   \Bigg)
\\
 &= {\Theta\big(s_Q(\bar{k}_1,...,\bar{k}_{n'}) -4m_f^2\big)\over  1+\bar{K}^0 + {1\over 4} \bar{K}^2},
\end{split}
\end{equation}
and from now on
\begin{equation}
  \label{eq:sQ}
  s_Q= s_Q(\bar{k}_1,...,\bar{k}_{n'}) \equiv  {s_X \over  1+\bar{K}^0 + {1\over 4} \bar{K}^2}.
\end{equation}
Also the single-photon distribution gets transformed:
\begin{equation}
  \begin{split}
 &{d^3 \tilde{k}'_j\over  {\tilde{k'}}_j^0}\; 
  \tilde{S}_f(\tilde{k}'_j)
  = {dx_j\over x_j}\; {d\phi_j \over 2\pi}\; d \cos\theta_j\; 
    {\alpha \over \pi}\; f\bigg(\theta_j, {m_f^2\over s_Q} \bigg),
\\
 & f\bigg(\theta_j, {m_f^2\over s_Q} \bigg)
                  = {1+\beta_f^2 \over\delta_{1j}\delta_{2j}  }
                   -{\mu_f^2\over 2}\; {1\over \delta_{1j}^2}
                   -{\mu_f^2\over 2}\; {1\over \delta_{2j}^2},\quad
      \delta_{1j} = 1-\beta_f \cos\theta_j,\;\;
      \delta_{2j} = 1+\beta_f \cos\theta_j,\;\;
\end{split}
\end{equation}
and the whole integral is transformed into the semi-factorized form:
\begin{equation}
  \begin{split}
    \Fmf_{n'}=
  & {1\over n'!}\; \prod_{j=1}^{n'} 
        \int\limits_{\varepsilon_f}^\infty {dx_j\over x_j}\; 
        \int\limits_0^{2\pi} {d\phi_j \over 2\pi}\; 
        \int\limits_{-1}^1   d \cos\theta_j\; 
        {\alpha \over \pi}\; f\bigg(\theta_j, {m_f^2\over s_Q} \bigg)\;\;
        { \Theta(s_Q -4m_f^2) \over 1+\bar{K}^0 +{1\over 4} \bar{K}^2}\;
        e^{ \gamma_f \ln\varepsilon_f }.
  \end{split}
\end{equation}
The reader should not, however, be misled by the apparent simplicity of the above integral --
it does not factorize yet into a product of  independent integrals, one per photon,
because the collective dependence on all photon momenta $\bar{k}_j$ is entering everywhere through
the variable $s_Q$, see eq.~(\ref{eq:sQ}).

Another complication due to the use of ${1\over 2}\sqrt{s_Q}$ as an energy scale is that,
in the case of the hard FSR photon,
the upper limit of $x_j$ extends to large values, not really to infinity because
of the $\Theta(s_Q -4m_f^2)$; nevertheless, this is not very
convenient for the MC integration. 
This is cured with the following change of variables:
\begin{equation}
  \begin{split}
   &y_i = {x_i \over 1+\sum x_j},\quad 
    x_i = {y_i \over 1-\sum y_j},
\\
   &1+\sum_j x_j = {1\over  1-\sum_j y_j} = 1+\bar{K}^0
    = 1+{ 2K'\cdot Q \over s_Q} = {s_X \over s_Q} \bigg( 1- {{K'}^2\over s_X} \bigg),
  \end{split}
\end{equation}
which leads to
\begin{equation}
  \begin{split}
    \Fmf_{n'}=
    {1\over n'!}\; \prod_{j=1}^{n'} \;
        \int\limits_{\varepsilon_f/(1+\bar{K}^0)}^1 {dy_j\over y_j}\; 
   &    \int\limits_0^{2\pi} {d\phi_j \over 2\pi}\; 
        \int\limits_{-1}^1   d \cos\theta_j\; 
        {\alpha \over \pi}\; f\bigg(\theta_j, {m_f^2\over s_Q} \bigg)\;
\\ &
        { 1+\bar{K}^0  \over 1+\bar{K}^0 +{1\over 4} \bar{K}^2}\;
        \Theta(s_Q -4m_f^2)\; e^{ \gamma_f \ln\varepsilon_f }.
  \end{split}
\end{equation}
With the new variables the condition $s_Q >4m_f^2$ (easily implementable in the MC)
translates approximately into $\sum_j y_j < 1$.
Furthermore, we have
\begin{displaymath}
{ 1+\bar{K}^0 \over 1+\bar{K}^0 +{1\over 4} \bar{K}^2} \leq 1,
\end{displaymath}
which is ideal for the MC\ %
\footnote{  In the case of ISR a similar factor is causing a lot of trouble because
  it has a negative coefficient in front of $\bar{K}^2$ and the corresponding contribution
  to the MC weight is not well bounded from  above.}.
The new IR limit $y_j>\varepsilon_f/(1+\bar{K}^0)$ is however inconvenient for the MC.
The solution is to substitute
\begin{equation}
  \varepsilon_f = \delta_f \; (1+\bar{K}^0)
\end{equation}
where $\delta_f\ll 1$ is from now on the new IR regulator for FSR real photons.
Note that this sets 
\begin{equation}
  E''_{\min}= \delta_f {1\over 2}\sqrt{s_Q} (1+\bar{K}^0)
            = \delta_f {1\over 2}\sqrt{s_Q} \bigg( 1+ {2K'\cdot \hat{Q} \over s_Q} \bigg)
\end{equation}
as a lower limit for the photon energy in the QMS, which is {\em higher}
than the previous one $E'_{\min}= {1\over 2}\sqrt{s_Q} \delta_f$ (for $\varepsilon=\delta_f$).
Consequently, we have to keep the value of $\delta_f$ very low,
in fact we need%
\footnote{ This should be listed as a disadvantage of the actual method
  of the MC treatment of the FSR.}
$\delta_f \ll m_f^2/s_X$, which can be a problem for $f=e$.

\subsubsection{Simplifications and MC generation}
Up to this point, the FSR integral of eq.~(\ref{eq:fsr-integral}) was transformed without
any approximations and the integral was conveniently parametrized
for the MC generation:
\begin{equation}
  \begin{split}
    \Fmf_{n'}=
    {1\over n'!}\; \prod_{j=1}^{n'} 
        \int\limits^1_{\delta_f} {dy_j\over y_j}\; 
  &     \int\limits_0^{2\pi} {d\phi_j \over 2\pi}\; 
        \int\limits_{-1}^1   d \cos\theta_j\; 
        {\alpha \over \pi}\; f\bigg(\theta_j, {m_f^2\over s_Q} \bigg)\;
\\&
        { 1+\bar{K}^0 \over 1+\bar{K}^0 +{1\over 4} \bar{K}^2}\;
        \Theta(s_Q -4m_f^2)\;  
        e^{ \gamma_f \ln(\delta_f (1+\bar{K}^0)) }.
  \end{split}
\end{equation}
There is also a one-to-one correspondence between the points
in the Lorentz-invariant phase space and the points in space of our new variables:
\begin{equation}
  \{ n',(\tilde{k}'_1,\dots, \tilde{k}'_{n'}) \} 
   \leftrightarrow \{ n', ( y_j,\theta_j,\phi_j), j=1,\dots, n'  \}.
\end{equation}
We can also write explicitly the differential distributions in the 
two {\em equivalent} parametrizations
\begin{equation}
  \label{eq:isr-equivalent}
  \begin{split}
   &{d\Fmf_{n'} \over 
    d s_Q \delta\Big(s_X - \big(\hat{Q} + \sum\limits_{l=0}^{n'} \tilde{k}'_l\big)^2 \Big)
    \prod\limits_{j=1}^{n'} {d^3 \tilde{k}'_j\over 2 {\tilde{k'}}_j^0}  }
 = {\Theta(s_Q-4m_f^2)\over n'!}\; 
   e^{ \gamma_f \ln\left( { 2E''_{\min}  \over \sqrt{s_X} } \right) }
   \prod\limits_{j=1}^{n'} 2\tilde{S}_f(\tilde{k}'_j) \Theta( \tilde{k}'_j-E''_{\min} ),
\\
  &{d\Fmf_{n'} \over \prod_{j=1}^{n'} dy_j\; d\cos\theta_j\; d\phi_j }
   = {\Theta(s_Q-4m_f^2)\over n'!}\; 
   e^{ \gamma_f \ln(\delta_f (1+\bar{K}^0)) }
   \bigg( {\alpha \over 2\pi^2} \bigg)^{n'}
   \prod_{j=1}^{n'} { \Theta(y_j-\delta_f) \over y_j}  f\bigg(\theta_j, {m_f^2\over s_Q} \bigg).
  \end{split}
\end{equation}

We are now ready to introduce the {\em simplifications} leading us to a {\em primary}
distribution, which can be integrated analytically and generated
using standard uniform random numbers.
The simplifications are%
\footnote{ We drop the mass term from $f(\theta_j)$ for the same reasons as in the case of ISR;
           see next subsection.}
\begin{equation}
  \begin{split}
   &f\bigg(\theta_j, {m_f^2\over s_Q} \bigg) \to 
    \bar{f}\bigg(\theta_j, {m_f^2\over s_X} \bigg) = 
    {1+\bar{\beta}_f^2 \over \bar{\beta}_f}\; {1\over 1 -\bar{\beta}_f^2 \cos^2\theta_j},
\\
   &{ 1+\bar{K}^0 \over 1+\bar{K}^0 +{1\over 4} \bar{K}^2}\; \Theta(s_Q -4m_f^2)\; \to 1,
\\
   &e^{ \gamma_f \ln(\delta_f (1+\bar{K}^0)) } \to e^{ \bar{\gamma}_f \ln(\delta_f)},
  \end{split}
\end{equation}
where
\begin{equation}
  \bar{\beta}_f = (1- (m_f^2 /s_X)^2)^{1/2},\quad
  \bar{\gamma}_f = Q_f^2\; {\alpha\over \pi}\; {1+\bar\beta_f^2 \over \bar\beta_f}\;
                   \ln { 1+\bar\beta_f \over 1-\bar\beta_f }.
\end{equation}
The main purpose of the above is to remove any complicated dependence on the momenta
of all photons through $s_Q$ -- it is achieved trivially by replacing $s_Q$ by $s_X$.
With this hard FSR photons, get stronger collinear peaks
at $\cos\theta_j=\pm 1$ in the primary differential distribution.
The resulting FSR {\em primary} differential distribution is:
\begin{equation}
  \label{eq:fsr-prim-dist}
  \begin{split}
  &{d\Fmf^{\rm Pri}_{n'} \over \prod_{j=1}^{n'} dy_j\; d\cos\theta_j\; d\phi_j }
   = 
     e^{ \bar\gamma_f \ln(\delta_f) }\;
     \bigg( {\alpha \over 2\pi^2} \bigg)^{n'}
     \prod_{j=1}^{n'} {\Theta(y_j-\delta_f) \over y_j} \bar{f}\bigg(\theta_j, {m_f^2\over s_X} \bigg),
  \end{split}
\end{equation}
and the compensating weight transforming the primary distribution into the  crude distribution is
\begin{equation}
  \begin{split}
    w^{\rm Cru}_{\rm FSR}
   ={d\Fmf_{n'} \over d\Fmf^{\rm Pri}_{n'} }
   ={ 1+\bar{K}^0 \over 1+\bar{K}^0 +{1\over 4} \bar{K}^2}\; \Theta(s_Q -4m_f^2)\;
    e^{ \gamma_f \ln(\delta_f (1+\bar{K}^0)) - \bar{\gamma}_f \ln(\delta_f)}\;
    \prod_{j=1}^{n'}\;
    {    f\bigg(\theta_j, {m_f^2\over s_Q} \bigg) \over
         \bar{f}\bigg(\theta_j, {m_f^2\over s_X} \bigg) }.
  \end{split}
\end{equation}
Events $\{ n',(y_j,\cos\theta_j,\phi_j),j=1,...,n' \}$ generated according to
$d\Fmf^{Prim}_{n'}$, defined in eq.~(\ref{eq:fsr-primary}) below
with the weight $w^{\rm Cru}_{\rm FSR}$,
will be distributed (if rejection is applied)  according
to the differential distribution,
that is the integrand in eq.~(\ref{eq:isr-equivalent}), as desired.
There is only one thing to be remembered:
removing $\Theta(s_Q -4m_f^2)$ in the primary distribution means
that we cannot map every event generated according to $d\Fmf^{\rm Pri}_{n'}$
into a Lorentz-invariant phase-space point, 
the unambiguous mapping exists in a strict sense in only one direction:
\begin{displaymath}
  \{ n',(\tilde{k}'_1,\dots, \tilde{k}'_{n'}) \} 
   \rightarrow \{ n', ( y_j,\theta_j,\phi_j), j=1,\dots, n'  \}.
\end{displaymath}
This is, however, not really a serious problem because it occurs only for the events with 
$w^{\rm Cru}_{\rm FSR}=0$, while for events with $w^{\rm Cru}_{\rm FSR} \neq 0$ we are able to map
\begin{displaymath}
  \{ n',(\tilde{k}'_1,\dots, \tilde{k}'_{n'}) \} 
   \leftarrow \{ n', ( y_j,\theta_j,\phi_j), j=1,\dots, n'  \}.
\end{displaymath}
Nevertheless, one should keep in mind  that some 0-weighted events
generated by the FSR algorithm do not have four-momenta assigned to them.

Finally we may check that, as advertised,
the integral over the FSR primary distribution can be evaluated analytically:
\begin{equation}
  \label{eq:fsr-primary}
  \begin{split}
    \sum_{n'=0}^\infty \Fmf^{\rm Pri}_{n'}&=
    \sum_{n'=0}^\infty {1\over n'!}\;
    \prod_{j=1}^{n'} 
        \int\limits^1_{\delta_f} {dy_j\over y_j}\; 
        \int\limits_0^{2\pi} {d\phi_j \over 2\pi}\; 
        \int\limits_{-1}^1   d \cos\theta_j\; 
        {\alpha \over \pi}\; \bar{f}\bigg(\theta_j, {m_f^2\over s_X} \bigg)\;
        e^{ \bar{\gamma}_f \ln(\delta_f) }
\\
   &= \sum_{n'=0}^\infty e^{ -\bar{\gamma}_f \ln{1\over \delta_f} } 
    {1\over n'!}\;
    \bigg( \bar{\gamma}_f \ln{1\over \delta_f} \bigg)^{n'}
    = \sum_{n'=0}^\infty e^{ -\langle n'\rangle }\; {  \langle n'\rangle ^{n'} \over n'!} =1.
  \end{split}
\end{equation}
The photon multiplicity for the primary distribution is the standard Poisson distribution,
with the average
\begin{equation}
  \label{eq:fsr-crude-multiplicity}
  \langle n'\rangle = \bar{\gamma}_f \ln{1\over \delta_f},
\end{equation}
and the overall normalization is trivially equal to 1,
which is a natural choice for the FSR anyway.

The MC generation of the distribution (\ref{eq:fsr-prim-dist}) is rather easy.
It is fully factorized 
-- variables $\cos\theta_j$,  $\phi_j$ and $y_j$ can be generated independently.
The distribution of  $\phi_j$ is just flat and the distribution of $y_j$ is trivial to generate:
\begin{equation}
    \phi_j = 2\pi r_{1j},\;\;\;
    y_j    = \delta_f^{r_{2j}},
\end{equation}
where $r_{ij}$ are the standard uniform random numbers $0<r_{ij}<1$.
The distribution of $\cos\theta_j$ requires applying the branching method:
it is split into two components
\begin{equation}
  \begin{split}
    {2\over 1-\bar\beta_f \cos^2\theta_j}
    = {1\over 1-\bar\beta_f \cos\theta_j} + {1\over 1+\bar\beta_f \cos\theta_j},
  \end{split}
\end{equation}
and $\cos\theta_j$ is generated according to one component, chosen with the equal probability
between the two.
For example, if the first component $1/(1-\bar\beta_f\cos\theta_j)$ is chosen then
\begin{equation}
   \cos\theta_j = {1\over \bar\beta_f}\;
      \bigg\{( 1-(1+\bar\beta_f)  \bigg( {1-\bar\beta_f \over 1+\bar\beta_f} \bigg)^{r_{3j}}\bigg\},
\end{equation}
where $r_{3j}$ is another uniform random number.

\subsection{ISR momenta}
In the following we shall  describe the MC algorithm of the generation of the ISR photon momenta.
The algorithm was already described%
\footnote{ The essential part of the algorithm was given in ref.~\cite{yfs-mpi:1987}.}
in  ref.~\cite{yfs2:1990},
and for the sake of completeness we shall describe it here, but without going into the fine details.
Let us consider the ISR part of the crude integral of eq.~(\ref{eq:ready-to-go})
for one final fermion type $f$:
\begin{equation}
  \label{eq:isr-integ}
    \FmI_n=
    {1\over n!} 
    \int d s_X\;  \sigma_{\born}^f(s_X) \;\;
    \prod_{j=1}^n   \int {d^3 k_j\over k^0_j}  \tilde{S}_e(k_j) \Theta(k^0_j-E_{\min})\;
    \delta\bigg(s_X - \Big(P- \sum_{j=0}^n k_j\Big)^2 \bigg)
    e^{ \gamma_e \ln\varepsilon_e}
\end{equation}
where $E_{\min}= \varepsilon_e {1\over 2} \sqrt{s}$ is the minimum energy of the real ISR
photon in the laboratory CMS.
In the first step we introduce the variable $v=1-s_X/s$ and order energies of the photons
\begin{equation}
  \begin{split}
    \FmI_n=
   &\int\limits_0^{v_{\max}} d v\;  \sigma_{\born}^f(s(1-v)) \;
    \prod_{j=1}^n   \int {d^3 k_j\over k^0_j}  \tilde{S}_e(k_j)
\\
   &\Theta(k^0_1-k^0_2)\; \Theta(k^0_2-k^0_3) \dots \Theta(k^0_n-E_{\min})\;
    \delta\bigg( v  - {2KP -K^2 \over s} \bigg)
    e^{ \gamma_e \ln\varepsilon_e},
  \end{split}
\end{equation}
where $K=\sum_{j=0}^n k_j$ and $v_{\max}=1-4m^2_f/s$.
Now we rescale all momenta and introduce a polar parametrization
\begin{equation}
k_i =\eta \bar{k}_i = \eta x_i (1,\sin\theta_i\sin\phi_i,\sin\theta_i\cos\phi_i,\cos\theta_i);
\end{equation}
we fix the scaling factor $\eta$ such that $\bar{k}^0_1 =x_1= v$:
\begin{equation}
  \begin{split}
    \FmI_n 
  =&\int d\eta\; \delta(\eta  -k^0_1/v )
    \int\limits_0^{v_{\max}} d v\;  \sigma_{\born}^f(s(1-v)) \;
    \prod_{j=1}^n   \int {d^3 k_j\over k^0_j}  \tilde{S}_e(k_j)
\\
   &\Theta(k^0_1-k^0_2)\; \Theta(k^0_2-k^0_3) \dots \Theta(k^0_n-E_{\min})\;
    \delta\bigg( v  - {2KP\over s} +{K^2 \over s} \bigg)
    e^{ \gamma_e \ln\varepsilon_e}
\\
  =&\int\limits_0^{v_{\max}} d v\;  \sigma_{\born}^f(s(1-v)) \;
    \prod_{j=1}^n  \int_0^1      { d x_j   \over x_j} 
                   \int_0^{2\pi} { d\phi_j \over 2\pi} 
                   \int_{-1}^1    d\cos\theta_j\; {\alpha\over \pi} f(\cos\theta_j)
\\
   &\delta(v -x_1 )\;
    \Theta(x_1-x_2)\; \Theta(x_2-x_3) \dots \Theta(\lambda_0 x_n-\varepsilon)\;
    e^{ \gamma_e \ln\varepsilon_e}\; {\cal J}(\bar{K},v),
  \end{split}
\end{equation}
where $\eta_0$ is the solution%
\footnote{Note that for a single photon $A=0$ and $\eta_0 = s^{1/2}/2$.}
of  the equation $v-{2\bar{K}P\over s}\eta +{\bar{K}^2\over s}\eta^2=0$ and
${\cal J}(\bar{K},v)$ is an overall Jacobian factor:
\begin{equation}
  \label{eq:isr-integ2}
  \begin{split}
  &{\cal J}(\bar{K},v)
    = {v\over \eta_0}\; {1\over {2\bar{K}P\over s} -{\bar{K}^2\over s}2\eta_0}
    = {1\over 2} \bigg( 1+ {1\over \sqrt{1-Av}} \bigg),\;\;
\\
  & \eta_0 = {\sqrt{s}\over 2} {v\over \bar{K}^0}\;  {2\over 1+\sqrt{1-Av}}
           \equiv {\sqrt{s}\over 2} \lambda_0,\;\;
    A={\bar{K}^2 P^2 \over (\bar{K} P)^2}= {\bar{K}^2\over(\bar{K}^0)^2} \leq 1,\;\;
    0\leq \lambda_0 \leq 1,\;\;
  \end{split}
\end{equation}
and the photon angular distribution is governed by
\begin{equation}
  f(\cos\theta_j) = {2\over (1-\beta\cos\theta_j)(1+\beta\cos\theta_j)} 
                   -{2m_e^2\over s} {1\over (1-\beta\cos\theta_j)^2}   
                   -{2m_e^2\over s} {1\over (1+\beta\cos\theta_j)^2}.
\end{equation}

\subsubsection{Simplifications and MC generation}
Up to this point the ISR integral of eq.~(\ref{eq:isr-integ})
was transformed without any approximation and we maintain, 
modulo ordering of the photon energies,
the one-to-one correspondence
of the points in the Lorentz-invariant phase space and in the space of the new variables:
\begin{equation}
  \{ n,(\tilde{k}_1,\dots, \tilde{k}_{n}) \} 
   \leftrightarrow \{ n, ( x_j,\theta_j,\phi_j), j=1,\dots, n  \}.
\end{equation}
Before we define the {\em primary} differential distribution to be generated in the MC,
let us write once again explicitly
the two {\em equivalent} (modulo energy ordering)
parametrizations of the ISR {\em crude} differential distribution:
\begin{equation}
  \begin{split}
  &{d\FmI_n \over ds_X \prod\limits_{j=1}^n {d^3 k_j\over 2k^0_j} }
   = {1\over n!} \sigma_{\born}^f(s_X) \;
     \prod_{j=1}^n 2\tilde{S}_e(k_j) \Theta(k^0_j-E_{\min})\;
     e^{ \gamma_e \ln\varepsilon_e},\; n>0,
\\
  &{d\FmI_n \over dv \prod_{j=1}^n  dx_j d\cos\theta_j d\phi_j}
   = \sigma_{\born}^f(s(1-v)) \;
    \bigg({\alpha\over 2\pi^2}\bigg)^n
    \delta(v -x_1 )\;
    {\Theta(\lambda_0x_n-\varepsilon) \over x_n }
\\ &\qquad\qquad\qquad\qquad
    \prod_{j=1}^{n-1}  {\Theta(x_j-x_{j-1}) \over x_j }\;
    \prod_{j=1}^n      f(\cos\theta_j)\;\;
    e^{ \gamma_e \ln\varepsilon_e}\; {\cal J}(\bar{K},v),\; n>0,
\\
   &{d\FmI_0 \over ds_X } = \sigma_{\born}^f(s)\; \delta(s_X),\quad
    {d\FmI_0 \over dv } = \sigma_{\born}^f(s)\; \delta(v),\; n=0.
  \end{split}
\end{equation}
The {\em simplifications} leading to the ISR {\em primary} differential distribution
are the following:
\begin{equation}
  \label{eq:isr-simplify}
  \begin{split}
    &f(\cos\theta_j) \to \bar{f}(\cos\theta_j) = {2\over (1-\beta\cos\theta_j)(1+\beta\cos\theta_j)},
\\
    &{\cal J}(\bar{K},v) \to {\cal J}_0(v)
    = {1\over 2} \bigg( 1+ {1\over \sqrt{1-v}} \bigg),\;
\\
    &\Theta(\lambda_0x_n-\varepsilon) \to \Theta(x_n-\varepsilon),
  \end{split}
\end{equation}
where 
\begin{equation}
  \label{eq:bar-gamma-e}
  \bar\gamma_e = 2(\alpha/\pi) \ln(s/ m_e^2).
\end{equation}

The resulting ISR {\em primary} differential distribution is
\begin{equation}
  \label{eq:isr-prim-dist}
  \begin{split}
  &{d\FmI^{\rm Pri}_n \over dv \prod_{j=1}^n  dx_j d\cos\theta_j d\phi_j}
   = \sigma_{\born}^f(s(1-v)) \;
    \bigg({\alpha\over 2\pi^2}\bigg)^n
    \delta(v -x_1 )\;
    {\Theta(x_n-\varepsilon) \over x_n }
\\ &\qquad\qquad\qquad\qquad
    \prod_{j=1}^{n-1}  {\Theta(x_j-x_{j-1}) \over x_j }\;
    \prod_{j=1}^n      \bar{f}(\cos\theta_j)\;\;
    e^{ \gamma_e \ln\varepsilon_e}\; {\cal J}_0(v),\; n>0,
\\
   &{d\FmI^{\rm Pri}_0 \over dv } = \sigma_{\born}^f(s)\; \delta(v),\; n=0.
  \end{split}
\end{equation}
and the corresponding weight is
\begin{equation}
  \label{eq:wtcrud-isr}
  \begin{split}
    w_{\rm ISR}^{\rm Cru} = {d\FmI_n \over d\FmI^{\rm Pri}_n}
    = 
    \Theta(\lambda_0x_n-\varepsilon)\;
    {{\cal J}(\bar{K},v) \over {\cal J}_0(v)}\;
    \prod_{j=1}^n { f(\cos\theta_j) \over \bar{f}(\cos\theta_j)}\;.
  \end{split}
\end{equation}

Let us explain and justify the simplifications of eqs.~(\ref{eq:isr-simplify}).
The replacement $f(\cos\theta)\to \bar{f}(\cos\theta)$ 
is not really necessary in the present context of building an efficient MC algorithm for ISR. 
We could do without it, because the $f(\cos\theta)$ distribution is rather simple.
The problem is really in the model weight at \Order{\alpha^1} and higher orders.
As is well known in the soft limit, the 
helicity-non-conserving spin-amplitude contribution vanish, on the other hand,
a perfect helicity conservation contradicts the angular-momentum conservation for a photon
emitted exactly parallel to the respective emitting fermion, and this is reflected in the
photon distribution
\begin{equation}
  f(\cos\theta_j) = {2\sin^2\theta_j  \over [(1-\beta\cos\theta_j)(1+\beta\cos\theta_j)]^2},
\end{equation}
which has explicit zeros at $\cos\theta_j=\pm 1$.
When we admit the exact \Order{\alpha^1} hard photon emission matrix element,
then for hard photons these regions close to $\cos\theta_j=\pm 1$
will be filled in by the helicity-non-conserving contributions,
and the model weight based on $f(\cos\theta)$ would fluctuate wildly
when we approach in division by $f(\cos\theta)$, division by 0.
The solution is not to have these zeros at all, at the level of the primary distribution,
and this is why we opted for $\bar{f}(\cos\theta)$.
In such a case the product of the model and crude weight will be regular at \Order{\alpha^1}
and beyond.

The other two simplifications are introduced for purely technical reasons.
The simplification ${\cal J}\to {\cal J}_0$ is especially costly in terms of the MC
efficiency because, for $v\to 1$, it introduces the ``spurious'' singularity $(1-v)^{-1/2}$.
Together with the usual $(1-v)^{-1}$ from $\sigma_{\born}^f(s(1-v))$,
it builds up strong singularity $(1-v)^{-3/2}$ in the primary differential distribution,
and huge primary integrated cross section
$ \int^{1-4m_f^2/s} dv (1-v)^{-3/2} \sim  s^{1/2}/m_f$.
It is almost completely compensated by the very high rejection rate
of events close to $v=1$ due to the ratio
${\cal J}(\bar{K},v) / {\cal J}_0(v)$ in $w_{\rm ISR}^{\rm Cru}$.
The rejection rate is $\sim (m_f/s^{1/2})\ln(4m_f^2/s)$.
For muons 99\% of events are rejected.
However, in most of the applications, this effect can be easily eliminated by setting
$v_{\max}=0.999$ or lower.

The above problem is unfortunately unavoidable in the actual MC algorithm for the ISR.
It can be traced back to the fact that the present MC algorithm is not very well
suited for the emission of the two photons of large effective mass, 
such as the simultaneous emission of two hard photons along two beams.
The present algorithm ``folds in'' together the energies of photons emitted from both beams.
A more sophisticated algorithm, 
in which photons emitted from two beams are generated independently, 
is needed in order to eliminate this problem.

The $\Theta(\lambda_0x_n-\varepsilon)$ contribution to the weight has interesting consequences.
As discussed in refs.~\cite{yfs-mpi:1987,yfs2:1990} it leads directly%
\footnote{ As shown in refs.~\cite{yfs2:1990} this is not completely
  straightforward. In fact $F(\gamma_e)$ is not present for $v<\varepsilon$;
  it is nevertheless present in the integrated cross section, since the corresponding
  negative contribution is located just above $v=\varepsilon$.}
to a characteristic factor $F(\gamma_e)=e^{-C\gamma_e}/\Gamma(1+\gamma_e)$
in the $d\sigma/dv$ and in the total cross section.

Finally, let us integrate analytically the ISR {\em primary} differential distribution
\begin{equation}
  \label{eq:isr-prim-vdist}
  \begin{split}
    \FmI^{\rm Pri}
   &=\sum_{n=0}^\infty  \FmI^{\rm Pri}_n
    =\sum_{n=0}^\infty 
    \int\limits_0^{v_{\max}} d v\;  \sigma_{\born}^f(s(1-v)) \;
    \prod_{j=1}^n  \int_0^1      { d x_j   \over x_j} 
                   \int_0^{2\pi} { d\phi_j \over 2\pi} 
                   \int_{-1}^1     d\cos\theta_j\; {\alpha\over \pi} \bar{f}(\cos\theta_j)
\\  &\qquad\qquad\qquad\qquad
    \delta(v -x_1 )\;
    \Theta(x_1-x_2)\; \Theta(x_2-x_3) \dots \Theta(x_n-\varepsilon)\;
    e^{ \gamma_e \ln\varepsilon_e}\; {\cal J}_0(v)
\\
   &= \int\limits_0^{v_{\max}} d v\;  
    \sigma_{\born}^f(s(1-v)) \; {\cal J}_0(v)\; 
    e^{ \gamma_e \ln\varepsilon_e}\;
    \Bigg( \delta(v) 
           +\Theta(v-\varepsilon) {1\over v} \sum_{n=1}^\infty {1\over (n-1)!}
            \bigg( \bar\gamma_e \ln{v\over \varepsilon} \bigg)^{n-1}
    \Bigg)
\\
   &= \int\limits_0^{\varepsilon} d v\; \gamma_e v^{\gamma_e-1} \sigma_{\born}^f(s)
     +\int\limits_\varepsilon^{v_{\max}} d v\;
            \sigma_{\born}^f(s(1-v)) \; {\cal J}_0(v)\; 
            \bar\gamma_e v^{\bar\gamma_e-1} \varepsilon^{\gamma_e -\bar\gamma_e}.
  \end{split}
\end{equation}

How do we generate the primary differential distribution $d\FmI^{\rm Pri}$?
We start with the generation of $v$ according to
\begin{equation}
  {d\FmI^{\rm Pri} \over dv}= \sigma_{\born}^f(s(1-v)) \; {\cal J}_0(v)\;
            \bar\gamma_e v^{\bar\gamma_e-1} \varepsilon^{\gamma_e -\bar\gamma_e}. 
\end{equation}
This is done by using a general-purpose MC tool such as Vesko1 or Foam;
care is taken of any possible resonance or threshold in
the $\sigma_{\born}^f(s(1-v))$.
In the next step, photon multiplicity $n$ is generated.
For $v<\varepsilon$ we have simply $n=0$, and for $v>\varepsilon$
the photon multiplicity distribution is:
\begin{equation}
  \label{eq:isr-multiplicity}
  \FmI^{\rm Pri}_n 
  = {\rm const}\times\; {1\over (n-1)!} \bigg( \bar\gamma_e \ln{v\over \varepsilon} \bigg)^{n-1},
\end{equation}
which is just the shifted-by-one Poisson distribution   $P_{n-1}$, 
with the average $ \angle n-1\rangle = \bar\gamma_e \ln(v/ \varepsilon)$.
The angles $\cos\theta_j$ and $\phi_j$ are generated in the same way as in the previously
discussed case of FSR.

\subsection{Getting common IR boundary for FSR and ISR}
\label{sec:hiding}

Let us consider the case of EEX:
\begin{equation}
  \sigma^{(r)}_{\rm EEX} 
  = \int W^{(r)}_{\rm EEX}\; d\sigma^{\rm Cru}\;, 
\end{equation}
where the model weight  $W^{(r)}_{\rm EEX}$ is defined in eq.~(\ref{eq:wtmot-eex})
in terms of the \Order{\alpha^r} EEX differential distribution of eq.~(\ref{eq:rho-eex3}).
Using eq.~(\ref{eq:crude-start}) with the later substitution
\begin{equation}
  \varepsilon_f = \delta_f \bigg(1+ {2QK'\over s_Q} \bigg),\quad 
  K' =\sum_{i=0}^{n'} k'_i,
\end{equation}
which was introduced in order to facilitate the MC generation, we obtain
\begin{equation}
\label{eq:observable}
\begin{split}
  &\sigma^{(r)}_{\rm EEX} \{A\}
  =  \sum_{n=0}^\infty  \sum_{n'=0}^\infty \int
     d\sigma^{\rm Cru}_{[n,n']}(\Omega_I,\Omega_F)\;\; 
     A(n,k_1,\dots,k_n; n',k'_1,\dots,k'_{n'}; p_i,q_i)\;
\\ & \qquad\qquad\qquad\qquad\qquad\qquad \times 
     W^{(r)}_{\rm EEX}(n,k_1,\dots,k_n; n',k'_1,\dots,k'_{n'}; p_i,q_i)\;
\\ &
     d\sigma^{\rm Cru}_{[n,n']}(\Omega_I,\Omega_F)
     \equiv
     d s_X \; { \sigma_{\born}(s_X) \over 4\pi }\; 
     d\tau_{n+1}  ( P; k_1,\dots,k_n, X)\;
     e^{ \gamma_e \ln\varepsilon_e} 
     {1\over n!} \prod_{j=1}^n    2\tilde{S}_e(k_j)  \bar\Theta(\Omega_I,k_j)\;
\\ & \qquad \times
     d\tau_{n'+2} ( X; k'_1,\dots,k'_{n'}, q_1,q_2)\;
     {s_X\over s_Q}\; {2 \over \beta_f } \;
     e^{ \gamma_f \ln\big( \delta_f {s_Q+2K'Q\over s_Q} \big) }\;
     {1\over n'!}\prod_{l=1}^{n'} 2\tilde{S}_f(k'_l) \bar\Theta(\Omega_F,k'_l)\;,
\end{split}
\end{equation}
where for the sake of the discussion of the IR cancellations we have introduced
a general {\em acceptance function} $A$.
Every physical, i.e. IR-safe, observable corresponds uniquely to one
or more such acceptance functions.
Just to give an example: the total cross section corresponds to $A\equiv 1$,
the forward--backward asymmetry is related to $A$ expressed in terms
of final fermion momenta like $A=\Theta(q_1^3)$,
the cross section for the production of exactly two photons above $E_0=1$ GeV
corresponds to $A= \sum_{i,j} \Theta(k_i^0-E_0)\Theta(k_j^0-E_0)$, and so on.
The acceptance function corresponding to a physically meaningful, IR-safe,
observable has to obey one important rule
\begin{equation}
  \label{eq:ir-safe-rule}
  \lim_{k_i\to 0}  A(  n,k_1,\dots,k_{i-1},k_i,k_{i+1},\dots,k_n)
                  =A(n-1,k_1,\dots,k_{i-1}    ,k_{i+1},\dots,k_n),
\end{equation}
and a similar rule should hold for FSR photons
(in EEX we can make a distinction between ISR and FSR photons
because we neglect ISR--FSR interference).

So far we kept the IR domains different for ISR and FSR, for ISR 
$\Omega_I$ was defined by: $k^0_j< \varepsilon_e {1\over 2} \sqrt{s}$
in the laboratory CMS system where $\vec{p}_1+\vec{p}_2=0$,
while for FSR $\Omega_F$ was defined by
${k'}^0_j< \delta_f  ((s_Q+2K'Q)/ s_Q) {1\over 2} \sqrt{s_Q} $
in the QMS system where $\vec{q}_1+\vec{q}_2=0$.
Our task is now to bring the two IR domains together.

We know that the total cross section and
any IR-safe observable are completely independent of $\Omega_F$ and $\Omega_I$.
The self-suggesting solution is, loosely speaking,
to set $\delta_f$ so small that we always have  $\Omega_F \subset \Omega_I$,
and simply neglect all FSR photons 
$k'_i \in \delta\Omega = \Omega_I \setminus \Omega_F $,
that is just remove them from the list of the generated momenta in the MC.
Note that because $(s_Q+2K'Q)/ s_Q \sim s_X/s_Q \ll s_X/(4m_f^2)$
we may need $\delta_f/\varepsilon \ll s/(4m_f^2)$.

Let us work out the details of the above method, providing a formal proof of its validity.
The above prescription definitely leads to a certain new
crude distribution
\begin{displaymath}
       d\sigma^{\rm Cru^*}_{[n,n']}(\Omega_I,\Omega_I)
\end{displaymath}
in which the IR-domain $\Omega=\Omega_I$ is common for ISR and FSR photons.
The question is: 
What is the above new crude distribution?

It turns out to be calculable analytically. 
(In the calculation we follow closely the algebra of 
the formal proof of the independence of the physical observables on the IR domain
$\Omega$ as given in ref.~\cite{ceex2:1999}.)
Let us consider the internal FSR subintegral in
eq.~(\ref{eq:observable}), that is all ISR photon momenta are fixed;
\begin{equation}
\begin{split}
  &\FmI \{A\}
  =\sum_{n'=0}^\infty
  \int d\tau_{n'+2} ( X; k'_1,\dots,k'_{n'}, q_1,q_2)\;
  {1\over n'!}\prod_{l=1}^{n'} 2\tilde{S}_f(k'_l) \bar\Theta(\Omega_F,k'_l)\;
  b(k'_1,\dots,k'_{n'}; p_i,q_i)
\\ &
  b(k'_1,\dots,k'_{n'}; p_i,q_i) \equiv
  e^{ \gamma_f \ln\big( \delta_f {s_Q+2K'Q\over s_Q} \big) }\;
  {s_X\over s_Q}\; {2 \over \beta_f } \;
\\ & \qquad\qquad
  \times W^{(r)}_{\rm EEX}(n,k_1,\dots,k_n; n',k'_1,\dots,k'_{n'}; p_i,q_i)\;
                         A(n,k_1,\dots,k_n; n',k'_1,\dots,k'_{n'}; p_i,q_i).
  \end{split}
\end{equation}
Following $\Omega_I=\Omega_F \bigcup \delta\Omega$ 
we split every photon integral into two parts and reorganize the sum
factorizing out the integral over $\delta\Omega$:
\begin{equation}
\label{eq:binomial}
\begin{split}
  \FmI \{A\}
  =&\sum_{n'=0}^\infty
  {1\over n'!}\prod_{l=1}^{n'}
  \bigg\{
     \int {d^3k'_l\over {k'}_l^0}\; \Theta(\delta\Omega,k'_l)\; \tilde{S}_f(k'_l)
    +\int {d^3k'_l\over {k'}_l^0}\; \bar\Theta(\Omega_I,k'_l)\; \tilde{S}_f(k'_l)
  \bigg\}
\\ &
  \int d\tau_{n'+2} ( X, k'_i; q_1,q_2)\;
  b(k'_1,\dots,k'_{n'}; p_i,q_i)
\\
  =&\sum_{n'=0}^\infty
  {1\over n'!}
    \sum_{s=0}^{n'} \left( {n'}\atop s \right)
  \bigg\{
     \int {d^3k\over 2k^0}\; \Theta(\delta\Omega,k')\; \tilde{S}_f(k')
  \bigg\}^s
\\ &
  \int d\tau_{n'+2-s} ( X-\sum^s_1; k'_1,\dots,k'_{n'-s},q_1,q_2)\;
  \prod_{l=1}^{n'-s} \bar\Theta(\Omega_I,k'_l)\; \tilde{S}_f(k'_l)\;
  b(k'_1,\dots,k'_{n'-s}; p_i,q_i),
  \end{split}
\end{equation}
where $\Theta(\delta\Omega,k')=1$ for $k'\in \delta\Omega$ and $=0$ otherwise.
The most important ingredient in the above algebraic transformation was
that the model weight $W^{(r)}_{\rm EEX}$,
due to the particular expansion of $\rho^{(r)}_{\rm EEX}$ into $\bbeta$-components, 
see eq.~(\ref{eq:rho-eex3}),
also fulfils the ``IR-safeness'' condition
\begin{equation}
  \label{eq:ir-safeness}
  \lim_{k'_i\to 0} 
   W^{(r)}_{\rm EEX}(n',  k'_1,\dots,k'_{i-1},k'_{i},k'_{i+1},\dots,k'_{n'})
  =W^{(r)}_{\rm EEX}(n'-1,k'_1,\dots,k'_{i-1},       k'_{i+1},\dots,k'_{n'}),
\end{equation}
and consequently the function  $b(k'_1,\dots,k'_{n'}; p_i,q_i)$ as well.
The resulting integral
\begin{equation}
  \label{eq:new-integral}
  \begin{split}
  \FmI \{A\}
  =&\sum_{n'=0}^\infty
  \int d\tau_{n'+2} ( X; k'_1,\dots,k'_{n'}, q_1,q_2)\;
  {1\over n'!}\prod_{l=1}^{n'} 2\tilde{S}_f(k'_l) \bar\Theta(\Omega_I,k'_l)\;
\\ &
  \exp\bigg( \int {d^3k\over 2k^0}\; \Theta(\delta\Omega,k)\; 2\tilde{S}_f(k) \bigg)
  b(k'_1,\dots,k'_{n'}; p_i,q_i)  
  \end{split}
\end{equation}
gets an additional exponential factor, which is easy to interpret. 
It can be expressed in terms of the function
\begin{displaymath}
\tilde{B}(\Omega,q_1,q_2)
 = -{ 1 \over 8\pi^2} \int {d^3k\over k^0} \Theta(\Omega;k) 
   \bigg({q_1\over kq_1} - {q_2\over kq_2} \bigg)^2
\end{displaymath}
as follows
\begin{displaymath}
  \exp\bigg( \int {d^3k\over 2k^0}\; \Theta(\delta\Omega,k)\; 2\tilde{S}_f(k) \bigg)
 =\exp\big( 2Q_f^2\alpha\tilde{B}(\Omega_I,q_1,q_2) 
           -2Q_f^2\alpha\tilde{B}(\Omega_F,q_1,q_2) \big).
\end{displaymath}

We have therefore found out by explicit calculation that in the proposed method,
in which for $\Omega_F$ much smaller than $\Omega_I$
we skip photons that fall into 
$\delta\Omega = \Omega_I \setminus \Omega_F$,
the distribution of the remaining photons is the following
\begin{equation}
\label{eq:crude-star}
\begin{split}
  &d\sigma^{\rm Cru^*}_{[n,n']}(\Omega_I,\Omega_I)
   = d s_X \; { \sigma_{\born}(s_X) \over 4\pi }\; 
  d\tau_{n+1}  ( P; k_1,\dots,k_n, X)\;
  e^{ \gamma_e \ln\varepsilon_e} 
  {1\over n!} \prod_{j=1}^n    2\tilde{S}_e(k_j)  \bar\Theta(\Omega_I,k_j)\;
\\ &\qquad \qquad \times
  d\tau_{n'+2} ( X; k'_1,\dots,k'_{n'}, q_1,q_2)\;
  {s_X\over s_Q}\; {2 \over \beta_f } \;
  e^{ R_F(\Omega_I)  }\;
  {1\over n'!}\prod_{l=1}^{n'} 2\tilde{S}_f(k'_l) \bar\Theta(\Omega_I,k'_l)\;
\\ &
  R_F= \gamma_f \ln\left( \delta_f {s_Q+2K'Q\over s_Q} \right)
       +2Q_f^2\alpha\tilde{B}(\Omega_I,q_1,q_2) -2Q_f^2\alpha\tilde{B}(\Omega_F,q_1,q_2).
\end{split}
\end{equation}
Also note that,  {\em by construction}, the integral value is preserved
\begin{displaymath}
 \sum_{n,n'} \int d\sigma^{\rm Cru^*}_{[n,n']}(\Omega_I,\Omega_I)
=\sum_{n,n'} \int d\sigma^{\rm   Cru}_{[n,n']}(\Omega_I,\Omega_F).
\end{displaymath}

Now, since the IR boundary in the above new distribution has changed for FSR photons,
we cannot continue to use the  $\rho^{(r)}_{EEX}$ of eq.~(\ref{eq:rho-eex3}).
We have to use another $\rho^{*(r)}_{EEX}$ in which we replace 
$\tilde{B}(\Omega_F)$ by $\tilde{B}(\Omega_I)$
in the YFS form factor, see eq.~(\ref{eq:Y-function}):
\begin{displaymath}
  \rho^{*(r)}_{\rm EEX} = \rho^{(r)}_{\rm EEX} 
   e^{2Q_f^2\alpha( \tilde{B}(\Omega_I,q_1,q_2)
                   -\tilde{B}(\Omega_F,q_1,q_2))},
\end{displaymath}
and consequently, 
since the model weight is the ratio of  the model and crude distributions, 
the new exponential factors cancel out,
and the new model weight is functionally exactly the same
\begin{displaymath}
  W^{*(r)}_{\rm EEX} = W^{(r)}_{\rm EEX}. 
\end{displaymath}
In the new MC calculation 
\begin{equation}
  \sigma^{(r)}_{\rm EEX} 
  = \int W^{*(r)}_{\rm EEX}\; d\sigma^{\rm Cru^*}\; 
  = \int W^{(r)}_{\rm EEX}\; d\sigma^{\rm Cru^*}\; 
\end{equation}
both the product of the weights and the normalization is the same.
The above result is so trivial that, in fact, in the MC program
for the EEX model we change almost nothing -- 
we are only omitting hidden photons in the calculation of the model weight.
This simplicity reflects the basic fact that very soft photons are unimportant for 
all IR-safe integrand functions.

The term $\gamma_f \ln(\dots)$ in $R_F$ is cancelled by $\tilde{B}(\Omega_F)$
and there is in fact no dependence on $\Omega_F$ nor $\delta_f$ 
in $d\sigma^{\rm Cru^*}_{[n,n']}(\Omega_I,\Omega_I)$ any more.
The IR cancellation is now assured by the term
\begin{displaymath}
  2{\alpha\over\pi} \left( \ln{2q_1q_2\over m_f^2} -1 \right) \ln \varepsilon,
\end{displaymath}
which is implicitly present in $\tilde{B}(\Omega_I,q_1,q_2)$.

The situation is, however, not as good as we described above.
There is one important complication due
to the use of the weighted events, at the level of the crude distribution.
Let us go back again to the case of EEX
\begin{equation}
  \sigma^{(r)}_{\rm EEX} 
  = \int W^{(r)}_{\rm EEX}\;  W^{\rm Cru}_{\rm FSR}\; W^{\rm Cru}_{\rm ISR}\;  d\sigma^{\rm Pri}.
\end{equation}
Now the problem is that photons in $\delta\Omega$ cannot be ``hidden'',
because $W^{\rm Cru}_{\rm FSR}$ does not obey the ``IR-safeness'' condition
\begin{displaymath}
  \lim_{k'_i\to 0} 
   W^{\rm Cru}_{\rm FSR}(n',  k'_1,\dots,k'_{i},    \dots,k'_{n'})
  =W^{\rm Cru}_{\rm FSR}(n'-1,k'_1,\dots,k'_{i-1},  k'_{i+1},\dots,k'_{n'})
   {f\bigg(\theta_i, {m_f^2\over s_Q}\bigg) \over \bar{f}\bigg(\theta_i, {m_f^2\over s}\bigg) }.
\end{displaymath}
Even the softest photons contribute the finite ratio $(f/\bar{f})$, and this contribution
is essential for the IR-cancellations and for the overall normalization.

There is, however, a way of saving our method of replacing $\Omega_F$ with  $\Omega_I$.
Let us repeat again the calculation of eq.~(\ref{eq:binomial})
assuming that photons  ``hidden'' inside $\delta\Omega$
do not contribute the factor $(f/\bar{f})$ to the overall weight.
We are able to carry out the calculation as before, obtaining the modified exponential factor
\begin{equation}
  \begin{split}
  \FmI' \{A\}
  =&\sum_{n'=0}^\infty
  \int d\tau_{n'+2} ( X; k'_1,\dots,k'_{n'}, q_1,q_2)\;
  {1\over n'!}\prod_{l=1}^{n'} 2\tilde{S}_f(k'_l) \bar\Theta(\Omega_I,k'_l)\;
\\ &
  \exp\left( \int_{\delta\Omega} {d^3k\over k^0}\; \tilde{S}_f(k) 
            {\bar{f}(\theta, m_f^2/s) \over f(\theta, m_f^2/s_Q) }
      \right)
  b(k'_1,\dots,k'_{n'}; p_i,q_i),
  \end{split}
\end{equation}
It is very important that the effect due to the omission of $(f/\bar{f})$ in the overall weight
can be evaluated analytically, 
and therefore {\em corrected for} analytically.
In other words we shall be able to compensate analytically for the
missing average contribution to $W^{\rm Cru}_{\rm FSR}$ from the hidden photons.
The evaluation of the integral over $\delta\Omega$ 
is based on the observation that
\begin{displaymath}
  \tilde{S}^*_f(k) = \tilde{S}_f(k) {\bar{f}(\theta, m_f^2/s) \over f(\theta, m_f^2/s_Q) }
                   = - Q_f^2 {\alpha \over 4\pi^2}
                     \Bigg( {q^*_1\over kq^*_1} -{q^*_2\over kq^*_2} \Bigg)^2,
\end{displaymath}
where $q^*_i,i=1,2$, are defined such that $(q^*_i)^2=m_f^2 (s_Q/s)$.
Furthermore, in the QMS, they have the same directions as the original $\vec{q}_i$
and the same total energy,  $q^{*0}_1+q^{*0}_2=\sqrt{s_Q}$.
With the help of the above we get
\begin{displaymath}
  I_{\delta\Omega}
  =\int_{\delta\Omega} {d^3k\over k^0}\; \tilde{S}_f(k) 
            {\bar{f}(\theta, m_f^2/s) \over f(\theta, m_f^2/s_Q) }
  = 2\alpha Q_f[ \tilde{B}(\Omega_I,q_1^*,q_2^*) - \tilde{B}(\Omega_F,q_1^*,q_2^*)].
\end{displaymath}
We have at our disposal an analytical representation of the function $\tilde{B}(\Omega,q_1,q_2)$,
for spherical $\Omega$ and regularized with $m_\gamma$,
in terms of logarithmss and dilogarithms, for arbitrary  $q_i$, not necessarily antiparallel.
Although $I_{\delta\Omega}$ is IR-finite by definition, it is useful to keep $m_\gamma$
and evaluate separately $\tilde{B}(\Omega_I)$ in the CMS and $\tilde{B}(\Omega_F)$ in QMS,
where the corresponding IR-boundaries are spherical,
and subtract the results afterwards.
We may calculate $\tilde{B}$'s in any frame, 
because they are (when regularized with $m_\gamma$) essentially Lorentz-invariant --
we only need to transform IR-boundaries $\Omega$ correctly from one frame to another.

Summarizing: in the realistic case of the weighted events (with non-IR-safe weights)
the method in which we hide photons in $\delta\Omega = \Omega_I \setminus \Omega_F$
leads to a new crude distribution similar to that in eq.~(\ref{eq:crude-star}) with the new
\begin{equation}
  R_F= \gamma_f \ln\left( \delta_f {s_Q+2K'Q\over s_Q} \right)
       +2Q_f^2\alpha\tilde{B}(\Omega_I,q^*_1,q^*_2) -2Q_f^2\alpha\tilde{B}(\Omega_F,q^*_1,q^*_2).
\end{equation}
As a consequence, the above exponential factor
does not cancel out exactly in the model weight
with the correction to the YFS form factor as before,
and we have the following additional correcting factor in the model weight:
\begin{equation}
  \label{eq:wt-correction}
  \begin{split}
   W_{\rm hide}=   
     e^{
          -2\alpha Q_f[ \tilde{B}(\Omega_I,q_1^*,q_2^*) - \tilde{B}(\Omega_F,q_1^*,q_2^*)]
          +2\alpha Q_f[ \tilde{B}(\Omega_I,q_1,q_2)     - \tilde{B}(\Omega_F,q_1,q_2)]
        }.
  \end{split}
\end{equation}
It should be really present in the model weight, 
but in the program, for historical reasons and for convenience, 
it is included in the crude weight.

The important profit from the above method is that
with the above fix  we can now make our calculation for the CEEX
model with the ISR-FSR interference (IFI) switched on.

Note that the above treatment is more elaborate than the analogous one in BHLUMI because
it is valid for finite $m_f$, while in BHLUMI we use the approximation $m_f \ll \sqrt{s}$.
The correcting weight in BHLUMI is a simple one-line expression while here it expressed by a long
series of logarithms and dilogarithms.

Let us finally add a side remark:
another valid method of realizing a hypothetical Monte Carlo with common $\Omega$
for ISR and FSR is to generate photons using small $\Omega_F$ and to apply
a brute-force rejection of all events with one or more photons falling into $\Omega_I$.
This was used in the early version of BHLUMI.
We do not like this method because it may lead to an excessive number of events with zero
weight, lowering substantially the efficiency of the MC.

\subsection{Photon multiplicity enrichment}
\label{sec:enrichment}
Let us finally describe yet another complication of the {\em primary}
distribution, which is introduced for technical reasons, i.e. in order to get
better total weight distribution, and a smaller rejection rate
in the process of turning weighted events into unweighted events.
This modification is not necessary for weighted events.
(It can be switched off by adjusting input data).

The problem is essentially due to the introduction of the ISR-FSR interference, which we have already
called IFI.
As already known from the \Order{\alpha^1} case, the weight that introduces IFI
is sharply peaked around 1 and has a strict upper bound $W_{\rm IFI}\leq 2$.
The destructive interference with the weights $W_{\rm IFI}\sim 0$ occurs
in the backward scattering (in the fermion scattering angle) and there is 
a little constructive interference $W_{\rm IFI}\sim 2$, mostly in the forward direction.
The maximum weight $=2$ or a factor of $2$ in the {\em primary} primary cross section
solves the problem, at the expense of the factor of 2 rejection 
rate~\cite{mustraal-cpc:1983,koralb:1985}.

In the case of $n$ photons, however, 
the same leads to: the maximum weight $=2^n$ or equivalently
the increase of the {\em primary} cross section by a factor of $2^n$,
and consequently increase of the generated photon multiplicity by a factor of 2.
Of course, almost all of this increase is artificial, and it is compensated by $W_{\rm IFI}$,
leaving only some small net effect due to IFI in the MC events.
A more sophisticated method would be to increase the {\em primary} primary cross section
(and photon multiplicity) more selectively, that is in a way dependent on the
fermion scattering angle.
For the moment we do not do it. It may be done in the future.
What we do is the following:
we increase the {\em primary} cross section by a factor of $2^\lambda$,
where $\lambda$ is not equal to 2, but is
adjusted empirically so that the tail of the total MC weight is  acceptable. 
We have found that the value $\lambda\sim 1.25$ is the optimal one.

The introduction of the above $\lambda$ factor  affects all formulas 
for the {\em primary} cross section and for $W^{\rm Cru}$ in a rather trivial way,
so we do not write it explicitly.
The only non-trivial modification is in the compensating factor for the {\em hidden}
photons in the previous section.
This can be understood and implemented as a modification of the electric
charge of the final fermion $Q_f\to \lambda Q_f$
in the {\em primary} cross section.

\subsection{Entire MC algorithm top-to-bottom}

Specializing to the CEEX model,
we can summarize the results of the three previous sections as follows
\begin{equation}
  \sigma^{(r)}_{\rm CEEX} \{A\}
  = \sum_{f=\mu,\tau,d,u,s,c,b}\;
    \sum_{n=0}^\infty \sum_{n'=0}^\infty \int
    A\; W^{(r)}_{\rm CEEX}\;  W^{\rm Cru}_{\rm FSR}\; W^{\rm Cru}_{\rm ISR}\; W_{\rm hide}\;
    d\sigma^{\rm Pri^*}_{[n,n']}(\Omega_I).
\end{equation}
Here $d\sigma^{\rm Pri^*}_{[n,n']}(\Omega_I)$ 
is obtained from the product of the ISR and FSR
primary differential distributions
\begin{equation}  
  d\sigma^{\rm Pri}_{[n,n']}(\Omega_I,\Omega_F)
  = d\FmI^{\rm Pri}_n(\Omega_I)\;
    d\Fmf^{\rm Pri}_{n'}(\Omega_F),
\end{equation}
see eqs.~(\ref{eq:fsr-prim-dist}) and (\ref{eq:isr-prim-dist}),
by means of hiding/ignoring FSR photons in $\delta\Omega$.
Consequently,
in the evaluation of $W^{(r)}_{\rm CEEX}$ and of all other weights,
only momenta outside the common IR-domain $\Omega_I$ enter.

The value of the integrated cross section 
with the acceptance function $A$
is obtained in the MC run in a standard way
\begin{equation}
  \sigma^{(r)}_{\rm CEEX} \{A\}
  = \langle A\; W^{(r)}_{\rm CEEX}\;  W^{\rm Cru}_{\rm FSR}\; W^{\rm Cru}_{\rm ISR}\; W_{\rm hide}\rangle
    \sigma^{\rm Pri^*}.
\end{equation}
The acceptance function $A$ may for instance define
the entire cross section ($A\equiv 1$), or just a single bin in the histogram 
of $\cos\theta$ for the outgoing fermion, 
or any other IR-safe observable.
The overall normalization is based on
\begin{equation}
  \begin{split}
    \sigma^{\rm Pri^*}
   &= \sum_{f=\mu,...,b}\;
      \sum_{n=0}^\infty \sum_{n'=0}^\infty  \int
            d\sigma^{\rm Pri^*}_{[n,n']}(\Omega_I)
    =  \sum_{n=0}^\infty  \int  d\FmI^{\rm Pri}_n(\Omega_I)\;
       \sum_{n'=0}^\infty \int  d\Fmf^{\rm Pri}_{n'}(\Omega_F)
\\ &
    =  \sum_{f=\mu,...,b}\;
       \sum_{n=0}^\infty  \int  d\FmI^{\rm Pri}_n(\Omega_I)\; 
    =  \sum_{f=\mu,...,b}\;
       \int\limits_0^1 d v\;
            \sigma_{\born}^f(s(1-v)) \; {\cal J}_0(v)\; 
            \bar\gamma_e v^{\bar\gamma_e-1} \varepsilon^{\gamma_e -\bar\gamma_e},
  \end{split}
\end{equation}
where we have exploited the property $\int \sum d\Fmf^{\rm Pri}(\Omega_F)\equiv 1$
of eq.~(\ref{eq:fsr-primary}),
and the ISR part is taken from eq.~(\ref{eq:isr-prim-vdist}).
Note that we have put $v_{\max}=1$, understanding that 
$\sigma_{\born}^f(s)=0$ below the threshold, $s<4m_f^2$.

We shall now describe the entire generation of the MC event according to 
$d\sigma^{\rm Pri^*}_{[n,n']}(\Omega_I)$
from the top to the bottom, as is done in the program,
starting from the generation of $v$ describing the total energy loss due to ISR,
the type of final fermion $f$ and the photon multiplicities $n$ and $n'$.
Generation of photon energies and angles comes later, 
using methods already described in detail in the previous section.

\subsection{ISR spectrum and fermion type}
First comes the important practical question: Shall we 
(a) generate first the fermion type $f$ and later $v$ or (b) vice versa? 
Both options are technically realizable.
In case (a) we would calculate numerically
\begin{equation}
    \sigma^{\rm Pri^*}_f=
       \int\limits_0^1 d v\;
            \sigma_{\born}^f(s(1-v)) \; {\cal J}_0(v)\; 
            \bar\gamma_e v^{\bar\gamma_e-1} \varepsilon^{\gamma_e -\bar\gamma_e},
\end{equation}
for $f=\mu,\tau,d,u,s,c,b$, and generate fermion type $f$ first; 
and later on, for a given $f$, we would generate the variable $v$ according to
\begin{equation}
    {d\sigma^{\rm Pri^*}_f \over dv}=
            \sigma_{\born}^f(s(1-v)) \; {\cal J}_0(v)\; 
            \bar\gamma_e v^{\bar\gamma_e-1} \varepsilon^{\gamma_e -\bar\gamma_e}.
\end{equation}
This looks as a natural solution; however, the generation and integration
of the ISR spectrum $d\sigma^{\rm Pri^*}_f / dv$ is done numerically in a MC module
that creates a look-up matrix, which memorizes very precisely the shape of
the distribution during the initialization phase of the MC run
(before MC event generation).
In this method we would need several initializations, creating several tables of
this kind.
This is feasible, but not very convenient.
The situation is much worse, when beamstrahlung is switched on because in this case
the 1-dimensional problem of the generation of $v$ is replaced with the 3-dimensional
problem of generating $v,z_1,z_2$, and consequently we would need to manage
several sets of 3-dimensional look-up matrices.
This would make the initialization phase rather long in CPU time, and the tables
would occupy a lot of processor memory.

We think that the above scenario is still technically realizable, 
even in the presence of beamstrahlung.
Nevertheless, we decided for option (b), which is in our opinion more economical.
In this case, we generate first the $v$ variable (the case of beamstrahlung is described below)
according to a distribution summed up over final--state flavour:
\begin{equation}
    \label{eq:primary-v}
    {d\sigma^{\rm Pri^*} \over dv}=
            \sum_{f=\mu,...,b}\;
            \sigma_{\born}^f(s(1-v)) \; {\cal J}_0(v)\; 
            \bar\gamma_e v^{\bar\gamma_e-1} \varepsilon^{\gamma_e -\bar\gamma_e};
\end{equation}
next, for a given $v$, we generate the final-state flavour $f$
according to a probability
\begin{equation}
  \label{eq:flavour-dist}
  P_f= {\sigma_{\born}^f(s(1-v))  \over \sum\limits_{g=\mu,...,b} \sigma_{\born}^g(s(1-v))}.
\end{equation}

\subsection{Inclusion of beamstrahlung}
In the presence of beamstrahlung, 
the flavour-summed three-dimensional distribution to be generated
in the very beginning of the MC algorithm is
\begin{equation}
  \label{eq:beamstrahlung-primary}
   {d\sigma^{\rm Pri^*} \over dv dz_1 dz_2}=
            \sum_{f=\mu,...,b}\;
            \sigma_{\born}^f(s(1-v)z_1 z_2) \; {\cal J}_0(v)\; 
            \bar\gamma_e v^{\bar\gamma_e-1} \varepsilon^{\gamma_e -\bar\gamma_e}
            {\cal D}(z_1,z_2,\sqrt{s});
\end{equation}
see also eq.~(\ref{eq:beamstrahlung}).
The above 3-dimensional
distribution is explored and memorized in the look-up matrices in the initialization phase
of the MC run.
This allows us to generate $v,z_1,z_2$ in a very efficient way for arbitrary
$\sigma_{\born}$ and arbitrary beamstrahlung structure function ${\cal D}(z_1,z_2,\sqrt{s})$.
As discussed previously, we admit in ${\cal D}(z_1,z_2,\sqrt{s})$
$\delta$-like singularities in $z_i$ and, as a result,
the MC integration of $d\sigma^{\rm Pri^*} / dv dz_1 dz_2$ is split into three
branches with three separate look-up matrices.
The above organization assures that the beamstrahlung structure function can
be a completely arbitrary ``user function''.

Next, for a given set $(v,z_1,z_2)$, the final fermion flavour $f$ is generated 
with the probability
\begin{equation}
  \label{eq:flavour-dist2}
  P_f= {\sigma_{\born}^f(s(1-v)z_1z_2)  
       \over \sum\limits_{g=\mu,...,b} \sigma_{\born}^g(s(1-v)z_1z_2)}.
\end{equation}

\subsection{Photon multiplicities and momenta}

Having defined the total ISR loss variable $v$ or $vz_1z_2$,
the ISR photon momenta are generated first, and the FSR photon momenta are generated second.
For $v<\varepsilon$  the ISR photon multiplicity is zero and for $v>\varepsilon$
it is generated according to the shifted Poisson distribution $P_{n-1}$ with
$ \langle n\rangle=\bar\gamma_e \ln(v/ \varepsilon)$,
see eq.~(\ref{eq:isr-multiplicity}), 
where, in the presence of beamstrahlung, the modified $s'=sz_1z_2$ enters instead of $s$
into the definition of $\bar\gamma_e$,
see eq.~(\ref{eq:bar-gamma-e}).
Next, all ISR photons are generated according to the distribution 
$d\FmI^{\rm Pri}_n$ of eq.~(\ref{eq:isr-prim-dist}),
with the methods already described.
The crude weight of eq.~(\ref{eq:wtcrud-isr}) is calculated.

Having generated ISR photons (and optionally beamstrahlung) we now know the total
four-momentum of the final fermions plus FSR momenta
\begin{equation}
  X = p_a z_1 + p_b z_2 -\sum_{j=1}^n k_j,
\end{equation}
but to start generation of the FSR momenta we need to know only $s_X=X^2$.
First, the FSR photon multiplicity $n'$ is generated according to a Poisson distribution
with the average defined in eq.~(\ref{eq:fsr-crude-multiplicity}).
Then, FSR momenta are generated in the rest frame of $Q=q_1+q_2$ (QMS).
More precisely,
their dimensionless energy parameters and angles are generated,
according to the corresponding FSR primary distribution of eq.~(\ref{eq:fsr-prim-dist}),
such that the $s_Q=Q^2$ needs not be known.
The $s_Q$ is determined as a fraction of $s_X$ with the help of eq.~(\ref{eq:sQ}),
such that photon four-momenta can be constructed in absolute (GeV) units
in QMS.

At this point we need to generate angles $\psi$ and $\omega$
in the transformation  form QMS down to CMS defined
in eq.~(\ref{eq:LA-transform}).
Knowing the momentum $\hat{X} = \hat{Q}-\sum_j \tilde{k}'_j$ in the QMS
we may apply this transformation and calculate final fermion momenta in the
CMS where $\vec{p}_1+\vec{p}_2=\vec{p}_a+\vec{p}_b=0$.
In the case of beamstrahlung this transformation bring us to a frame
where $\vec{p}_1z_1+\vec{p}_2z_2=0$, and we need an additional boost along beams to
brings generated momenta to the laboratory system.
The same boost is done for ISR photons.

Removal (hiding) of the FSR photons in $\delta\Omega$ is done at the end
of the generation of the FSR photons.
All remaining photons have the common IR-domain $\Omega_I$ defined in
the CMS where 
$\vec{p}_1+\vec{p}_2=\vec{p}_a+\vec{p}_b=0$ or in the presence of beamstrahlung
in the frame where $\vec{p}_1z_1+\vec{p}_2z_2=0$.

\newpage
\section{Structure of the program}
\label{sec:structure}
In the following we shall describe the topography of the distribution directory,
then the programming rules which we follow, and finally we shall  briefly describe
the functionality of the principal modules of the program.

\subsection{Topography of the distribution}
The program source code is organized into {\em modules},
also called {\em pseudo-classes}, which are located in several
Unix-type subdirectories of the distribution directory {\sf KK-all}.
The distribution directory also contains
one additional subdirectory, {\sf ffbench},
with demonstration (template) programs
and one subdirectory, {\sf dok}, with the documentation.
The essential part of the source code of the \KK\ Monte Carlo event generator
is located in the two subdirectories {\sf KK2f} and {\sf bornv}.
The tool box of various utilities is located in the subdirectory {\sf glibk},
the electroweak library is located in {\sf dizet}, the
MC library of the $\tau$ lepton decays TAUOLA 
is in the subdirectory {\sf tauola}, PHOTOS in the subdirectory 
{\sf photos}  and the hadronization package JETSET is in
the subdirectory {\sf jetset}.

\begin{figure}[!ht]
\centering
\setlength{\unitlength}{0.1mm}
\begin{picture}(1200,1200)
\thicklines
\put(     00,1150){\makebox(0,0)[b]{\framebox{\hbox{\tt  KK2f-all}}}}
\drawline( 00,1150)(  00, 100)        
\put(100,1055){\makebox(0,0)[b]{\hbox{\tt  KK2f}}}
\put( 00,1050){\line(1,0){200}}      
\put(200,1050){\makebox(0,0)[l]{\framebox{\parbox{75mm}{\scriptsize
              \setlength{\rightskip}{0pt plus 2cm}
              {\it Event generator: } 
              {\bf  KK2f, KarLud, KarFin,  GPS, QED3, HepEvt, PseuMar, TauPair}
}}}}
\put(100, 905){\makebox(0,0)[b]{\hbox{\tt  bornv}}}
\put( 00, 900){\line(1,0){200}}      
\put(200, 900){\makebox(0,0)[l]{\framebox{\parbox{75mm}{\scriptsize
              \setlength{\rightskip}{0pt plus 2cm}
              {\it Event generator: } 
              {\bf BornV, MBrA, MBrB, Vesk1, BStra, FoamA, FoamB, FoamC, VegasA,B,C}
}}}}
\put(100, 805){\makebox(0,0)[b]{\hbox{\tt  glibk}}}
\put( 00, 800){\line(1,0){200}}      
\put(200, 800){\makebox(0,0)[l]{\framebox{\parbox{75mm}{\scriptsize
              {\it Library of utilities:} {\bf GLK, MathLib} }}}}
\put(100, 705){\makebox(0,0)[b]{\hbox{\tt  dizet}}}
\put( 00, 700){\line(1,0){200}}      
\put(200, 700){\makebox(0,0)[l]{\framebox{\parbox{75mm}{\scriptsize
              {\it Library electroweak corrections:} {\bf DZface}, DIZET 6.21 \\
              {\it sub-directory:} {\tt tabtest}
}}}}
\put(100, 575){\makebox(0,0)[b]{\hbox{\tt  tauola}}}
\put( 00, 570){\line(1,0){200}}      
\put(200, 570){\makebox(0,0)[l]{\framebox{\parbox{75mm}{\scriptsize
              {\it Tau-lepton decays:} TAUOLA 2.6 }}}}
\put(100, 475){\makebox(0,0)[b]{\hbox{\tt  photos}}}
\put( 00, 470){\line(1,0){200}}      
\put(200, 470){\makebox(0,0)[l]{\framebox{\parbox{75mm}{\scriptsize
              {\it Radiative corrections in decays:} PHOTOS 2.02 }}}}
\put(100, 375){\makebox(0,0)[b]{\hbox{\tt  jetset}}}
\put( 00, 370){\line(1,0){200}}      
\put(200, 370){\makebox(0,0)[l]{\framebox{\parbox{75mm}{\scriptsize
              {\it Hadronization of quarks:} JETSET 7.2 }}}}
\put(100, 255){\makebox(0,0)[b]{\hbox{\tt  dok}}}
\put( 00, 250){\line(1,0){200}}      
\put(200, 250){\makebox(0,0)[l]{\framebox{\parbox{75mm}{\scriptsize
              {\it Documentation in postscript:} KKcpc.ps.gz }}}}
\put(100, 105){\makebox(0,0)[b]{\hbox{\tt  ffbench}}}
\put( 00, 100){\line(1,0){200}}      
\put(200, 100){\makebox(0,0)[l]{\framebox{\parbox{75mm}{\scriptsize
              {\it Demo programs, example input data, benchmark outputs}  \\
              {\it sub-directories:} {\tt demo, Inclusive, Mu, Tau, Down, Up, Botom, Beast}
}}}}
\end{picture} 
\caption{\sf\small
Topography of the distribution directory}
\label{fig:topography}
\end{figure}
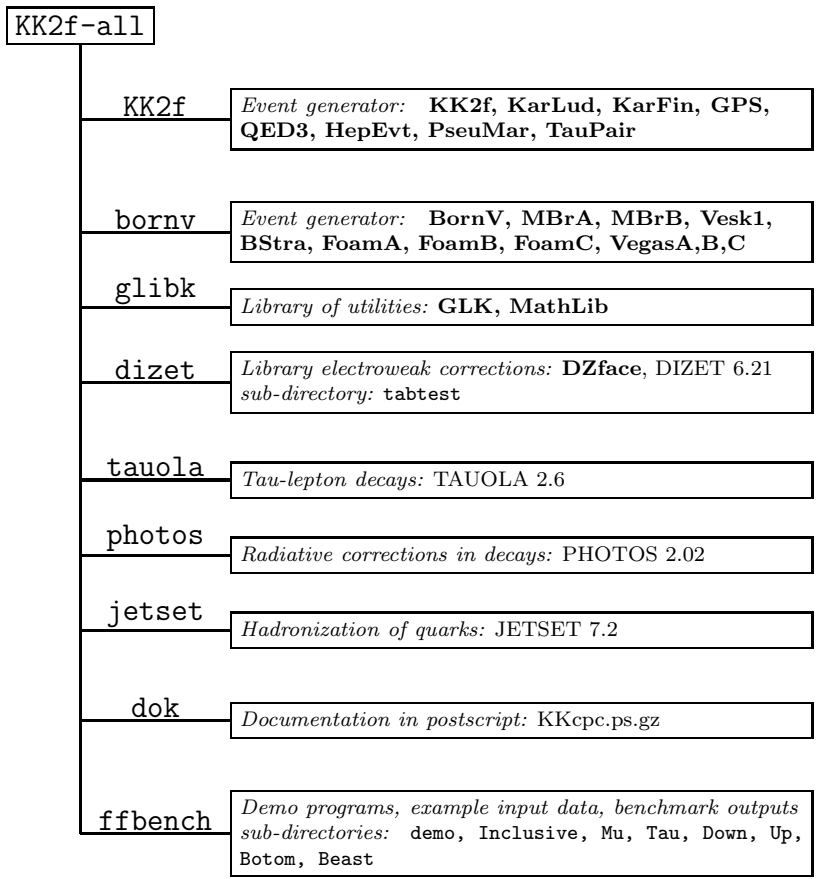

\subsection{Programming rules}
\label{sec:prog-rules}
The program 
is written in Fortran77 with popular extensions such as long variable names, long source lines, etc.,
which are available on all platforms.
In the {\em Makefile} in the main directory {\sf KK-all},
there is a collection, of f77 compilation flags, for  Linux, AIX,
HPUX and ALPHA compilers, which should be used to activate these extensions.
The program is written in such a way that its translation to 
an object-oriented language such as C++ should not be very difficult.
In fact the program is divided into modules,
which have the structure of the C++ classes,
as far as it is possible to do it within f77.
Below we characterize the rules according to which the program was written.

Each pseudo-class with the name {\sf Module} consists of a separate source file
{\sf Module.f} and the header file {\sf Module.h}.
Each module obeys the following rules:
\begin{itemize}
\item
  There is only one common block {\tt /c\_{}Module/ } which contains all class member variables,
  which is placed in the header file {\tt Module.h}. 
  Each subroutine in the {\tt Module.f} source file includes an {\tt INCLUDE 'Module.h'} statement.
  The outside programs should never include directly {\tt /c\_{}Module/}.
  All input/output communication is done with the help of dedicated, easy to use,  subroutines.
\item
  Variables in {\tt /c\_{}Module/} are  {\em class members} and all have the special prefix
  ``{\tt m\_{}}'' in their name, for example {\tt m\_{}Iterat} is the number of iterations.
\item
  The user has access to some class members through ``getters'' and ``setters''; see below.
\item
  Strong typing is imposed with the help of {\tt IMPLICIT NONE}.
\item
  Initializator with the name {\tt Module\_{}Initialize}
  performs initialization. Typically it initializes variables in {\tt /c\_{}Module/}.
\item
  Finalizer with the name {\tt Module\_{}Finalize}, summarizes the whole run, sets output
  values in {\tt /c\_{}Module/}, prints output, etc.
\item
  Maker with the name {\tt Module\_{}MakeSomething}, or a similar one, does the essential 
  part of the job, for instance a maker {\tt Module\_{}MakeEvent} generates
  a single MC event.
\item
  Setter with the name {\tt Module\_{}SetVariable} 
  is called from the outside world to set {\tt m\_{}Variable} in {\tt /c\_{}Module/}.
  For example {\tt  CALL BornV\_{}SetCMSene(100d0)} sets the variable {\tt m\_{}CMSene=100d0}.
  Only certain privileged variables have a right to be served by their own setter, the other
  ones are in principle ``private''.
\item
  Getter with the name {\tt Module\_{}GetVariable} is called from the outside world to get 
  the {\tt m\_{}Variable} from {\tt /c\_{}Module/}.
  It is a preferred way of sending output information to the outside world.
  For example, with the help of {\tt CALL KK2f\_{}GetXsecMC(xSecPb, xErrPb)} 
  one gets the MC cross section
  {\tt xSecPb} and its error {\tt xErrPb} in the user program.
\end{itemize}
In the following we shall describe all pseudo-classes and their role.

\subsection{KK2f: Top-level class}
\label{sec:KK2f}
The main purpose of this top-level
pseudo-class is to provide the user interface, see section \ref{sec:usage}
on the usage of the program.
Let us list and explain the main entries in this class:
\begin{itemize}
  \item
    {\sf KK2f\_ReaDataX('data\_file',iReset,imax,xpar)} reads the input data file.
    It should be called twice, once with {\sf INTEGER iReset=1}, for the default data file
    {\sf .KK2f\_defaults}, which is placed in the main directory {\sf KK-all}. 
    The user has to provide a link to this file, or absolute path in the name of the file.
    (Copying it to the local directory is not recommended.)
    It should be called  for the second time for the user data {\sf ./user\_data} with 
    {\sf INTEGER iReset=0}, in order to modify some entries in the input of the program.
    The variable {\sf INTEGER imax} is the dimension of the {\sf DOUBLE PRECISION xpar(imax)}.
    For the moment {\sf imax$\leq$ 3000} is required, but we reserve {\sf imax= 10000}
    for future use.
    The flag {\sf INTEGER iReset=0} is for reading data with resetting all undefined
    values to zero, while with {\sf iReset=1} only entries listed in the data file are modified.
  \item
    {\sf KK2f\_Initialize(xpar)} initializes the whole MC generator.
    This initializer calls initializers of other classes like 
    {\sf BornV\_Initialize}, 
    {\sf KarLud\_Initialize},
    {\sf KarFin\_Initialize},
    {\sf QED3\_Initialize},
    {\sf GPS\_Initialize},
    {\sf TauPair\_Initialize}.
    It initializes also the database for 
    the branching over final fermion flavours in the  class {\bf MBrA}.
    Note that {\sf BornV\_Initialize} reads from the disk look-up tables for the electroweak corrections
    and {\sf KarLud\_Initialize} manages initialization of the ISR energy spectrum either
    with the help of {\bf Vesk1} class or, in the presence of the beamstrahlung, with the help of
    the {\bf Bstra} class, which in turn initializes three copies of the Foam package,
    {\bf FoamA, FoamB, FoamC} (or of Vegas package, {\bf VegasA, VegasB, VegasC}).
  \item
    {\sf KK2f\_Make} generates a single event.  
    It calls {\sf KarLud\_Make} to make ISR photons, 
    {\sf KarFin\_Make} to make FSR photons and {\sf KK2f\_Merge} to merge all photons in a single list.
    It invokes {\sf KK2f\_MakePhelRand} to generate the photon helicities randomly,
    calculates the EEX model weight using {\sf CALL QED3\_Make}
    and/or the CEEX model weight using {\sf CALL GPS\_Make}.
    Optional rejection is performed to produce weight-1  events and the
    weight book-keeping is done separately for each final fermion type using {\sf MBrA\_Fill}.
    Finally, quarks are hadronized using {\sf HepEvt\_Hadronize} (interface to JETSET),
    or $\tau$ decays are simulated with all spin effects
    (including all spin correlations). This is done using subprograms
    of the {\bf TauPair} class (interface to TAUOLA).
  \item
    {\sf KK2f\_GetPhotAll(Nphot,PhoAll)} provides the user with the momenta of all photons:
    {\sf DOUBLE PRECISION PhoAll(100,4)} and photon multiplicity {\sf INTEGER  Nphot}.
    Alternatively, {\sf Nphot} is provided by {\sf KK2f\_GetNphot(Nphot)}
    and the $i$-th photon momentum by  {\sf KK2f\_GetPhoton1(iPhot,Phot)},
    with {\sf DOUBLE PRECISION Phot(4)}.
  \item
    {\sf KK2f\_GetFermions(q1,q2)} provides the user with the momenta of the final fermions
    {\sf DOUBLE PRECISION   q1(4),q2(4)}.
  \item
    {\sf KK2f\_GetBeams(p1,p2)} provides the user with the momenta of the beams
    {\sf DOUBLE PRECISION   p1(4),p2(4)}.
  \item
    {\sf KK2f\_GetWtAll(WtMain,WtCrud,WtSet)} can be used to get access to
    the main MC weight {\sf WtMain} and the list of all alternative weights {\sf WtSet(1000)}.
    The weight for the crude differential cross section {\sf WtCrud} is also provided.
    All of them are {\sf DOUBLE PRECISION} type.
    Alternatively, the getter {\sf KK2f\_GetWt(WtMain,WtCrud)} may be more convenient.
  \item
    {\sf KK2f\_Finalize} may be called at the end of the MC run, in order
    to perform the final book-keeping and printing.
  \item
    {\sf KK2f\_GetXSecMC(XSecPb,XErrPb)} should be called after calling {\sf KK2f\_Finalize} 
    in order to get the total cross section (in picobarns) and its absolute error:
    {\sf DOUBLE PRECISION XSecPb, XErrPb}.
  \item
    {\sf  KK2f\_GetVersion(Version)} and 
    {\sf  KK2f\_GetDate(Date)} provide the user with the version number
    {\sf  DOUBLE PRECISION   Version} and the release data
    {\sf  CHARACTER*14   Date} information. 
    This should help to keep track of the development of the program.
  \item
    {\sf  KK2f\_Print(ie1,ie2)} can be used to print the actual MC events, 
    limiting their serial
    number to stay between {\sf INTEGER ie1} and {\sf INTEGER ie2}.
\end{itemize}
There are several other getters in the {\bf KK2f} class, which are mainly for internal use.

Let us briefly list other subroutines in the {\bf KK2f} class, which are not called
by the user of the program:
\begin{itemize}
  \item
    {\sf  KK2f\_WignerIni(KFbeam,CMSene,PolBeam1,PolBeam2, Polar1,Polar2)}
    does Wigner rotation for spin polarization vectors of beams.
    Beam polarization vectors (in input data) are defined
    in the beam particle rest frames, which are reached from the CMS by a simple
    $z$-boost without any rotation. (The first beam is parallel to the $z$-axis.)
  \item
    {\sf  KK2f\_ZBoostAll(exe)}
    performs a $z$-boost on all momenta of the event.
    This $z$-boost corresponds to beamstrahlung or beam spread
    and is done at the very end of generation, after the calculation of the matrix element.
  \item
    {\sf  KK2f\_DsigOverDtau(mout,Rho)}
    is only for documentation and testing purposes.
    It calculates the distribution $d\sigma/d\tau$ corresponding to {\sf WtCrud},
    normalized with respect to $d\tau$ = Lorentz invariant phase space.
  \item
    {\sf KK2f\_Merge}
    merges lists of ISR and FSR photon momenta.
    The resulting merged photons are ordered according to their energy.
  \item
    {\sf  KK2f\_MakePhelRand} generates the photon helicities randomly.
\end{itemize}

\subsection{HepEvt: HEP event record class}
\label{sec:HepEvt}
The pseudo-class  {\bf HepEvt} has the double purpose of 
(a) being another user interface, alternative to getters in {\bf KK2f},
and (b) managing also the hadronization of quarks.
The user may also traditionally put the common block
of the {\bf HepEvt} class directly into its code.
It is a {\sf DOUBLE PRECISION} version of 
the standard PDG {\sf /HEPEVT/} common block, for a maximum of 2000 particles.
{\small
\begin{verbatim}
*----------------------------------------------------------------------
      INTEGER nmxhep         ! maximum number of particles
      PARAMETER (nmxhep=2000)
      DOUBLE PRECISION   phep, vhep
      INTEGER nevhep, nhep, isthep, idhep, jmohep, jdahep
      COMMON /d_HepEvt/
     $     nevhep,           ! serial number
     $     nhep,             ! number of particles
     $     isthep(nmxhep),   ! status code
     $     idhep(nmxhep),    ! particle ident KF
     $     jmohep(2,nmxhep), ! parent particles
     $     jdahep(2,nmxhep), ! childreen particles
     $     phep(5,nmxhep),   ! four-momentum, mass [GeV]
     $     vhep(4,nmxhep)    ! vertex [mm]
      SAVE  /d_hepevt/
*----------------------------------------------------------------------
\end{verbatim}}

\noindent
Let us now list and explain the subprograms in the {\bf HepEvt} class.
\begin{itemize}
  \item
    {\sf  HepEvt\_Fill} 
    fills in all of common block {\sf /d\_HepEvt/} using the subprogram {\sf HepEvt\_Fil1}.
    Momenta are provided by getters from {\bf KarLud} and {\bf KarFin}.
  \item
    {\sf HepEvt\_Fil1} writes a single particle record into the common block {\sf /d\_HepEvt/}.
  \item
    {\sf  HepEvt\_Hadronize(HadMin)}
    arranges jets with the help of {\sf LuJoin} and {\sf LuShow},
    and hadronizes quarks using  {\sf LuExec}.
  \item
    {\sf CALL HepEvt\_GetBeams(p1,p2)}
    provides four-momenta of the two beams
    {\sf DOUBLE PRECISION  p1(4),p2(4)}.
  \item
    {\sf CALL HepEvt\_GetFfins(q1,q2)} provides four-momenta
    of the two final fermions
    {\sf DOUBLE PRECISION   q1(4),q2(4)}.
  \item
    {\sf CALL HepEvt\_GetPhotAll(NphAll,PhoAll)} 
    provides the photon multiplicity  {\sf INTEGER NphAll}
    and the photon four-momenta {\sf DOUBLE PRECISION PhoAll(100,4)}.
  \item
    {\sf HepEvt\_GetPhotBst(nPhot,Phot)} 
    provides the multiplicity {\sf INTEGER nPhot} (=0,1,2) and
    four-momenta {\sf DOUBLE PRECISION Phot(100,4)}
    of the beamstrahlung photons.
  \item
    {\sf HepEvt\_LuHepc}
    This is the double-precision version of {\sf LUHEPC} of JETSET.
    It translates {\sf DOUBLE PRECISION}  {\sf /c\_HepEvt/} 
    into the old style {\sf REAL*4} Lund commons.
\end{itemize}

\subsection{KarLud: Crude level MC for ISR}
The simulation of ISR together with the beamstrahlung and generation of the
type of final-fermion flavour is implemented in the {\bf KarLud} pseudo-class.

\noindent
Let us now list and explain all subprograms in the {\bf KarLud} class:
\begin{itemize}
   \item
     {\sf KarLud\_Initialize(xpar\_input,XCrude)} is the initializer of the class.
     It initializes the generation of $v$ with the help of {\sf Vesk1\_Initialize}
     or of $v,z_1,z_2$ with the help of {\sf BStra\_Initialize}.
     It defines the primary integrated cross section on which the entire normalization is based.
   \item
     {\sf KarLud\_SmearBeams} implements beam spread.
     This is correct only for a small spread $<$ 2~GeV.
     It should not be used together with beamstrahlung, since this has not been tested yet.
     The distribution is Gaussian $\rho(X)=N\exp( (X-{\tt CMSene}/2)^2/(2 {\tt DelEne}^22))$
     ({\tt DelEne} is the dispersion of the beam energy {\tt Ebeam}, 
      not of {\tt CMSene}).
   \item
     {\sf KarLud\_Make(PX,wt\_ISR)} generates ISR photons with the help of other subprograms;
     {\sf PX(4)} is the four-momentum left after photon emission,
     {\sf wt\_ISR} is the ISR component of the ``crude weight''.
     The other subroutines called here are:
     {\sf  KarLud\_SmearBeams} (see below),
     {\sf  BornV\_SetCMSene(XXXene)}, which resets the CMS four-momentum 
     in {\bf BornV} in the case of beam smearing,
     {\sf  Vesk1\_Make}, which generates the variable $v$, in the absence of beamstrahlung
     (alternatively it is done with {\sf  VegasA\_Generate} for {\sf KeyFix=2} ),
     or {\sf  BStra\_Make}, which generates $v,z_1,z_2$ in case of beamstrahlung.
     The ISR photons are generated with the help of
     {\sf KarLud\_YFSini}, see below, 
     and the type of final-state fermion {\sf KF} is generated with the help of  {\sf MBrA\_GenKF}. 
     Finally, if FSR is off, then final momenta are produced locally
     with the help of  {\sf KinLib\_phspc2}
   \item
     {\sf KarLud\_Finalize(Mode, XKarlud, KError)}
     calculates the crude cross section  {\sf XKarlud} and its error {\sf KError},
     and prints out final statistics.
     The crude cross section is coming from the {\sf Vesk1\_Finalize}
     (alternatively from {\sf VegasA\_GetIntCrude})
     or, in case of beamstrahlung, from {\sf BStra\_GetXCrude},
   \item
     {\sf KarLud\_YFSini(XXXene,vv, PX,WtIni)} is generating the ISR photon momenta.
     Its input is the total energy available {\sf XXXene}, and {\sf vv}$=v$.
     {\sf WtIni} is the ISR component of the ``crude weight''
     and {\sf PX(4)} is the four-momentum left after photon emission.
     {\sf KarLud\_YFSini} calls {\sf BornV\_GetAvMult} to get the average ISR multiplicity,
     {\sf KarLud\_PoissGen} and {\sf KarLud\_AngBre}, see below.
   \item
     {\sf KarLud\_PoissGen} generates photon multiplicity.
   \item
     {\sf KarLud\_AngBre} generates photon angle.
   \item
     {\sf KarLud\_ZBoostAll} performs $z$-boosts of all photons.
   \item
     {\sf KarLud\_GetPhotons(nphot,sphot)} provides all ISR photons.
   \item
     {\sf KarLud\_GetPhoton1(iphot,phot)} provides single ISR photons.
   \item
     {\sf KarLud\_GetPX(PX)} provides four-momentum {\sf PX}, see above.
   \item
     {\sf KarLud\_GetBeams(p1,p2)} provides the beam momenta.
     In the case of beamstrahlung {\sf p1,p2)} are beams {\em after} beamstrahlung.
   \item
     {\sf KarLud\_GetBeasts(p1,p2)}
     provides the collinear photons of the  beamstrahlung.
\end{itemize}

\subsection{KarFin: Crude level MC for FSR}
The FSR is implemented in the {\bf KarFIN} class.
This package was already used in the KORALZ~\cite{koralz4:1994} program for some time.
Thanks to recent improvements it now works
properly without any approximations close to $\tau$ threshold.
The $m_f\ll \sqrt{s}$ approximation is not used anymore.

\noindent
Let us now list and explain all subprograms in the {\bf KarFin} class:
\begin{itemize}
  \item
    {\sf KarFin\_Initialize}
    initializes some internal variables (weight book-keeping).
  \item
    {\sf KarFin\_Make(PX,amfi1,amfi2,CharSq,WtFin)} generates FSR photons
    with the help of other subprograms, see below.
    {\sf PX} is the four-momentum of the entire FSR system (fermions + photons),
    {\sf amfi1,amfi2} are masses of the final charged pair (not necessarily equal),
    {\sf CharSq} is the final-state fermion charge squared and 
    {\sf WtFin} is the FSR part of the crude weight.
  \item
    {\sf KarFin\_YFSfin(PX,amfi1,amfi2,CharSq,WtFin)} generates momenta of the FSR photons.
    {\sf PX, amfi1, amfi2, CharSq, WtFin} are as defined above. 
    It calls {\sf KarFin\_PoissGen} and {\sf  KarFin\_AngBre} to generate photon multiplicity
    and angles, then {\sf KarFin\_Kinf1} and
    and {\sf KarFin\_Piatek} see below.
  \item
    {\sf  KarFin\_Kinf1(PX,...,phsu)}
    transforms from the rest frame of {\sf Q=q1+q2} QMS down to the laboratory
    through the intermediate rest frame of {\sf PX=q1+q2+ phsu}.
  \item
    {\sf KarFin\_Piatek( Mas1,Mas2,CharSq,WtMlist, Wt3)}%
    \footnote{ Written in CERN, Piatek$\equiv$Friday, 22.IX.1989  (S.J.)}
    optionally removes photons below $E_{\min}$ from the list of photons, 
    appropriately modifying the crude weight.
    {\sf Mas1,2}  = fermion masses,
    {\sf WtMlist}  = list of mass weights $(f/\bar{f})$ for all photons.
    {\sf Wt3} = product of $(f/\bar{f})$ for the alive (not hidden) photons.
    The correcting weight is calculated with the help of 
    {\sf  BVR\_Btildc}, calculating $\tilde{B}$ for $q^*_i$ and
    {\sf  BVR\_Btilda} for $q_i$.
  \item
    {\sf KarFin\_PoissGen} generates the photon multiplicity randomly.
  \item
    {\sf KarFin\_AngBre} generates the photon angles randomly.
  \item
    {\sf KarFin\_Kinf1}
    transforms the final fermions and all photons from QMS through the $Z$-frame to CMS.
    Random rotation with angles $\psi,\omega$ is applied in the intermediate rest frame 
    of {\sf PX} ($Z$ boson) using {\sf KarFin\_BostEul}.
  \item
    {\sf KarFin\_BostEul}
    performs Lorentz transformations consisting of:
    (1) parallel boost from the final fermions rest frame to fermions + photons rest frame ($Z$ frame);
    (2) two rotations with angles $\psi,\omega$;
    (3) parallel boost to the laboratory system CMS.
  \item
    {\sf KarFin\_ZBoostAll(exe)} does an additional $z$-boost of all particles in case of beamstrahlung.
  \item
    {\sf KarFin\_Finalize} prints final statistics.
  \item
    {\sf KarFin\_GetNphot(nphot)} provides the FSR photon multiplicity.     
  \item
    {\sf KarFin\_GetPhoton1(iphot,phot)} provides the four-momentum of a single FSR photon.
  \item
    {\sf KarFin\_GetPhotons(nphot,sphot)} provides the four-momenta of all FSR photons.
  \item
    {\sf KarFin\_GetFermions(qf1,qf2)}  provides the four-momenta of the final fermions.
  \item
    {\sf KarFin\_WtMass(WtMass)}  provides the product of $(f/\bar{f})$ for the alive (not hidden) photons.
\end{itemize}

\subsection{BornV class: particle data base and ISR spectrum}
Class {\bf BornV} is serving as a data-base for fermion properties
such as mass, charge, isospin, colour and other fermion-type dependent
parameters relevant to MC generation, like the maximum weight for rejection.
It also reads from the disk and keeps the EW form factors produced 
by the interface to DIZET 6.21.

All other classes use the data-base of the {\bf BornV} class through its getters, see below.
The data-base is located in the class common block {\sf c\_BornV},
which is initialized by {\sf BornV\_Initialize}
from the default input data file {\sf KK-all/.KK2f\_defaults} passed 
by arguments from {\sf KK2f\_Initialize}. Optionally, only after user modifications, 
see section \ref{sec:usage} on the usage of the program.

For this particular class it is instructive to look into the list of the 
{\em class member} variables in the class common block{\sf /c\_BornV/}.
Below we quote part of the {\sf BornV.h} source code:
{\small
\begin{verbatim}
*----------------------------------------------------------------------
      COMMON /c_BornV/
* Tables of EW formfactors
     $  m_cyy(m_poin1+1,7,16),           ! formfactor, table
     $  m_czz(m_poin2+1,7,16),           ! formfactor, table
     $  m_ctt(m_poin3+1,m_poinT+1,7,16), ! formfactor, table
     $  m_clc(m_poin4+1,m_poinT+1,7,16), ! formfactor, table
     $  m_syy(m_poin1+1,16),             ! QCD correction, table
     $  m_szz(m_poin2+1,16),             ! QCD correction, table
     $  m_stt(m_poin3+1,m_poinT+1,16),   ! QCD correction, table
     $  m_slc(m_poin3+1,m_poinT+1,16),   ! QCD correction, table
     $  m_GSW(100),    ! form-factors,   at the actual energy/angle
     $  m_QCDcor,      ! QCD correction, at the actual energy/angle
*--------------------- EVENT ------------------------------------------
     $  m_CMSene,      ! Initial value of CMS energy
     $  m_XXXene,      ! CMS energy after beamstrahlung or beam spread
     $  m_x1,          ! 1-z1 = x1 for first  beam(strahlung)
     $  m_x2,          ! 1-z2 = x2 for second beam(strahlung)
     $  m_vv,          ! v = 1-sprim/s
     $  m_AvMult,      ! Average photon multiplicity CRude at given v
     $  m_YFSkon,      ! YFS formfactor finite part
     $  m_YFS_IR,      ! YFS formfactor IR part
* ---------------------------------------------------------------------
     $  m_vvmin,       ! minimum v, infrared cut
     $  m_vvmax,       ! maximum v
     $  m_HadMin,      ! minimum hadronization mass [GeV]
* Basic QED------------------------------------------------------------
     $  m_alfinv,      ! 1/alphaQED, Thomson limit
     $  m_alfpi,       ! alphaQED/pi
     $  m_Xenph,       ! Enhancement factor for Crude photon multipl.
* EW parameters
     $  m_MZ,          ! Z mass
     $  m_amh,         ! Higgs mass
     $  m_amtop,       ! Top mass
     $  m_swsq,        ! sin(thetaW)**2
     $  m_gammz,       ! Z width
     $  m_amw,         ! W mass
     $  m_gammw,       ! W width
     $  m_Gmu,         ! Fermi constant (from muon decay)
* Table of fermion parameters, quarks (1->6) and leptons (11->16)
     $  m_KFferm(20),  ! fermion KFcode (1->6) and (11->16)
     $  m_NCf(20),     ! number of colours
     $  m_Qf(20),      ! electric charge
     $  m_T3f(20),     ! isospin, L-hand component
     $  m_helic(20),   ! helicity or polarization
     $  m_amferm(20),  ! fermion mass
     $  m_auxpar(20),  ! auxiliary parameter
     $  m_IsGenerated(20),  ! Generation flag, only for SemiAn.! 
* Normalisation
     $  m_gnanob,      ! GeV^(-2) to nanobarns
* Initial/final fermion types
     $  m_KFini,       ! KF code of beam
* Test switches
     $  m_KeyINT,      ! ISR/FSR intereference switch
     $  m_KeyElw,      ! Type of Electroweak Library
     $  m_KeyZet,      ! Z-boson on/off
     $  m_KeyWtm,      ! Photon emission without mass terms
     $  m_out          ! output unit for printouts $
*----------------------------------------------------------------------
\end{verbatim}}

\noindent
Let us now list and explain all subprograms in the {\bf BornV} class:
\begin{itemize}
  \item
    {\sf  BornV\_Initialize(xpar)} initializes data members in {\sf /c\_BornV/}.
  \item
    {\sf  BornV\_StartEW(xpar)} initializes electroweak formfactors in {\sf /c\_BornV/}.
    There are two versions of this routine: one in {\sf KK-all/bornv/BornV\_StartEW.f}
    which reads electroweak formfactors from the disk file and another one in
    {\sf KK-all/dizet/BornV\_StartEW.f} which calculates it using DIZET library.
    See section 4.13 for more details.
  \item
    {\sf  BornV\_ReadAll}
    reads from the disk-file  pretabulated EW form factors for $\mu$ and $\tau$ leptons,
    and for $d,u,b$ quarks. For $s,c$ quarks the form factors of $d,u$ are used.
  \item
    {\sf  BornV\_ReadFile(DiskFile,KFfin)}
    reads from the disk a single file for a single final fermion.
  \item
    {\sf  BornV\_StartDZ(xpar)}
    Initialized DIZET library using current input data in {\sf xpar}.
  \item
    {\sf  BornV\_ReBin1, BornV\_ReBin1a, BornV\_ReBin2, BornV\_ReBin2a}
    subroutines map the variable $r\in (0,1)$ (random number) into $v\in (0,v_{\max})$.
    Various methods are used to do it, with various kinds of the mapping function.
  \item
    {\sf  DOUBLE PRECISION FUNCTION BornV\_RhoFoamC(xarg)} is the
    integrand for {\sf FoamC} in the 3-dimensional mode for beamstrahlung.
    Remember that {\sf BornV\_Crude} and {\sf BornV\_MakeRho} use the hidden input {\sf m\_XXXene}.
    {\sf BornV\_Crude} is in the R-units (point-like cross-section at  $\sqrt{s}$={\sf m\_XXXene}).
    It defines  {\sf m\_vv}, which is later on exported to {\bf KarLud}.
  \item
    {\sf  DOUBLE PRECISION FUNCTION BornV\_RhoFoamB(xarg)} is the
    integrand for {\sf FoamB} in the two-dimensional mode for beamstrahlung
    (it defines  {\sf m\_vv}).
  \item
    {\sf  DOUBLE PRECISION FUNCTION BornV\_RhoFoamA(xarg)} is the
    integrand for {\sf FoamA} in the one-dimensional mode for beamstrahlung off and on
    (it defines  {\sf m\_vv}).
  \item
    {\sf  DOUBLE PRECISION  FUNCTION BornV\_RhoVesko1(R)} is the
    integrand of {\sf Vesko1}.
    (The comment about hidden input {\sf m\_XXXene} applies.)
    In the case of beamstrahlung the additional normalization
    factor {\sf Circee(1d0,1d0)} is added inside {\sf BStra\_Initialize}
    (it defines  {\sf m\_vv}).
  \item
    {\sf  BornV\_MakeGami(CMSene,gamiCR,gami)} calculates
    {\sf GamiCR}$=\bar\gamma_e$ and {\sf gami}$=\gamma_e$ as functions of {\sf CMSene}.
  \item
    {\sf  BornV\_MakeISR(Rho)}
    This procedure is tightly related to ISR photon generation in {\bf KarLud}.
    It provides {\sf Rho(m\_vv, m\_XXXene)}, the primary distribution of $v$.
    It also calculates {\sf m\_AvMult}, which is later used in {\sf KarLud\_YFSini};
    {\sf m\_YFSkon ,m\_YFS\_IR}, which are later used in {\sf GPS\_Make} and {\sf QED3\_Make}.
  \item
    {\sf  DOUBLE PRECISION  FUNCTION BornV\_Crude(vv)}
    calculates the crude total Born cross section  summed over fermion types.
    It exploits the fact that the Born differential distribution reads 
    $a + b \cos\theta + d \cos^2\theta$.
    (Hidden input is {\sf m\_XXXene}).
    It is used in {\sf BornV\_RhoVesko1, BornV\_RhoFoamA}, etc.
  \item
    {\sf  DOUBLE PRECISION  FUNCTION BornV\_Differential(Mode,KFf,svar,CosThe,...)}
    is the Born differential distribution.
    For {\sf Mode=0} it is a crude version of pure Born, no spin, no EW corrections.
    For {\sf Mode=1} it is the full result with EW corrections spin, etc.
    In this mode it is used in {\bf QED3}, and for all kinds of tests.
    For {\sf Mode=3} it is used in the tests of pretabulation.
    In this case {\sf GSW(s,theta)} has to be provided from
    the outside, with the help of {\sf BornV\_SetGSW}%
    \footnote{
        Note that in the test mode {\sf KeyEwl=0} and {\sf Mode=1} we use {\sf BornV\_Simple},
        which will perhaps have to be changed in the future, because of the lack of spin effects.
        At this stage, however, we are bound to use it because the KeyZet etc.
        are implemented only in {\sf BornV\_Simple} and not in {\sf BornV\_Dizet}.}.
  \item
    {\sf  DOUBLE PRECISION  FUNCTION BornV\_Simple(KFi,KFf,svar,costhe)}
    provides for \;\;\;\; {\sf BornV\_Differential} 
    an unsophisticated Born differential distribution without EW corrections,
    with the $Z$ and $\gamma$ $s$-channel exchange.
  \item
    {\sf  DOUBLE PRECISION  FUNCTION BornV\_Integrated(KFfin,svar)}
    {\em is used only in semianalytical programs.}
    It calculates the total Born cross section.
    For {\sf KFfin = 0} it sums over all allowed flavours; otherwise,
    for {\sf KFfin.NE.0}, it calculates the cross section for the actual value of {\sf m\_KFfin}.
  \item
    {\sf  DOUBLE PRECISION FUNCTION BornV\_Dizet(Mode,KFi,KFf,svar,CosThe,...)}
    provides for {\sf BornV\_Differential} the
    differential Born cross section with/without EW corrections.
    For {\sf Mode=0} it provides pure Born and for {\sf Mode=1} electroweak corrections are added.
    {\sf KFi,KFf} can also be negative for an antiparticle; in this case it is
    important to produce tables with the correct input {\sf KFini, KFfin}.
  \item
    {\sf  BornV\_InterpoGSW(KFf,svar,CosThe)}
    calculates EW form factors from look-up tables, using linear interpolation.
  \item
    {\sf  BornV\_givizo(idferm,ihelic,sizo3,charge,kolor)}
    provides electric charge, weak isospin and colour of the fermion,
    where {\sf idferm =1,2,3,4}
    denotes:    neutrino,   lepton, up, down quark;
    negative {\sf idferm=-1,-2,-3,-4}, denotes the corresponding antiparticle;
    {\sf ihelic =     +1,  $-$1}   denotes  right- and left-handedness (chirality)
    {\sf sizo3} is the third projection of weak isospin ($\pm 1/2$),
    {\sf charge} is the electric charge 
    (in units of magnitude of the electron charge),
    {\sf kolor} is the QCD colour, 1 for lepton, 3 for quarks.
  \item
    {\sf  DOUBLE PRECISION  FUNCTION BornV\_Sig0nb(CMSene)}
    provides the point-like muon cross section in nanobarns
    for the normalization purpose.
\end{itemize}

\noindent
Communication subprograms (setters and getters) used by all other classes
are the following:
\begin{itemize}
  \item
    {\sf  BornV\_GetParticle(KFferm, mass, Qf, T3f, NCf)};
    for the fermion type {\sf INTEGER KFferm} provides
    its QCD colour {\sf INTEGER NCf},
    mass, electric charge and weak isospin {\sf DOUBLE PRECISION   mass, Qf, T3f}.
  \item
    {\sf  DOUBLE PRECISION  FUNCTION BornV\_GetMass(KFferm)};
     for the fermion type {\sf INTEGER KFferm} provides its mass.   
  \item
    {\sf  DOUBLE PRECISION  FUNCTION BornV\_GetCharge(KFferm)};
     for the fermion type {\sf INTEGER KFferm} provides its electric charge.  
  \item
    {\sf  INTEGER FUNCTION BornV\_GetColor(KFferm)};
     for the fermion type {\sf INTEGER KFferm} provides its QCD colour.  
  \item
    {\sf  DOUBLE PRECISION  FUNCTION BornV\_GetAuxPar(KFferm)};
     for the fermion type {\sf INTEGER KFferm} it provides its auxiliary parameter.  
  \item
    {\sf  BornV\_SetKeyElw(KeyElw)} sets the EW switch {\sf KeyElw}.
  \item
    {\sf  BornV\_GetKeyElw(KeyElw)} gets the EW switch {\sf KeyElw}.
  \item
    {\sf  BornV\_GetKeyZet(KeyZet)} sets the $Z$ boson switch {\sf KeyZet}.
  \item
    {\sf  BornV\_SetKeyZet(KeyZet)} gets the $Z$ boson switch {\sf KeyZet}.
  \item
    {\sf  BornV\_SetCMSene(CMSene)} sets the CMS total energy  {\sf CMSene}.
  \item
    {\sf  BornV\_SetMZ(MZ)} sets the $Z$ boson mass.
    \item
    {\sf  BornV\_GetMZ(MZ)} gets the $Z$ boson mass.
  \item
    {\sf  BornV\_GetGammZ(GammZ)} gets the $Z$ boson width.
  \item
    {\sf  BornV\_GetGmu(Gmu)} gets the $G_{\rm Fermi}$.
  \item
    {\sf  BornV\_GetSwsq(Swsq)} gets the electroweak mixing angle.
  \item
    {\sf  BornV\_GetAlfInv(AlfInv)} gets  the $\alpha_{\rm QED}/\pi$.
  \item
    {\sf  BornV\_GetAvMult(AvMult)} provides the average ISR multiplicity.
  \item
    {\sf  BornV\_GetYFSkon(YFSkon)} provides
    the finite part of the YFS form factor. Used in {\bf QED3}.
  \item
    {\sf  BornV\_GetYFS\_IR(YFS\_IR)} provides
    IR (cut-off-dependent) part of the ISR YFS form factor. Used in {\bf QED3}.
  \item
    {\sf  BornV\_GetQCDcor(QCDcor)} provides
    the QCD correction factor, defined by DIZET.
  \item
    {\sf  BornV\_GetVV(vv)}  provides $v$={\sf vv}.
  \item
    {\sf  BornV\_GetVXX(vv,x1,x2)}  provides $v$={\sf vv}, $x_i=1-z_i=$ {\sf x1,x2}.
  \item
    {\sf  BornV\_GetGSW(GSW)} provides
    the EW form factors {\sf GSW(k) k=1,...,7}. 
    It is used in {\bf GPS} and in {\sf BornV\_Dizet}.
    Note that {\sf BornV\_InterpoGSW} has to be called before, in order to interpolate properly.
  \item
    {\sf  BornV\_SetGSW(GSW)}.
    For special tests of pretabulation the values of EW form factors
    {\sf GSW(k) k=1,...,7} can be set with this subprogram from outside.
\end{itemize}

\subsection{Bstra, IRC and MBrB classes for beamstrahlung}
As was already described, the MC integral for beamstrahlung and ISR
has three components: $\int dv dz_1 dz_2$, $\int dv dz_1$
and  $\int dv$ ($\int dv dz_2$ is obtained by symmetrization).
The corresponding three-fold  branching method is managed by the class {\bf MBrB}.
The class {\sf Bstra} contains mainly the interface to the {\sf Foam} and {\sf Vegas}
packages, which generate in each branch the corresponding subset of the variables $v, z_1, z_2$.

Since {\sf Foam} makes look-up tables for the distribution
to be generated,
we need three copies of the {\sf Foam}, which are {\bf FoamA, FoamB, FoamC}.
Similarly for {\sf Vegas}.
There are only five subprograms in the {\bf Bstra} class:
\begin{itemize}
  \item
    {\sf BStra\_Initialize(KeyGrid,Xcrude)} initializes
    {\bf FoamA, FoamB, FoamC} (or {\bf VegasA, VegasB, VegasC})
    and the {\bf MBrB} for book-keeping in the three-fold branching.
    It also provides the primary integrated cross section used
    to establish the overall normalization in {\bf Karlud} and {\bf KK2f}.
  \item
    {\sf BStra\_Make(vv, x1, x2, MCwt)} randomly chooses one of the
    branches with the help of {\sf MBrB\_GenKF}.
  \item
    {\sf BStra\_Finalize(Integ,Errel)} is calculating the total cross section
    using the average provided by  {\sf MBrB\_MgetAve}. 
    This is for control only.
  \item
    {\sf BStra\_GetXCrude(XCrude)}
    provides the primary integrated cross section that enters the overall normalization.
  \item
    {\sf BStra\_GetIntegMC(IntegMC,ErRelMC)}
    provides the value of the primary integrated cross section from the entire MC run.
    This is for control only.
\end{itemize}

The {\tt CIRCE} library of the beamsstrahlung structure functions \cite{circe:1996}
is placed in the {\bf IRC} module.
The only changes are: adding prefix {\tt IRC\_} to names of all subprograms
and the common block {\tt /circom/} is  renamed as {\tt  /c\_IRC/}.
It is exploited in the {\bf BStra} class.

\subsection{BVR class: virtual corrections}
The class {\bf BVR} is a collection of the complex functions
used in the calculation of the virtual corrections.
It provides also the $\tilde{B}$ function necessary to calculate the compensating
weight in the procedure of the removal of the FSR photons, see subsection \ref{sec:hiding}.
It has its own library of  complex logarithms and dilogarithms.

\subsection{QED3 class: EEX distributions}
The EEX differential distributions is implemented in the {\bf QED3} module.
This module is rather monolithic.
It contains the initializer {\sf QED3\_Initialize}, 
the maker {\sf QED3\_Make} and several small functions
for calculating virtual corrections and up to third-order
leading-logarighmic structure functions.
The basic ingredient in the  EEX differential distributions
is the Born differential distribution that comes from the {\bf BornV} class.

\subsection{GPS class: CEEX Matrix element}
The CEEX matrix element is programmed in the {\bf GPS} class.
It calculates spin amplitudes for the $e^-e^+\to f\bar{f} n\gamma$ process.
It has grown to a very large module (almost 5000 lines of code) and will therefore
be split in the next version into a low-level library of GPS tools and the
module {\bf CEEX}, which calculates solely the CEEX spin amplitudes.

The main subprogram in the {\bf GPS} class is {\sf GPS\_Make}, which calculates
\Order{\alpha^r} $r=0,1,2$ CEEX spin amplitudes 
{\sf m\_AmpExpo0(4,4,4,4), m\_AmpExpo1(4,4,4,4), m\_AmpExpo2(4,4,4,4)}
of the {\sf DOUBLE COMPLEX} type.
Photon helicities are generated randomly in the upper class {\bf KK2f}
and provided with the getter {\sf KK2f\_GetPhel}.
Virtual corrections are provided by subprograms from the {\bf BVIR} class.
Masses, charges, isospin of the particles are provided by the getters of the {\bf BornV} class
and the electroweak form factor come from DIZET 6.21 through the interface 
subprogram {\sf BornV\_GetGSW}.
The three model weights \Order{\alpha^r}, $r=0,1,2$, are  calculated
in {\sf GPS\_Make} using GPS\_MakeRho for polarized beams and {\em unpolarized} final fermions.
Subprogram {\sf GPS\_MakeRho2} is calculating model weights for 
polarized beams and {\em polarized} final fermions.
It is used by {\sf Taupair\_ImprintSpin} to implement spin effects in $\tau$ decays.

Let us now list and explain three groups of subprograms in the {\bf GPS} class,
(a) the main subprograms calculating the spin amplitudes, 
(b) the library of basic tools and
(c) the communication subprograms (setters and getters).
The first group includes:
\begin{itemize}
\item
  {\sf GPS\_Initialize} initializes of the class.
  It sets some coupling constants, Pauli matrices, the axial gauge vector $\beta$.
\item
  {\sf GPS\_Make} is the main routine that calculates spin amplitudes
  {\sf m\_AmpExpo0}, {\sf m\_AmpExpo1} and {\sf m\_AmpExpo2}.
  Spin amplitudes are calculated in such a way that they are first set to zero and
  then, in the sum over partitions, they are incremented 
  by $\beta^{(r)}_0$ with the help of {\sf GPS\_BornPlus},
  by $\beta^{(r)}_1$ with the help of {\sf GPS\_HiniPlus} and {\sf GPS\_HfinPlus},
  and by $\beta^{(r)}_2$ with help of {\sf GPS\_HiiPlus, GPS\_HffPlus} and {\sf  GPS\_HifPlus}.
  Three model weights are calculated  using {\sf GPS\_MakeRho}
  and set to  {\sf m\_WtSet(i), i=1,2,3}
  (or to {\sf m\_WtSet(i),i=51, 52, 53} if ISR--FSR interference is switched off).
  The best weight is set as {\sf m\_WtBest = m\_WtSet(3)} (or {\sf m\_WtSet(53)}).
  Weights are available through getter {\sf GPS\_GetWtSet}, see below.
\item
  {\sf GPS\_MakeRho(ExpoNorm)}
  calculates differential distributions (normalized to Lorentz-invariant phase space) 
  from spin amplitudes
  {\sf m\_AmpExpo}$i$, $i=1,2,3$, for polarized beams and
  {\em unpolarized} final fermions.
  Beam polarizations are set from outside with the help of {\sf GPS\_SetPolBeams}.
\item
  {\sf GPS\_MakeRho2(wt0,wt1,wt2)} is
  used in {\sf  Taupair\_ImprintSpin} and it
  calculates the differential distributions (normalized to Lorentz-invariant phase space) 
  from spin amplitudes
  {\sf m\_AmpExpo}$i$, $i=0,1,2$, for polarized beams and
  {\em polarized} final fermions.
  Final-state polarimeter vectors are set from outside with the help of 
  {\sf GPS\_SetHvectors}.
\item
  {\sf GPS\_BornPlus} calculates the spin amplitudes of $\beta^{(r)}_0$.
  It is optimized for summation over partitions.
  Virtual corrections (boxes and vertices) are included.
\item
  {\sf GPS\_Born} provides the Born spin amplitudes
  used in the construction of the hard non-IR parts: in 
  {\sf GPS\_HiniPlus, GPS\_HfinPlus} and other subprograms.
  It is essentially a simplified clone of {\sf GPS\_BornPlus}.
\item
  {\sf GPS\_EWFFact}
  creates form factors for electro-weak corrections.
  They are in vector couplings (multiplied by the correcting factors).
  Because of the $\cos\theta$ dependence of WW boxes, 
  we had to introduce {\sf CosThetD} parameter.
\item
  {\sf GPS\_HiniPlus} calculates
  the IR-finite part of 1-photon amplitudes for ISR $\beta^{(r)}_{1\{1\}}$.
  (It is equivalent to a testing subprogram {\sf GPS\_Hini}).
\item
  {\sf GPS\_HfinPlus} calculates
  the IR-finite part of 1-photon amplitudes for FSR $\beta^{(r)}_{1\{0\}}$.
  (It is equivalent to a testing subprogram {\sf GPS\_Hfin}).
\item
  {\sf GPS\_HffPlus} calculates
  the IR-finite part of 2-photon amplitudes for FSR $\beta^{(r)}_{1\{00\}}$.
\item
  {\sf GPS\_HiiPlus} calculates
  the IR-finite part of 2-photon amplitudes for ISR $\beta^{(r)}_{1\{11\}}$.
\item
  {\sf GPS\_HifPlus} calculates
  the IR-finite part of 2-photon amplitudes $\beta^{(r)}_{1\{10\}}$ 
  for one ISR and one FSR photon.
\end{itemize}

\noindent
Let us now list and explain the subprograms that play a role
of the library of basic tools. (In the future version to be isolated
as a separate class, or even several classes).
This group includes:
\begin{itemize}
\item
  {\sf GPS\_PartitionStart(nphot,last)}
  initializes the first partition in the sum over partitions.
\item
  {\sf GPS\_PartitionPlus}
  updates the partition vector {\sf m\_isr}, checks if it is the last partition.
\item
  {\sf GPS\_BornZero(AmpBorn)}
  sets {\sf AmpBorn} to zero.
\item
  {\sf GPS\_BornCopy(AmpBorn,AmpBorn2)}
  copies {\sf AmpBorn} into {\sf AmpBorn2}.
\item
  {\sf GPS\_BornSumSq(AmpBorn,Sum)}
  sums up {\sf AmpBorn} amplitudes squared.
\item
  {\sf GPS\_TralorPrepare}
  prepares transformation for {\sf Tralor}, according to GPS rules.
  The resulting Lorentz transformation matrix is stored for multiple use.
\item
  {\sf GPS\_GPS(xi,eta,Rot)}
  defines the basis vectors $e_1,e_2,e_3$ from $\xi$ and $\eta$ according to GPS rules.
  Columns in the matrix {\sf Rot} are $e_1,e_2,e_3$.
  This subprogram is called in {\sf GPS\_TralorPrepare}.
\item
  {\sf GPS\_TralorDoIt(id,pp,q)}
  transforms the four-vector {\sf pp} from rest frame of fermion {\sf id} to LAB, 
  {\sf q} is the result.
  It uses a Lorentz transformation, prepared and memorized in
  the subprogram {\sf GPS\_TralorPrepare}, which has to be called first.
  This organization saves CPU time in the case of multiple calls for several $\tau$
  decay products.
\item
  {\sf GPS\_TralorUnDo(id,pp,q)}
  is the inverse of {\sf GPS\_TralorDoIt}.
  It transforms {\sf pp} from the laboratory to the rest frame of the 
  final fermion, {\sf q} is the result.
  It uses a Lorentz transformation, prepared and memorized in 
  the subprogram {\sf GPS\_TralorPrepare}, which has to be called first.
\item
  {\sf GPS\_TraJacobWick(Mode,QQ,pp,rr)} is for tests only.
  It is a {\tt Tralor}-type transformation for the classical Jacob--Wick quantization axes.
  Not optimized. 
\item
  {\sf GPS\_RmatMake} is for tests only.
  It translates Born spin amplitudes into a double-spin density matrix
  {\sf m\_AmpBorn} $\to$ $R_{ab}$
\item
  {\sf GPS\_MakeU(ph,sigma,p1,m1,p2,m2,U)} builds
  the transition matrix $U$, ($\bar{u}\not\!{\epsilon}^*u$).
\item
  {\sf GPS\_MakeV(ph,sigma,p1,m1,p2,m2,V)} builds
  the transition matrix $V$, ($\bar{v}\not\!{\epsilon}^*v$).
\item
  {\sf GPS\_MakeUb(ph,sigma,p1,m1,p2,m2,U)} builds
  the transition matrix $U$, ($\bar{u}\not\!{\epsilon}^*u$).
\item
  {\sf GPS\_MakeVb(ph,sigma,p1,m1,p2,m2,V)} builds
  the transition matrix $V$, ($\bar{v}\not\!{\epsilon}^*v$).
\item
  {\sf GPS\_MatrU(Cfact,ph,sigma,p1,m1,p2,m2,U)} builds
  the transition matrix $U$, ($\bar{u}\not\!{\epsilon}^*u$).
\item
  {\sf GPS\_MatrV(Cfact,ph,sigma,p1,m1,p2,m2,V)} builds
  the transition matrix $V$, ($\bar{v}\not\!{\epsilon}^*v$).
\item
  {\sf GPS\_MatrUb(Cfact,ph,sigma,p1,m1,p2,m2,U)} builds
  the transition matrix $U$, ($\bar{u}\not\!{\epsilon}^*u$).
\item
  {\sf GPS\_MatrVb(Cfact,ph,sigma,p1,m1,p2,m2,V)} builds
  the transition matrix $V$, ($\bar{v}\not\!{\epsilon}^*v$).
\item
  {\sf DOUBLE COMPLEX FUNCTION GPS\_Sof1(sigma,ph,pf)} calculates
  the single soft photon contribution to the $\sfac$-factor.
\item
  {\sf DOUBLE COMPLEX FUNCTION GPS\_Sof1b(sigma,ph,pf,mf)} calculates
  the single soft photon contribution to the $\sfac$-factor.
\item
  {\sf DOUBLE COMPLEX  FUNCTION GPS\_soft(sigma,ph,p1,p2)} calculates
  the two-fermion $\sfac$-factor.
\item
  {\sf DOUBLE COMPLEX  FUNCTION GPS\_bfact(sigma,phot,pferm)} calculates the
  diagonal element of the $U$-matrix for the massive fermion (the numerator in the $\sfac$-factor).
\item
  {\sf DOUBLE COMPLEX  FUNCTION GPS\_softb(sigma,ph,p1,m1,p2,m2)} calculates
  the $\sfac$-factor.
\item
  {\sf DOUBLE COMPLEX  FUNCTION GPS\_bfacb(sigma,phot,pferm,mass)} calculates
  the diagonal element of the $U$-matrix for massive fermion (the numerator in the $\sfac$-factor).
\item
  {\sf DOUBLE COMPLEX  FUNCTION GPS\_iProd1(L,p,q)} calculates
  the basic inner product of spinors $s_{\lambda}(p,q)=\bar{u}_{\lambda}(p) u_{-\lambda}(q)$.
  We exploit the identity $s_{-}(p,q) = -[s_{+}(p,q)]^*$.
\item
  {\sf DOUBLE COMPLEX  FUNCTION GPS\_iProd2(Lamp,p,mp,Lamq,q,mq)} calculates
  the general spinor product $s_{\lambda_{1},\lambda_{2}}(p,q)$ for massive spinors $u$ and/or $v$;
  {\sf mp} and {\sf mq} are the masses of four-vectors {\sf p} and {\sf  q}. 
  Negative mass means an antiparticle.
\item
  {\sf DOUBLE PRECISION  FUNCTION GPS\_XiProd(p,q)} is the
  auxiliary function called in {\sf  GPS\_iProd2}.
\end{itemize}

\noindent
The last group includes communication subprograms (setters and getters)
and some miscellaneous routines for debugging:
\begin{itemize}
\item
  {\sf GPS\_BPrint(nout,word,AmpBorn)}
  prints 16 spin amplitudes of {\sf AmpBorn}
  in a nice format on output unit {\sf nout}.
\item
  {\sf GPS\_GetXi(xi,eta)} provides
  $\xi$,  the basic light-like vector in the laboratory frame, entering the definition of all spinors
  (called $k_0$ in Kleiss--Stirling papers).
\item
  {\sf GPS\_SetKeyArb(KeyArb)}
  {\sf GPS\_GetKeyArb(KeyArb)} sets
  {\sf KeyArb}, which is switching on/off the use of {m\_b},   
  {\sf KeyArb=0} means $\beta\to \xi$.
\item
  {\sf GPS\_Setb1}
  {\sf GPS\_Setb2} switches the axial gauge vector $\beta${\sf =b} to another predefined value.
  This is for testing the gauge invariance of the spin amplitudes.
\item
  {\sf GPS\_GetWtSet(WtBest,WtSet)}
  provides a complete list of weights.
\item
  {\sf GPS\_SetKeyINT(KeyINT)}
  sets the  IFI switch {\sf KeyINT}
\item
  {\sf GPS\_SetPolBeams(PolBeam1,PolBeam2)}
  sets the beam polarization vectors.  
  One should not forget the Wigner rotation to the GPS frame!
\item
  {\sf GPS\_SetHvectors(HvecFer1,HvecFer2)}
  sets the final-fermion polarimeter vectors.
\end{itemize}

\subsection{TAUOLA and PHOTOS}
TAUOLA and PHOTOS are placed in {\sf KK-all/tauola} and {\sf KK-all/photos}.
They communicate with the rest of the program
through an interface class {\bf Taupair} located in {\sf KK-all/KK2f}. The initialization is performed in the { \bf Tauface} class 
as well.
The other, very important role of {\bf Taupair} is to implement
spin effects in the decays of both $\tau$'s, including all spin correlations
with the rejection method according to the special spin weght;
as in KORALB~\cite{koralb:1985}.
The spin weight is:
\begin{equation}
  \label{eq:spin-weight}
  \begin{split}
    W_{\rm spin}=
       {
        \sum\limits_{\sigma_r,\lambda_A,\bar{\lambda}_A}\;
        \sum\limits_{i,j,l,m}\;
        \hat{\varepsilon}^i_1                  \hat{\varepsilon}^j_2\;
        \sigma^i_{\lambda_a \bar{\lambda}_a}   \sigma^j_{\lambda_b \bar{\lambda}_b}
        \Mmf^{(r)}_n \left(\st^{p}_{\lambda} \st^{k_1}_{\sigma_1} \st^{k_2}_{\sigma_2} 
                                                            \dots \st^{k_n}_{\sigma_n} \right)
        \left[
        \Mmf^{(r)}_n \left(\st^{p}_{\bar{\lambda}}\st^{k_1}_{\sigma_1}
                                            \dots \st^{k_n}_{\sigma_n} \right)
        \right]^\star
        \sigma^l_{\bar{\lambda}_c \lambda_c }   \sigma^m_{\bar{\lambda}_d \lambda_d }
        \hat{h}^l_3                             \hat{h}^m_4
       \over
        \sum\limits_{\sigma_r,\lambda_A}\;
        \sum\limits_{i,j}\;
        \left|
        \Mmf^{(r)}_n \left(\st^{p}_{\lambda} \st^{k_1}_{\sigma_1} \st^{k_2}_{\sigma_2} 
                                                            \dots \st^{k_n}_{\sigma_n} \right)
        \right|^2
       }.
  \end{split}
\end{equation}
Note that in the present version of the program we include at this step not only the final-state
spin effects but also beam-polarization effects.
This is not a very economical solution, especially for strongly polarized beams, when
we may get large rejection rates
(roughly equal to the ratio of polarized to unpolarized Born cross section).
The radical solution of this problem is
to introduce longitudinal polarizations in the Born cross section
as used in the {\em crude} and {\em primary} integrated cross section.
At the moment they are completely unpolarized.

In order to save CPU time a special method of ``recycling'' the $\tau$ decay events
is devised (see below).
The interface supplies also the subroutine {\sf TRALO4},
which is required by {\sf TAUOLA}
in order to transform $\tau$ decay products to the laboratory frame.

Some additional subroutines, necessary for the proper
functioning of TAUOLA and PHOTOS, are placed in {\sf KK-all/KK2f/Tauface.f}.
\begin{itemize}
   \item
     {\sf Taupair\_Initialize(xpar) } initializes
     {\sf TAUOLA} and {\sf PHOTOS} packages with the help of
     {\sf INIMAS, INITDK, INIPHY} and {\sf DEKAY}.
     It initializes the book-keeping for the spin weight, that is the weight
     used to introduce all spin effects in $\tau$ decays.
   \item
     {\sf Taupair\_Finalize } prints the average spin weight.
   \item
     {\sf Taupair\_Make1 }
     generates in the first step the unpolarized $\tau$ decays using {\sf DEKAY}.
     The polarimeter vectors {\sf m\_HvecTau1} and  {\sf m\_HvecTau2} are determined.
   \item
     {\sf Taupair\_ImprintSpin }
     introduces spin effects with the help of rejection using spin weight.
     The polarimeter vectors are sent to {\bf GPS} with the help of {\sf GPS\_SetHvectors}
     and the spin weight is
     calculated with the help of {\sf GPS\_MakeRho2};
     the event is then rejected or accepted.
     For the rejected event
     the $\tau$-pair event is ``recycled'', that is each $\tau$
     decay product is Euler-rotated and reused in the rejection method.
     The procedure is repeated until the event is accepted.
     The whole procedure is correct because
     we know exactly the average of the spin weight.
   \item
     {\sf Taupair\_Make2 }
     transforms accepted $\tau$ decay products to the CMS by calling 
     {\sf DEKAY(11)} and {\sf DEKAY(12)}.
     The transformation is defined according to the GPS rules,
     for each $\tau$ by {\sf GPS\_tralorPrepare},
     and is performed with the help of {\sf GPS\_TralorDoIt} hidden
     inside the {\sf TRALO4} routine. 
   \item
     {\sf Taupair\_Clone }
     performs the ``recycling'' of a $\tau$-pair by means of the Euler rotation in the rest frame
     of each $\tau$.
   \item
     {\sf Tralo4(Kto,P,Q,AM) }, see above.
   \item
     {\sf FILHEP(N,IST,ID,JMO1,JMO2,JDA1,JDA2,P4,PINV,PHFLAG) }
     writes single particles in $\tau$-decay into {\sf HepEvt}
     class. For historical reasons {\sl HepEvt\_Fil1} is not 
     used directly.
   \item
     {\sf Taupair\_SetKeyClone(KeyClone) } sets {\sf KeyClone}.
     {\sf KeyClone} switches between two operational modes of {\sf Taupair\_Clone}.
     Both of the modes implement a valid solution.
   \item
     {\sf Taupair\_GetIsInitialized(IsInitialized) }
     gets to know the outside world if TAUOLA is active (IsInitialized=1).
   \item
     {\sf Taupair\_GetHvectors(HvecFer1,HvecFer2) }
     provides the polarimetric $h$-vectors.
\end{itemize}

\subsection{Electroweak library}
\label{sec:ew-lib}
The library of electroweak (EW) corrections is placed in the {\tt KK-all/dizet} subdirectory.
In the initialization phase the EW form factors dependent on $s'$, fermion type and some of them
(electroweak boxes) also on the scattering angles,
are placed in the look-up tables.
During the event generation they are interpolated in $s'$ and $\cos\theta$ and
provided to the {\bf GPS} module, where the CEEX spin amplitudes are calculated,
or used in the {\sf BornV\_Dizet} being used (through  {\sf BornV\_Differential})
in the {\bf QED3}, where the EEX distributions are calculated.
The main aim of the above organization it
to speed up the MC by using the EW form factor from look-up tables instead
of calculating them, in fact many times, for each MC event.
This reason may be even more important in the future version of the EW corrections 
which will be slower, owing to the inclusion of more genuine two-loop corrections.

There are two modes of the initialization of the EW look-up tables.
In the default mode they are calculated and stored in several disk files, 
each for one fermion type, and in the initialization
of the \KK\ MC run these tables are read by the {\bf BornV} module.
This mode is more conservative (safer), because 
the Fortran77 program providing electroweak corrections
does not need to be linked and executed together with the proper MC event generator.
We do not therefore need to worry about the clashes of the names of the procedures and
common blocks, and the possible problems with re-initialization of the EW library
for different types of the fermions is avoided%
\footnote{
  This reason seems to be now less important than in the early stages of the development
  of the \KK\ MC, because most of the \KK\ MC code now fulfils the rules of programming
  in sections \ref{sec:prog-rules}.
}.
In the actual implementation the EW library DIZET
is run separately, under a special main program
{\tt KK-all/dizet/TabMain.f}, together with
an interface module {\bf DZface}, which acts as an interface
to the {\tt Dizet} library, properly setting the input data to {\tt Dizet}
and writing the EW form factors in the disk file.

One important disadvantage of the above method is that the input parameters of the EW corrections,
such as the Higgs mass, cannot be changed easily (for fitting), 
because it requires re-producing new look-up tables of the EW corrections.
This is why we also implemented the second interface to EW library DIZET in which the look-up
tables are calculated in the initialization phase of the MC run 
(in this case the EW library is linked with the entire MC program).

How are the EW tables produced in the default method?
This is done by invoking in the {\sf KK-all/dizet/}
one of the commands:
{\small
\begin{verbatim}
   make tables        # it makes all tables
   make table.mu      # it makes ./table.mu     using ./input.mu
   make table.tau     # it makes ./table.tau    using ./input.tau
   make table.down    # it makes ./table.down   using ./input.down
   make table.up      # it makes ./table.up     using ./input.up
   make table.botom   # it makes ./table.botom  using ./input.botom
\end{verbatim}}
The input data should be used the same as in the MC run.
In fact, the {\bf BornV} module is checking if the important EW
input data used to generate tables match the actual data provided
by the user for the MC run. If not then the program stops.

How does one avoid  producing EW tables on the disk (and apply the second method)?
For the instructions see {\tt KK-all/ffbench/Makefile}.

The interface {\bf DZface} to {\tt DIZET} of the Dubna--Zeuthen EWRC group version 6.21
is based on the analogous interface in KORALZ 4.x.
Let us now list and explain the subprograms in the interface module {\bf DZface}
\begin{itemize}
   \item
     {\tt DZface\_Initialize( KFfin, xpar)}
     is the class initializer.
     Initialization of {\tt DIZET} is done with {\tt CALL DIZET(NPAR,...)}.
     {\tt NPAR} and other input parameters are defined in the {\sf xpar} vector,
     see Table \ref{tab:KK-input3}.
   \item
     {\tt DZface\_ReaDataX(DiskFile,iReset,imax,xpar)} is functionally the same as\\
     {\tt KK2f\_ReaDataX}.
   \item
     {\tt DZface\_Tabluj} 
     fills in the EW form factors into look-up tables in {\sf /c\_BornV/}.
     The factors are provided by {\tt DZface\_MakeGSW} (see below).
     Tabulation is done in the three ranges of $\sqrt{s}$ with different numbers of points.
     The energy ranges and numbers of points are defined in {\tt BornV.h}.
   \item
     {\tt DZface\_WriteFile(DiskFile)}
     writes tables of the EW form factors into a disk file.
   \item
     {\tt DZface\_Clone(KFfin)}
     copies tables calculated for the actual {\tt m\_KFfin} into tables for {\tt KFfin}.
     It is used to create tables for $c$ and $s$ quarks using tables of $u$ and $d$ quarks
     (saving a little bit of CPU time).
   \item
     {\tt  DZface\_MakeGSW(Mode,ww,cosi,GSW,QCDcorN)} gets
     the EW form-factors {\tt GSW} and QCD corrections {\tt QCDcorN}
     out of DIZET, at $\sqrt{s}=${\tt ww} and $\cos\theta=${\tt cosi}.
     EW form-factors are obtained with  the {\tt CALL rokanc(...)},
     while QCD corrections come from {\tt DZface\_QCDtab} (see below).
   \item
     {\tt DZface\_QCDtab(Mode,ww,QCDcorN)} makes QCD corrections with the\\
     {\tt CALL qcdcof(...)}. It is done in an iterative way
     in order to find out the QCD corrections at a given $\sqrt{s}$.
\end{itemize}
Note that the QED coupling constant the {\tt alfinv} is separate from 
the {\tt alfinv} used in the bremsstrahlung part of \KK\ and 
the fermion masses in {\tt Dizet} are isolated from those in \KK.

\subsection{Random number generators}

The \KK\ program in the present version uses exclusively the
{\sf RANMAR} random number generator~\cite{ranmar:1990,marsaglia:1987}.
It is reprogrammed as the  pseudo-class {\bf  PseuMar}.
Its single-precision generator is accessible with the help of the double-precision interface routine 
{\sf PseuMar\_MakeVec}.
In this way we avoid possible interference with libraries of
JETSET, PHOTOS and TAUOLA, which have their own independent 
random-number generators, and typically also have their own version of the {\sf RANMAR} generator.

\subsection{Other modules}

Every MC program of this size has to have its own tools for book-keeping of the MC weights
and for making histograms of the weight distribution.
In the \KK\ MC the built-in histogramming package {\bf GLK} plays this role.
Histogramming entries are similar to those of the CERN library {\tt HBOOK}.
Apart from histogramming, it also has the capability of ``measuring'' several properties
of the MC weight.
This weight monitoring is done with the entries
{\tt GLK\_Mbook,  GLK\_Mfill, GLK\_MgetAll, GLK\_MgetNtot, GLK\_MgetAve} and  {\tt GLK\_Mprint}.
The {\bf GLK} module features also simple, though versatile, graphical capabilities --
it can plot histograms by exploiting the \LaTeX environment {\sf picture}. 

The other auxiliary package is {\bf MathLib.f}, which includes
subprograms for Gaussian integration and some transcendental functions.
The \KK\ MC does not need any external mathematical library.

\begin{figure}[!ht]
\centering
\setlength{\unitlength}{0.1mm}
\begin{picture}(1600,850)
\put( 600,600){\makebox(0,0)[b]{\LARGE (a)}}
\put(1400,600){\makebox(0,0)[b]{\LARGE (b)}}
\put(  -20, 0){\makebox(0,0)[lb]{
\epsfig{file=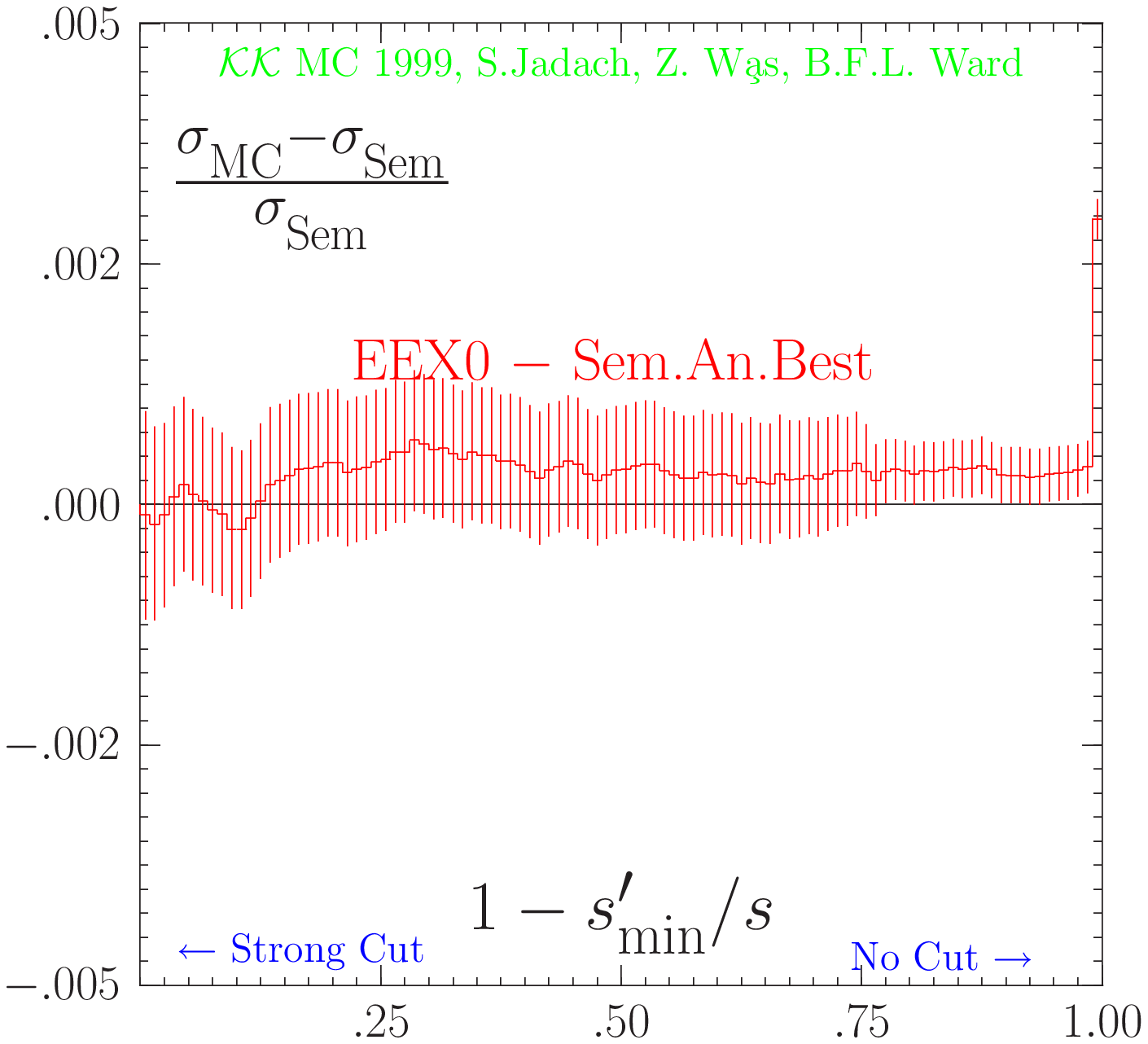,width=80mm,height=80mm}}}
\put(  800, 0){\makebox(0,0)[lb]{
\epsfig{file=chi-mcan-O0tech.eps.999,width=80mm,height=80mm}}}
\end{picture}
\caption{\small\sf
  Test of the normalization of the low-level part of the Monte Carlo,
  for simplified QED multiphoton distribution.
  The difference of the MC result and semianalytical result divided by semianalytical
  is plotted as a function of $v_{\max}=1-s'_{\min}/s$.
  Test is done for $\mu^+\mu^-$ final state at $\protect\sqrt{s}=189$GeV.
  In case (a) $v_{\max}=1-4m_\mu^2/s$ is taken;
  the last bin represent the entire phase phase space.
  In case (b) $v_{\max}=0.999$.
}
\label{fig1}
\end{figure}

\section{Semianalytical ``normalization anchor''}

In the complicated MC program aspiring to a precission of order $10^{-4}$,
it is critical to master the overall normalization at this precision level.
This can be done by comparing the program results with another MC program or with
a semi-analytical calculation, that is the calculation in which
as many integrations as possible are carried out analytically and the remaining ones
are done with the Gaussian method%
\footnote{This is the way the {\tt Zfitter} phase space integration is done.}.
Such a test of MC normalization is of critical importance -- this is why we call it a
{\em normalization anchor}.

It is not trivial to analytically integrate the multiphoton phase space;
nevertheless, for simple or simplified exponentiated distributions such as EEX
and simple or absent kinematical cuts, such an integration is possible,
see refs.~\cite{bhlumi-semi:1996,ceex2:1999}.

Here, since our aim is essentialy limited to the very precise numerical test
of the MC phase-space integration, we have chosen the \Order{\alpha^0} EEX model,
that is the Born differential cross section multiplied by the soft factors,
see eqs.~(\ref{sigma-eex3}) and (\ref{eq:rho-eex3}).

In this case it is relatively easy to obtain by analytical integration 
the \Order{\alpha^0} EEX cross section,
keeping terms $L^0\alpha^1,L^1\alpha^1,L^2\alpha^2$,
where $L$ is the big logarithm, for both ISR and FSR.
For the ISR, however, due to the $Z$ resonance, and $\gamma^*$ peak at $s'=0$,
in order to reach the necessary precision of order $10^{-4}$,
it is necessary to analytically calculate two additional terms of orders
$L^1\alpha^2$ and $L^3\alpha^3$.
This was done in ref.~\cite{ceex2:1999},
and the relevant semianalytical formula looks as follows:
\begin{equation}
  \label{eq:eex0-best}
  \begin{split}
    &\sigma^f_{\rm SAN}
   = \int_0^{v_{\max}} dv\;
     \sigma^f_{\rm Born}(s(1-u)(1-v))\;
     D_{ISR}(v)\; D_{FSR}(u),\\
    &D_{ISR}(v) = e^{ -C\gamma_e) \over \Gamma(1+\gamma_e)} \;
                 e^{ {1\over 4} \gamma_e +{\alpha\over\pi}\big({1\over 2} +{\pi^2\over 3}\big)}\;
                 \gamma_e v^{\gamma_e-1}\;
                 \bigg(1 -{1\over 4}\gamma_e \ln(1-v) 
                         -{1\over 2} {\alpha\over\pi} \ln^2(1-v) 
                         +0\;\gamma_e^2
                 \bigg),\\
    &D_{FSR}(u) = e^{ -C\gamma_f) \over \Gamma(1+\gamma_f)} \;
                 e^{ {1\over 4} \gamma_f -{1\over 2} \gamma_f \ln(1-u)
                     +{\alpha\over\pi}\big({1\over 2} +{\pi^2\over 3}\big)}\;
                 \gamma_f u^{\gamma_f-1}\;
                 \bigg(1 -{1\over 4}\gamma_f \ln(1-u) \bigg),
  \end{split}
\end{equation}
Note that the coefficient in front of the \Order{L^3\alpha^3} term is zero,
as marked explicitly.
We have checked numerically that the ISR term of \Order{L^1\alpha^2} is worth
several per cent for the cross section located close to $v=1$.

In fig.~\ref{fig1} we present the comparison of the \KK\ MC with the above semianalytical formula.
The difference between the MC result and the semianalytical result 
is divided by the semianalytical result.
The comparison is done for the $\mu^+\mu^-$ final state at $\sqrt{s}=189$ GeV,
as a function of $v_{\max}$.
In the last point (bin) 
the entire phase space is covered, $v_{\max}=1-4m_\mu^2/s$.
We conclude that we control the phase-space integration at the level of 
$2\times 10^{-4}$ for $v_{\max}< 0.999$,
including the $Z$ radiative return,
and  at the level of $3\times 10^{-3}$ for no cuts at all.

\begin{table}[!ht]
\centering
\begin{small}
\begin{tabular}{|l|p{13.0cm}|}
\hline
Parameter & Position and meaning  \\ 
\hline\hline
{\tt CMSene}          & {\tt xpar(1) (=100)}: $\sqrt{s}$, centre-of-mass (CMS) energy  [GeV]\\
{\tt DelEne}          & {\tt xpar(2) (=0d0)}: Beam energy spread [GeV]\\
{\tt Ninp}            & {\tt xpar(3) (=5)}:   Input unit number (unused)\\
{\tt Nout}            & {\tt xpar(4) (=16)}:  Output unit number\\
{\tt LevPri}          & {\tt xpar(5) (=0)}:   PrintOut Level 0,1,2\\
{\tt Ie1Pri}          & {\tt xpar(6) (=1)}:   PrintOut Start point\\
{\tt Ie2Pri}          & {\tt xpar(7) (=1)}:   PrintOut End   point\\
{\tt IdYFS}$^{*}$     & {\tt xpar(8) (=600)}: Pointer for internal histograms\\
{\tt WtMax}$^{*}$     & {\tt xpar(9)  (=1)}:  Maximum weight for rejection\\
{\tt KeyWgt}          & {\tt xpar(10) (=0)}:  Switch between constant =0 and variable =1 weight events\\
{\tt IdeWgt}$^{*}$    & {\tt xpar(11) (=74)}: Ident of the EEX principal weight\\
{\tt KeyELW}          & {\tt xpar(12) (=1)}:  Type of electroweak corrections, 
                        =0 only for tests, =1 default for DIZET\\
{\tt vvmin}$^{*}$     & {\tt xpar(16) (=1D-5)}: Minimum real photon energy in units of beam energy\\
{\tt vvmax}           & {\tt xpar(17) (=1d0)}:  Maximum value of $v=1-s'/s$-variable, where $s'$
                        is mass squared of $f\bar{f}$ system including FSR photons! 
                        See more comments in the text.\\
{\tt DelFac}$^{*}$    & {\tt xpar(18) (=1d-3)}: FSR cut eps=vvmin*DelFac\\
{\tt NphMax}$^{**}$   & {\tt xpar(19) (=100)}: Hard-wired maximum photon multiplicity\\
{\tt KeyISR}          & {\tt xpar(20) (=1)}: Test switch, KeyISR=0 swithes off the ISR\\
{\tt KeyFSR}          & {\tt xpar(21) (=1)}: Test switch, KeyFSR=0 switches off the FSR\\
{\tt KeyPia}$^{**}$   & {\tt xpar(22) (=1)}: Removal of FSR photons below 
                          {\tt Emin=Ene*Delta} in CMS, for KeyPia=0,1 removal is OFF,ON\\
{\tt mltISR}$^{**}$   & {\tt xpar(23) (=0)}: Special tests: fixed ISR multiplicity for {\tt mltISR}$>$0\\
{\tt mltFSR}$^{**}$   & {\tt xpar(24) (=0)}: Special tests: fixed FSR multiplicity for {\tt mltFSR}$>$0\\
{\tt KeyFix}          & {\tt xpar(25) (=0)}: Type of ISR, for KeyFix=0,1 QED without beamstrahlung, 
                             for KeyFix=2 beamstrahlung is ON, see also KeyGrid\\
{\tt KeyWtm}$^{**}$   & {\tt xpar(26) (=0)}: Special tests only: 
                             mass terms in ``crude'' MC photon distrib.\\
{\tt KeyINT}          & {\tt xpar(27) (=2)}: Switch of ISR-FSR Interference (IFI), 
                             for KeyINT=0 it is OFF, for KeyINT=2 it is ON,
                             KeyINT=1 is only for special tests\\
{\tt KeyGPS}          & {\tt xpar(28) (=1)}: Level of new exponentiation CEEX, note {\tt vmaxGPS} 
                             overrules {\tt KeyGPS} for each type of final fermion\\
{\tt KeyQSR}          & {\tt xpar(29) (=1)}: Photon emission from 
                             the final quarks is ON,OFF for KeyQSR=0,1\\
\hline
\end{tabular}
\end{small}
\caption{\sf 
  List of input parameters of the \KK\ generator.
  General and related to QED radiation input parameters.
  Default values in brackets.
  User may change, with precautions, the starred items, while the doubly starred ones
  should never be changed.
}
\label{tab:KK-input1}
\end{table}

\begin{table}[!ht]
\centering
\begin{small}
\begin{tabular}{|l|p{13.0cm}|}
\hline
Parameter & Position and meaning  \\ 
\hline\hline
{\tt AlfInv}$^*$       & {\tt xpar(30) (=137.0359895D0)}:  Inverse of $\alpha_{\rm QED}$\\
{\tt GNanoB}$^*$       & {\tt xpar(31) (=389.37966D3)}:    GeV$^{-2}\to$[nb]  conversion factor\\
{\tt Gfermi}$^*$       & {\tt xpar(32) (= 1.16639d-5)}:    Fermi coupling [GeV$^{-2}]$\\
\hline\multicolumn{2}{|c|}{ Technical cuts for non-IR real photon corrections etc.}\\
\hline
{\tt Xenph}$^{**}$ & {\tt xpar(40) (=1.25d0)}:    Enhancement factor for ``crude'' photon multiplicity\\
{\tt vcut1}$^*$    & {\tt xpar(41) (=1.d-9)}:     Techn. cut for single non-IR bremss. correction\\
{\tt vcut2}$^*$    & {\tt xpar(42) (=5.d-2)}:     Techn. cut for double non-IR bremss. correction\\
{\tt vcut3}$^*$    & {\tt xpar(43) (=0.1d0)}:     Techn. cut for triple non-IR bremss. correction\\
\hline\multicolumn{2}{|c|}{ QCD flags/data }\\
\hline
{\tt KeyHad} & {\tt xpar(50) (=1)}:  Hadronization/showering flag, 
                                for {\tt KeyHad=0,1} hadronization is OFF,ON.\\
{\tt HadMin}$^*$ & {\tt xpar(51) (=0.200d0)}:  Minimum mass [GeV] for hadronization/showering\\
{\tt alfQCD}$^*$ & {\tt xpar(52) (=0.118d0)}:  $\alpha_{\rm QCD}$\\
\hline\multicolumn{2}{|c|}{ Non zero beam polarization may require adjustment of WtMax}\\
\hline
{\tt spin1x} & {\tt xpar(61) (=0d0)}:  polarization vector beam 1, $x$-component\\
{\tt spin1y} & {\tt xpar(62) (=0d0)}:  polarization vector beam 1, $y$-component\\
{\tt spin1z} & {\tt xpar(63) (=0d0)}:  polarization vector beam 1, $z$-component\\
{\tt spin1x} & {\tt xpar(64) (=0d0)}:  polarization vector beam 2, $x$-component\\
{\tt spin1y} & {\tt xpar(65) (=0d0)}:  polarization vector beam 2, $y$-component\\
{\tt spin1z} & {\tt xpar(66) (=0d0)}:  polarization vector beam 2, $z$-component\\
\hline\multicolumn{2}{|c|}{ Beamstrahlung parameters for Thorsten Ohl's package CIRCE}\\
\hline
{\tt IRCroots}    & {\tt xpar(71) (=350d0)}: $\sqrt{s}$ [GeV] discrete values 350,500,800GeV\\
{\tt IRCacc}      & {\tt xpar(72) (=3d0)}:  accelerator type\\
{\tt IRCver}      & {\tt xpar(73) (=5d0)}:  version number\\
{\tt IRCdat}      & {\tt xpar(74) (=19980505d0)}: date\\
{\tt IRCxchat}    & {\tt xpar(75) (=1)}: printout level\\
{\tt KeyGrid}$^*$ & {\tt xpar(76) (=2)}: Activated by setting {\tt KeyFix=2}
                                 KeyGrid=2 invokes Foam, KeyGrid=$0,-1,+1$ invokes Vegas:
                                 KeyGrid=0 creates and writes Vegas grid on the disk, 
                                 KeyGrid=$-1$ creates and dumps grid on the disk, =+1 reads only\\
\hline
\end{tabular}
\end{small}
\caption{\sf 
  List of input parameters of the \KK\ generator
  in {\tt xpar} vector.
  General parameters and QED radiation.
  Default values in brackets.
  User may change, with precautions, the starred items, while the doubly starred ones
  should never be changed.
}
\label{tab:KK-input2}
\end{table}

\begin{table}[!ht]
\centering
\begin{small}
\begin{tabular}{|l|p{13.0cm}|}
\hline
Parameter & Position and meaning  \\ 
\hline\hline
{\tt KFini$^*$}    & {\tt xpar(400) (= 11)}:  Beam flavour code\\
\hline\multicolumn{2}{|c|}{ $j$-th fermion is included in MC generation if its Mask($j$)=1}\\
\hline
{\tt Mask( 1)} & {\tt xpar(401) (=1)}:   Mask variable for  $d$ quark\\
{\tt Mask( 2)} & {\tt xpar(402) (=1)}:   Mask variable for  $u$ quark\\
{\tt Mask( 3)} & {\tt xpar(403) (=1)}:   Mask variable for  $s$ quark\\
{\tt Mask( 4)} & {\tt xpar(404) (=1)}:   Mask variable for  $c$ quark\\
{\tt Mask( 5)} & {\tt xpar(405) (=1)}:   Mask variable for  $b$ quark\\
{\tt Mask(13)} & {\tt xpar(413) (=1)}:   Mask variable for  muon lepton\\
{\tt Mask(15)} & {\tt xpar415() (=1)}:   Mask variable for  tau  lepton\\
\hline\multicolumn{2}{|c|}{ Basic electroweak input data}\\
\hline
{\tt MZ}       & {\tt xpar(502) (=91.187D0)}:      Mass of $Z$-boson [GeV]   (PDG 1996)\\
{\tt SwSq}     & {\tt xpar(503) (=.22276773D0)}:  $\sin^2(\theta_W)$ where $\theta_W$ is EW mixing angle\\
{\tt GammZ}    & {\tt xpar(504) (= 2.50072032D0)}: $Z$ width  (from Dizet)\\
{\tt MasPhot}$^*$& {\tt xpar(510) (= 1D-60)}:        Photon mass used as IR regulator\\
\hline\multicolumn{2}{|c|}{ The data base record below is for $d$ quark, $j=1$}\\
\hline
{\tt KFferm(j)}$^*$ & {\tt xpar(501+10*j) (= 1)}:      Flavour code\\
{\tt NCf(j)}$^*$    & {\tt xpar(502+10*j) (= 3)}:      Number of colours\\
{\tt Qf(j)}$^*$     & {\tt xpar(503+10*j) (=-1)}:      3$\times$charge\\
{\tt T3f(j)}$^*$    & {\tt xpar(504+10*j) (=-1)}:      2$\times$T3L =2$\times$Isospin for left component\\
{\tt Helic(j)}$^*$  & {\tt xpar(505+10*j) (= 1)}:       2$\times$helicity,        not used\\
{\tt Mferm(j)}$^*$  & {\tt xpar(506+10*j) (= 0.010d0)}: Mass [GeV] (PDG)\\
{\tt MfCon(j)}$^*$  & {\tt xpar(506+10*j) (= 0.100d0)}: Constituent mass, not used\\
{\tt WtMax(j)}$^*$  & {\tt xpar(507+10*j) (= 5.0d0)}:   Maximum weight for rejection\\
{\tt AuxPar(j)}$^*$ & {\tt xpar(508+10*j) (= 0.99d0)}:  below {\tt vmaxGPS} CEEX, above EEX\\
\hline\multicolumn{2}{|c|}{ More electroweak input data}\\
\hline
{\tt Ibox}     & {\tt xpar(801) (= 1)}:            EW box flag, input for Dizet\\
{\tt MH}       & {\tt xpar(805) (= 100D0)}:        Higgs mass,  input for Dizet\\
{\tt Mtop}     & {\tt xpar(806) (= 175D0)}:        Top mass,    input for Dizet\\
{\tt NPAR(j)}  & {\tt xpar(900+j)},$j=1,21$:          Input flags for Dizet 6.21\\
\hline
\hline
\end{tabular}
\end{small}
\caption{\sf 
  List of input parameters of the \KK\ generator.
  Initial/final fermion properties and EW parameters. Default values in brackets.
  User may change, with precautions, the starred items, while the doubly starred ones
  should never be changed.
}
\label{tab:KK-input3}
\end{table}

\begin{table}[!ht]
\centering
\begin{small}
\begin{tabular}{|l|p{13.0cm}|}
\hline
Parameter & Position and meaning  \\ 
\hline\hline
{\tt Jak1}        & {\tt xpar(2001) (=0)}:         First  $\tau$ decay mask\\
{\tt Jak2}        & {\tt xpar(2002) (=0)}:         Second $\tau$ decay mask\\
{\tt idff}$^{**}$ & {\tt xpar(2003) (=15)}:        PDG ident of the first $\tau$\\
{\tt itdkRC}      & {\tt xpar(2004) (=1)}:         QED rad. switch in leptonic decays\\
{\tt xk0dec}$^{*}$& {\tt xpar(2005) (=0.001d0)}:   IR-cut for QED rad. in leptonic decays\\
{\tt KeyA1}       & {\tt xpar(2006) (=1d0)}:       Type of $a_1$ current\\
{\tt Cabib}$^{**}$& {\tt xpar(2007) (=0.975d0)}:   Cosine of  Cabibbo angle\\
{\tt GV}$^{*}$    & {\tt xpar(2008) (= 1d0)}:      Vector coupling $g_V$ in $\tau$ decay\\
{\tt GA}$^{*}$    & {\tt xpar(2009) (=-1d0)}:      Axial coupling  $g_A$ in $\tau$ decay\\
\hline
{\tt BRA1} & {\tt xpar(2010) (= 0.5d0)}:  In 3-pion decay BR of $\pi^+\pi^-\pi^-$ (vs $\pi^-\pi^0\pi^0$)\\
{\tt BRKS} & {\tt xpar(2011) (=0.6667d0)}: In $K^*$ decay BR of $K^+\pi^0$ (vs $\pi^+K^0$)\\
{\tt BRK0} & {\tt xpar(2012) (=0.5d0)}:          Probability of $K^0$   to be $K_S$\\
{\tt BRK0B} & {\tt xpar(2013) (=0.5d0)}:         Probability of $K^0_B$ to be $K_S$\\
\hline\multicolumn{2}{|c|}{ Branching ratios }\\
\hline
 {\tt BRAE}   & {\tt xpar(2101) (=17.810d-2)}:   Branching ratio $\tau^-  \to  e^-$.
                                                 IMPORTANT!
                   Entry 2101 set smaller than -1d0 will activate internal defaults of Tauola.
                   In such a case all input from 2008-2122 will be IGNORED \\
 {\tt BRAMU}  & {\tt xpar(2102) (=17.370d-2)}:   Branching ratio $\tau^-  \to  \mu^-$ \\
 {\tt BRAPI}  & {\tt xpar(2103) (=11.080d-2)}:   Branching ratio $\tau^-  \to  \pi^-$ \\
 {\tt BRA2PI} & {\tt xpar(2104) (=25.320d-2)}:   Branching ratio $\tau^-  \to  \pi^-, \pi^0$\\
 {\tt BRA3PI} & {\tt xpar(2105) (=18.380d-2)}:   Branching ratio $\tau^-  \to  a_1^-$\\
\hline\multicolumn{2}{|c|}{ Other branching ratios are in xpar(2106-2122), see .KK2f\_defaults}\\
\hline
\hline
\end{tabular}
\end{small}
\caption{\sf 
  Input parameters for the TAUOLA package. 
  For a complete description, see the Manual of TAUOLA~\protect\cite{tauola2.4:1993}.
  Default values in brackets.
  User may change, with precautions, the starred items, while the doubly starred ones
  should never be changed.
}
\label{tab:KK-input4}
\end{table}

\section{Use of the program}
\label{sec:usage}

In this section we will familiarize the reader with the 
input and output parameters, and the use of the present version of the \KK\ Monte Carlo.
We will present two simple demonstration main programs.
Their double role is to serve as a useful template for the user 
to create his/her own main program and to help the user to check quickly that
the newly installed \KK\ generator runs correctly.
We shall describe in detail all the input parameters of \KK.

\subsection{Principal entries of \KK }
The principal entries of the \KK\ package, which the user will call in
his/her application in order to generate a series of MC events, were
already listed and described briefly in Section~\ref{sec:structure}.
Here we shall add more information on their functionality.
The calling sequence constituting a typical Monte Carlo run
will look as follows:
{\small
\begin{verbatim}
 CALL KK2f_ReaDataX('./.KK2f\_defaults',1,10000,xpar)! reading default input
 CALL KK2f_ReaDataX('./user.input'     ,0,10000,xpar)! reading user's input
 CALL KK2f_Initialize(xpar)                          ! initialize generator 
 DO loop=1,10000                                     ! loop over MC events
   CALL KK2f_Make                                    ! generate single event
 ENDDO    
 CALL KK2f_Finalize                                  ! final book-keeping
 CALL KK2f_GetXSecMC(XSecPb,XErrPb)                  ! get total cross section
\end{verbatim}
}
\noindent
In the first call of {\tt KK2f\_ReaDataX}, default data are read into the array 
{\tt REAL*8 xpar(10000)}.
The \KK\ itself has almost no data hidden in the source code.
(This is not true for TAUOLA and JETSET).
The file {\tt .KK2f\_defaults} is read first into array {\tt xpar}.
This file of defaults is provided in the main distribution directory.
The user should {\em never modify it}. 
It can be copied to a local directory or, better, 
a symbolic link should be created to the original file.
The {\tt .KK2f\_defaults} is rather large and the user is usually interested
only in changing some subset of these data.
In the second call on  {\tt KK2f\_ReaDataX} the user can overwrite the default
data with his/her own smaller set of input data, which are placed in the
{\tt user.input} file.
See next subsection for more details on the input data.

The {\tt KK2f\_Initialize} is invoked to initialize the generator.
It reads input data from array {\tt xpar}, prints them and sends them down
to various modules and auxiliary libraries.
The program entries have to be called in strictly the same order
as in the above example.
At this point we are ready to generate a series of the MC events.
The generation of a single event is done with the help of {\tt KK2f\_Make}.
After the generation loop is completed, we may invoke {\tt KK2f\_Finalize},
which does final book-keeping, prints various pieces of information 
on the MC run, and calculates the total MC integrated cross section and its statistical error
in units of picobarn.
In order to access the total cross section the user should call the routine
{\tt KK2f\_GetXSecMC(XSecMC,XErrMC)}.

\subsection{Input data }
As we stated previously, in the second call on {\tt KK2f\_ReaDataX} the users can overwrite the default
with their own preferred values.
Note that the user should never modify certain data items
(without consulting authors of the program) and that the other ones can be changed
by the user, see below.
For example the simplest input data, which define only the CMS energy, look
as follows:
{\small
\begin{verbatim}
BeginX
*<ia><----data-----><-------------------comments------------->
    1          190d0 CmsEne  =CMS total energy [GeV]
EndX
\end{verbatim}
}\noindent
As we see, data cards start with the keyword {\tt BeginX} and end with  
the keyword {\tt EndX}.
The comment lines are allowed -- they start with {\tt *} in the first column.
In the comments we specify the meaning of the data, 
their range, and whether the user is allowed to modify them.
The data themselves are in a fixed format, with the address $i$ in 
{\tt xpar(i)} followed by the data value and trailing comment.
The four examples of input data sets for the two demonstration programs 
{\tt ffbench/demo.f} and {\tt ffbench/ProdMC.f}
in the subdirectories {\tt ffbench/Mu}, {\tt ffbench/Inclusive} and the other ones,
provide useful templates for the typical user data.
The complete set of all user data in {\tt KK2f\_defaults} is described in
detail in Tables~\ref{tab:KK-input1}--\ref{tab:KK-input4}.
Understandably, the user will manipulate, in most cases, only a small subset of the data
and, in most cases, will stick to the default values.

\subsection{MC events and other output}

The principal output of \KK\ is the Monte Carlo {\em event},
which is just a list of final-state four-momenta in [GeV] units and flavours,
encoded in the standard {\tt /d\_HepEvt/} common block, see section \ref{sec:HepEvt}.

All beam, photon and parton momenta before hadronization are available
alternatively through ``getter'' subroutines from class {\bf KK2f},
see section \ref{sec:KK2f} or {\bf HepEvt}:
{\small\begin{verbatim}
   DOUBLE PRECISION    p1(4),p2(4),p3(4),p4(4),PhoAll(100,4)
   INTEGER             NphAll
 ....
   CALL HepEvt_GetBeams(p1,p2)           ! get beam momenta
   CALL HepEvt_GetFfins(p3,p4)           ! get momenta of two final fermions
   CALL HepEvt_GetPhotAll(NphAll,PhoAll) ! get photon multiplicity and momenta
\end{verbatim} }
\noindent
where {\tt NphAll} is the total photon multiplicity
(see also the {\tt ffbench/ProdMC.f} example).
Note that beamstrahlung photons are added to the record as two zero-angle ISR photons,
so that total energy is conserved.
Alternatively, beamstrahlung photon momenta are also available through a dedicated getter:
{\small\begin{verbatim}
   DOUBLE PRECISION    PhoBst(100,4)
   CALL HepEvt_GetPhotBst(NphBst,PhoBst)
\end{verbatim}}
\noindent

\begin{table}[!ht]
\centering
\begin{small}
\begin{tabular}{|l|p{13.0cm}|}
\hline
Parameter & Position and meaning  \\ 
\hline\hline
{\tt WtSet(71)}        & EEX \Order{\alpha^0}\\
{\tt WtSet(72)}        & EEX \Order{\alpha^1}\\
{\tt WtSet(73)}        & EEX \Order{\alpha^2}\\
{\tt WtSet(74)}        & EEX \Order{\alpha^3}\\
{\tt WtSet(201)}       & CEEX \Order{\alpha^0}\\
{\tt WtSet(202)}       & CEEX \Order{\alpha^1}\\
{\tt WtSet(203)}       & CEEX \Order{\alpha^2}\\
{\tt WtSet(251)}       & CEEX \Order{\alpha^0} without ISR-FSR interference\\
{\tt WtSet(252)}       & CEEX \Order{\alpha^1} without ISR-FSR interference\\
{\tt WtSet(253)}       & CEEX \Order{\alpha^2} without ISR-FSR interference\\
\hline\hline
\end{tabular}
\end{small}
\caption{\sf 
  The meaning of the weights in the {\tt WtSet}.
}
\label{tab:wtset}
\end{table}

\subsection{Weighted events, alternative weights}
\label{sec:alternative}
Normally, the user will run the program in the mode with the weight equal to 1.
Running in the mode with weighted events may be useful for various tests.
It can be useful, for example as a cross check, in the situation when
one selects output events strongly, that is imposes cuts that eliminate
all but say 1 event in a 1000.
If at the same time it is seen from the output of {\tt KK2f\_finalize} that the
cross section corresponding to $w>w_{\max}$ is at the similar level of $10^{-3}$,
it is then necessary to cross check if the accepted events do not coincide,
by bad luck, with the ``overweighted'' events.
If it were true, then the cross section and the distribution of the accepted events
could be affected by factor of 2 or more.
In that sense the weighted events are ``safer''.

It should be kept in mind that, although we have set the maximum weights for the
rejection rather high, the user may try an untested configuration of the input
data for which the cross section corresponding to $w>w_{\max}$ is too high.
We recommend that the user always check, at the end of the run, the output from {\tt KK2f\_finalize},
the table in which the percentage of the ``spill over'' cross section
corresponding to $w>w_{\max}$ is given.

The other advantage of the weighted events is that 
in most cases one needs less CPU time to get the same statistical error in the
cross sections and in the histogram.
It can be profitable if one needs to perform many runs with various input parameters.

In the run with weighted events the user may access the principal weight {\tt WtMain}
and the auxiliary weights {\tt WtSet} through another getter:
{\small\begin{verbatim}
   DOUBLE PRECISION    WtSet(1000), WtMain, WtCrud
   CALL KK2f_GetWtAll(WtMain,WtCrud,WtSet)
\end{verbatim}}
\noindent
see also the {\tt ffbench/ProdMC.f} example.
The actual auxiliary weight should be defined as {\tt WtCrud*WtSet(i)}.
Note that events with {\tt WtCrud=0d0} may have undefined four-momenta,
so the user should protect his program against crashing upon an attempt of
working out the kinematics of such an event.

The weights {\tt WtCrud} and {\tt WtSet} are also defined in the run with {\tt WtMain=1d0}.
They can be recorded and used in the subsequent run in order to estimate the effects
that are included or excluded in the auxiliary weight {\tt WtSet(i)}.
The meaning of the most important weights in {\tt WtSet} 
is described in table~\ref{tab:wtset}.

How to get cross sections and distributions corresponding to {\tt WtSet(i)}
using an event generated with {\tt WtMain =1} and 
recorded on the tape?
\begin{itemize}
\item
  The user should {\em record on the tape}
  the vector {\tt WtSet} for each event, together with {\tt WtCrud} and {\tt WtMain}.
\item
  In the subsequent run, to weight events from the tape, 
  each event should be weighted with the ratio%
  \footnote{ This will not work for muons with $v>0.999$ and quarks with $v>0.99$
    where, for technical reasons we use {\tt WtMain=WtCrud*WtSet(74)}.
    This restriction is not important for most of practical purposes.}
  {\tt WtSet(i)/ WtSet(203)}, because in the standard case\\
  {\tt WtMain=WtCrud*WtSet(203)}.
\end{itemize}
A typical application of the above method could be to find out, 
for a given arbitrary distribution or cross section, 
the estimate of {\em physical precision} due to higher orders.
We recommend that the user take
half of the difference \Order{\alpha^2} $-$ \Order{\alpha^1}
as an estimate of the physical precision.
This can be calculated by applying the above method
with the following weight:
{\tt (WtSet(203)-WtSet(202))*WtCrud/Wtmain}.

The above method cannot be used for varying the input parameters of the SM, 
such as the  Higgs
mass, because this would require recalculating {\tt WtSet}.
We may provide such a capability in the next versions.

\subsection{Frequently asked questions on program use}
Some additional information, useful for practical use of the
program is collected as answers to ``frequently asked questions'':
\begin{itemize}
  \item
    {\em How does one properly normalize total cross section?} 
    Look into two demonstration programs in $\tt ffbench$ subdirectory.
  \item
    {\em How does one update tables of electroweak corrections?} 
\begin{verbatim}
     cd KK-all/dizet
     make all (or make table.tau etc.)
\end{verbatim}
    For more details see section \ref{sec:ew-lib}.
  \item
    {\em How does one switch on beamstrahlung?}
    Include  {\tt KeyFix=2} and {\tt KeyGrid=2} in the user input data.
    An example program is included in $\tt ffbench$ subdirectory.
  \item
    {\em How does one switch off radiation for quarks?}
    Include {\tt KeyQSR=0} in the user input data.
  \item
    {\em How does one switch from CEEX to EEX  for quarks?}
    Include  {\tt vmaxGPS=0} for all quarks in the user input data.
  \item
    {\em How does one update compilation flags everywhere?} Compilation flags are set for AIX.
    Examples of f77 flags for HPUX, Linux, ALPHA are in {\tt ./ffbench/Makefile}.
    In order to update centraly {\tt makefiles} in all subdirectories do the following:
\begin{verbatim}
    cd ./ffbench 
    make makflag
\end{verbatim}
    This causes the mapping {\tt makefile.template} $->$ {\tt makefile} in all subdirectories,
    updating compilation flags everywhere with the ones from
./ffbench/Makefile.
  \item
    {\em How does one calculate the QED physical error for a given observable?} 
    Calculate the difference between \Oceex{\alpha^2} and \Oceex{\alpha^1}
    and the difference between \Oceex{\alpha^2} and \Oeex{\alpha^3}.
    This can be done by running the MC with weighted events and taking the difference
    of the weights or with unweighted events, following instructions in the 
    previous section.
  \item
    {\em How can  one be sure about the technical precision?} 
    The problem may arise for strong selection cuts.
    In this case we advise the user to rerun the program with weighted events and check 
    whether the results are the same.
\end{itemize}

\section{Outlook and conclusions}
As is summarized in Table~\ref{tab:KK-status},
the present version of the \KK\ MC has almost the full functionality
of the older  KORALZ and KORALB event generators.
The most important new features in the present \KK\ are the
ISR-FSR interference, the second-order subleading corrections, and the exact matrix
element for two hard photons.
This makes \KK\ already a unique source of SM predictions for the LEP2 physics program.
The inclusion of the beamstrahlung makes it useful for the LC studies.
Note that for these the electroweak correction library has to be reexamined.
The most important omission in the present version is the lack of neutrino and electron
channels.
Let us stress that the present program is an excellent starting platform
for the construction of the second-order Bhabha MC generator based on CEEX exponentiation.
We hope to be able to include the Bhabha and neutrino channels soon, 
possibly in the next version.
The other important directions for the development are 
the inclusion of the exact matrix element for three hard photons, together
with virtual corrections up to \Order{\alpha^3L^3} and the
emission of the light fermion pairs.
The inclusion of the $W^+W^-$ and $t\bar{t}$ final states is still in a farther perspective.

\section*{Acknowledgements}
Two of us (SJ and BFLW) would like to thank the  CERN EP and TH Divisions. We are grateful
to all four LEP Collaborations and their members for support. 
In particular we would like to thank Dr. D. Schlatter of ALEPH for continuous support and help.
One of us (S.J.) would like to thank the DESY Directorate for its generous support,
and Dr. F. Dydak of CERN EP Division for his support 
in the critical stage of the beginning of this project.
We would like to express our gratitude to W.~P\l{}aczek, E.~Richter-W\c{a}s, M.~Skrzypek and 
S.~Yost for valuable comments.

\newpage
\section*{Appendix: Output of the demonstration  program}

\renewcommand{\baselinestretch}{0.2}
{\scriptsize
\begin{verbatim}
 ==============================================
 ============ Demo for KK MC ==================
 1000  requested events 
            *************************************************************
            *  ****   ****    ****  ****    ***       ***     ******    *
            *  ****   ****    ****  ****    ****     ****   **********  *
            *  ****   ****    ****  ****    *****   *****  *****   ***  *
            *  **********     *********     *************  ****         *
            *  *******        ******        *************  ****         *
            *  **********     ********      **** *** ****  *****   ***  *
            *  ****  *****    ****  ****    ****  *  ****   **********  *
            *  ****   *****   ****   ****   ****     ****     *******   *
            *************************************************************
 ***************************************************************************
 *                         KK Monte Carlo                                  *
 *            Version       4.13          25 Jan. 2000                     *
 *     200.00000000                 CMS energy average       CMSene     a1 *
 *        .00000000                 Beam energy spread       DelEne     a2 *
 *              100                 Max. photon mult.        npmax      a3 *
 *                1                 ISR switch               KeyISR     a4 *
 *                1                 FSR switch               KeyFSR     a5 *
 *                2                 ISR/FSR interferenc      KeyINT     a6 *
 *                1                 New exponentiation       KeyGPS     a7 *
 *                1                 Hadroniz.  switch        KeyHad     a7 *
 *        .20000000                 Hadroniz. min. mass      HadMin     a9 *
 *       1.00000000                 Maximum weight           WTmax     a10 *
 *              100                 Max. photon mult.        npmax     a11 *
 *               11                 Beam ident               KFini     a12 *
 *        .00100000                 Manimum phot. ener.      Ene       a13 *
 *    .10000000E-59                 Phot.mass, IR regul      MasPho    a14 *
 *    1.2500000                     Phot. mult. enhanc.      Xenph     a15 *
 *    .10000000E-08                 Vcut1                    Vcut1     a16 *
 *    .50000000E-01                 Vcut2                    Vcut2     a16 *
 *    .00000000E+00                 Vcut3                    Vcut2     a16 *
 *        .00000000                    PolBeam1(1)           Pol1x     a17 *
 *        .00000000                    PolBeam1(2)           Pol1y     a18 *
 *        .00000000                    PolBeam1(3)           Pol1z     a19 *
 *        .00000000                    PolBeam2(1)           Pol2x     a20 *
 *        .00000000                    PolBeam2(2)           Pol2y     a21 *
 *        .00000000                    PolBeam2(3)           Pol2z     a22 *
 ***************************************************************************
 ***************************************************************************
 *            BornV  Initializator                                         *
 *      91.18700000                 Z mass     [GeV]         amz        a1 *
 *     100.00000000                 Higgs mass [GeV]         amh        a2 *
 *     175.00000000                 Top mass   [GeV]         amtop      a3 *
 *       2.50072032                 Z width    [GeV]         gammz      a4 *
 *        .22276773                 sin(theta_w)**2          sinw2      a5 *
 *     137.03598950                 1/alfa_QED  at  Q=0      AlfInv     a6 *
 *        .20000000                 MassCut light qqbar      HadMin     a6 *
 *               11                 KF code of beam          KFini      a7 *
 *    1.0000000                     Input vvmax              vvmax      a8 *
 *    .99999888                     reduced vvmax in MC      vvmax      a9 *
 *         Test switches:                                                  *
 *                1                 Electroweak lib.         KeyElw     10 *
 *                1                 Z on/off   switch        KeyZet     11 *
 *                0                 mass terms on/off        KeyWtm     12 *
 ***************************************************************************
 ***************************************************************************
 *         BornV  Reading from disk file:                                  *
 *        ../../dizet/table.down.340pt                                     *
 *      91.18700000                 Z mass                   amz        a1 *
 *     100.00000000                 Higgs mass               amh        a2 *
 *     175.00000000                 Top mass                 amtop      a3 *
 *        .22302485                 sin**2(thetaW)           swsq       a3 *
 *       2.49925439                 Z width                  gammz      a3 *
 *      80.37787000                 W mass                   amw        a3 *
 *       2.08825838                 W width                  gammw      a3 *
 ***************************************************************************
 ***************************************************************************
 *         BornV  Reading from disk file:                                  *
 *        ../../dizet/table.up.340pt                                       *
 *      91.18700000                 Z mass                   amz        a1 *
 *     100.00000000                 Higgs mass               amh        a2 *
 *     175.00000000                 Top mass                 amtop      a3 *
 *        .22302485                 sin**2(thetaW)           swsq       a3 *
 *       2.49925439                 Z width                  gammz      a3 *
 *      80.37787000                 W mass                   amw        a3 *
 *       2.08825838                 W width                  gammw      a3 *
 ***************************************************************************
 ***************************************************************************
 *         BornV  Reading from disk file:                                  *
 *        ../../dizet/table.down.340pt                                     *
 *      91.18700000                 Z mass                   amz        a1 *
 *     100.00000000                 Higgs mass               amh        a2 *
 *     175.00000000                 Top mass                 amtop      a3 *
 *        .22302485                 sin**2(thetaW)           swsq       a3 *
 *       2.49925439                 Z width                  gammz      a3 *
 *      80.37787000                 W mass                   amw        a3 *
 *       2.08825838                 W width                  gammw      a3 *
 ***************************************************************************
 ***************************************************************************
 *         BornV  Reading from disk file:                                  *
 *        ../../dizet/table.up.340pt                                       *
 *      91.18700000                 Z mass                   amz        a1 *
 *     100.00000000                 Higgs mass               amh        a2 *
 *     175.00000000                 Top mass                 amtop      a3 *
 *        .22302485                 sin**2(thetaW)           swsq       a3 *
 *       2.49925439                 Z width                  gammz      a3 *
 *      80.37787000                 W mass                   amw        a3 *
 *       2.08825838                 W width                  gammw      a3 *
 ***************************************************************************
 ***************************************************************************
 *         BornV  Reading from disk file:                                  *
 *        ../../dizet/table.botom.340pt                                    *
 *      91.18700000                 Z mass                   amz        a1 *
 *     100.00000000                 Higgs mass               amh        a2 *
 *     175.00000000                 Top mass                 amtop      a3 *
 *        .22302485                 sin**2(thetaW)           swsq       a3 *
 *       2.49925439                 Z width                  gammz      a3 *
 *      80.37787000                 W mass                   amw        a3 *
 *       2.08825838                 W width                  gammw      a3 *
 ***************************************************************************
 ***************************************************************************
 *         BornV  Reading from disk file:                                  *
 *        ../../dizet/table.mu.340pt                                       *
 *      91.18700000                 Z mass                   amz        a1 *
 *     100.00000000                 Higgs mass               amh        a2 *
 *     175.00000000                 Top mass                 amtop      a3 *
 *        .22302485                 sin**2(thetaW)           swsq       a3 *
 *       2.49925439                 Z width                  gammz      a3 *
 *      80.37787000                 W mass                   amw        a3 *
 *       2.08825838                 W width                  gammw      a3 *
 ***************************************************************************
 ***************************************************************************
 *         BornV  Reading from disk file:                                  *
 *        ../../dizet/table.tau.340pt                                      *
 *      91.18700000                 Z mass                   amz        a1 *
 *     100.00000000                 Higgs mass               amh        a2 *
 *     175.00000000                 Top mass                 amtop      a3 *
 *        .22302485                 sin**2(thetaW)           swsq       a3 *
 *       2.49925439                 Z width                  gammz      a3 *
 *      80.37787000                 W mass                   amw        a3 *
 *       2.08825838                 W width                  gammw      a3 *
 ***************************************************************************
 ***************************************************************************
 *                         KarLud_Initialize START                         *
 *     200.00000000                  CMS energy average      CMSene     == *
 *        .00000000                  Beam energy spread      DelEne     == *
 *                1                  ISR on/off switch       KeyISR     == *
 *                0                  Type of ISR             KeyFix     == *
 *                1                  Elect_weak switch       KeyZet     == *
 *                0                  Fixed nphot mult.       MltISR     == *
 *               50                  Max. photon mult.       nmax       == *
 ***************************************************************************
 *    4370.16701351                 xs_crude  vesko          xcvesk        *
 *    4363.90727756                 xs_crude  gauss          xcgaus        *
 *        .00143443                 xcvesk/xcgaus-1                        *
 ***************************************************************************
 *                          KarLud_Initialize END                          *
 ***************************************************************************
 ***************************************************************************
 *                        KarFin Initialize  START                         *
 *                1                FSR radiation on/off      KeyFSR     a1 *
 *                1                radiation from quark      KeyQSR     a2 *
 *                1                removal    switch         KeyPia     a3 *
 *    .10000000E-02                infrared cut FACTOR       delfac     a4 *
 *    .10000000E-07                infrared cut itself       delta      a5 *
 *    .10000000E-02                EminCMS for removal       [GeV]      a6 *
 *               50                Max. photon mult.         nmax       a7 *
 *                        KarFin Initialize  END                           *
 ***************************************************************************
 ***************************************************************************
 *             GPS   Initializator                                         *
 *      91.18700000                 Z mass     [GeV]         MZ         a1 *
 *       2.49925439                 Z width    [GeV]         GammZ      a2 *
 *        .22302485                 sin(theta_w)**2          Sw2        a3 *
 *     137.03598950                 1/alfa_QED  at  Q=0      AlfInv     a4 *
 *         Test switches:                                                  *
 *                1                 Z on/off   switch        KeyZet     a5 *
 *                1                 Electroweak lib.         KeyElw     a6 *
 *                1                 CEEX level               KeyGPS     a7 *
 *                1                 ISR emission             KeyISR     a8 *
 *                1                 FSR emission             KeyFSR     a9 *
 *                2                 ISR*FSR interferenc      KeyINT    a10 *
 ***************************************************************************
 ***************************************************************************
 *                           KK2f: Initialization                          *
 *    9.4892226                     x-crude [nb]             Xcrunb     ** *
 *         List of final fermions:                                         *
 *                1                 KF of final fermion      KFfin      ** *
 *    .10000000                     mass of final ferm.      amferm     ** *
 *    1.3410294                     Xborn [R]                Xborn      ** *
 *    5.0000000                     WtMax sampling par.      WtMax      ** *
 *    .99000000                     vmax for CEEX           vmaxGPS     ** *
 *                2                 KF of final fermion      KFfin      ** *
 *    .10000000                     mass of final ferm.      amferm     ** *
 *    2.1445691                     Xborn [R]                Xborn      ** *
 *    5.0000000                     WtMax sampling par.      WtMax      ** *
 *    .99000000                     vmax for CEEX           vmaxGPS     ** *
 *                3                 KF of final fermion      KFfin      ** *
 *    .20000000                     mass of final ferm.      amferm     ** *
 *    1.3410294                     Xborn [R]                Xborn      ** *
 *    5.0000000                     WtMax sampling par.      WtMax      ** *
 *    .99000000                     vmax for CEEX           vmaxGPS     ** *
 *                4                 KF of final fermion      KFfin      ** *
 *    1.3000000                     mass of final ferm.      amferm     ** *
 *    2.1445691                     Xborn [R]                Xborn      ** *
 *    5.0000000                     WtMax sampling par.      WtMax      ** *
 *    .99000000                     vmax for CEEX           vmaxGPS     ** *
 *                5                 KF of final fermion      KFfin      ** *
 *    4.5000000                     mass of final ferm.      amferm     ** *
 *    1.3410294                     Xborn [R]                Xborn      ** *
 *    5.0000000                     WtMax sampling par.      WtMax      ** *
 *    .99000000                     vmax for CEEX           vmaxGPS     ** *
 *               13                 KF of final fermion      KFfin      ** *
 *    .10565830                     mass of final ferm.      amferm     ** *
 *    1.2225177                     Xborn [R]                Xborn      ** *
 *    8.0000000                     WtMax sampling par.      WtMax      ** *
 *    .99900000                     vmax for CEEX           vmaxGPS     ** *
 *               15                 KF of final fermion      KFfin      ** *
 *    1.7770000                     mass of final ferm.      amferm     ** *
 *    1.2225177                     Xborn [R]                Xborn      ** *
 *    8.0000000                     WtMax sampling par.      WtMax      ** *
 *    1.0000000                     vmax for CEEX           vmaxGPS     ** *
 ***************************************************************************
 ***************************************************************************
 *        KK interface of Tauola                                           *
 *                2                Cloning procedure       KeyClone    t01 *
 ***************************************************************************
 ***************************************************************************
 *        Parameters passed from KK  to Tauola:                            *
 *                0                dec. type 1-st tau        Jak1      t01 *
 *                0                dec. type 2-nd tau        Jak2      t02 *
 *                1                current type a1 dec.      KeyA1     t03 *
 *               15                PDG id 1-st tau           idff      t04 *
 *                1                R.c. switch lept dec      itdkRC    t05 *
 *    .10000000E-02                IR-cut for lept r.c.      xk0dec    t06 *
 ***************************************************************************
 ***************************************************************************
 *    TAUOLA Initialization SUBROUTINE INIMAS:                             *
 *    Adopted to read from KK                                              *
 *    1.7770000                        AMTAU tau-mass             **** *** *
 *    .51099900E-03                    AMEL  electron-mass        **** *** *
 *    .10565830                        AMMU  muon-mass            **** *** *
 ***************************************************************************
............... skipped output from TAUOLA ...............
***************************************************************************
 *                         *****TAUOLA LIBRARY: VERSION 2.6 ******         *
 *                         ***********August   1995***************         *
 *                         **AUTHORS: S.JADACH, Z.WAS*************         *
 *                         **R. DECKER, M. JEZABEK, J.H.KUEHN*****         *
 *                         **AVAILABLE FROM: WASM AT CERNVM ******         *
 *                         ***** PUBLISHED IN COMP. PHYS. COMM.***         *
 *                         *******CERN-TH-5856 SEPTEMBER 1990*****         *
 *                         *******CERN-TH-6195 SEPTEMBER 1991*****         *
 *                         *******CERN TH-6793 NOVEMBER  1992*****         *
 *                         **5 or more pi dec.: precision limited          *
 *                         ****DEKAY ROUTINE: INITIALIZATION******         *
 *                   0     JAK1   = DECAY MODE TAU+                        *
 *                   0     JAK2   = DECAY MODE TAU-                        *
 ***************************************************************************

 ***************************************************************************
                            Event listing (summary)
    I  particle/jet KS     KF orig    p_x      p_y      p_z       E        m
    1  !e-!         21     11    0     .000     .000  100.000  100.000     .001
    2  !e+!         21    -11    0     .000     .000 -100.000  100.000     .001
    3  (tau-)       11     15    1  -11.368   -8.313  -16.590   21.834    1.777
    4  (tau+)       11    -15    1   77.073   19.642   34.404   86.677    1.777
    5  gamma         1     22    1  -65.705  -11.329    6.665   67.007     .000
    6  gamma         1     22    1    -.001     .000  -24.480   24.480     .000
    7  gamma         1     22    2     .002     .001     .000     .002     .000
    8  nu_tau        1     16    3    -.030    -.121    -.003     .125     .010
    9  (rho-)       11   -213    3  -11.338   -8.192  -16.587   21.709     .690
   10  pi-           1   -211    9   -6.338   -4.212   -9.177   11.922     .140
   11  pi0           1    111    9   -5.000   -3.980   -7.410    9.787     .135
   12  nu_tau~       1    -16    4   34.604    8.167   14.712   38.478     .010
   13  pi+           1    211    4   42.469   11.475   19.692   48.199     .140
                   sum:   .00          .000     .000     .000  200.000  200.000

 ***************************************************************************
 *                      KarLud  final  report                              *
 *            90875                  total no of events     nevtot      == *
 *    4370.16701351                  ISRcru  [R]            ISRcru      == *
 *    4363.65383283  +-  .28966439   ISRbest [R],ISRerr     ISRbest     == *
 *    4370.1670                      XKarlud [R]            XKarlud     == *
 *    .00000000E+00                  KError  [R]            KError      == *
 ***************************************************************************
 ***************************************************************************
 *                   Report on wt_ISR of KarLud                            *
 *            90875                 total no of events      nevtot      == *
 *                0                 wt<0        events      nevneg      == *
 *        .01551546  +-  .00666845  <wt>                    wt_ISR      == *
 *      67.80514951  +-  .45215550  sigma of KarLud [R]     xskarl      == *
 ***************************************************************************
 ***************************************************************************
 *        .99850963  +-  .00006638  Average WT of Vesk1      AVesk1     == *
 *    4363.90727756  +-  .04363907  xs_est gauss    [R]      xcgaus     == *
 *       -.00005808  +-  .00006738  xcve/xcgs-1                         == *
 ***************************************************************************

 ***************************************************************************
 *                        KarFin Finalize    START                         *
 *           180070                  generated events        nevgen     a2 *
 *        .99710113  +-  .00017970   kinematics, smin        wt1        a5 *
 *        .99980533  +-  .00002691   jacobian                wt2        a6 *
 *        .99929387  +-  .00061237   photon ang. dist.       wt3        a7 *
 *                            ON MASS WEIGHTS                              *
 *        .95525378  +-  .00052108  removal wgt wtrem                   b1 *
 *            89774                 no. of raw events                   b2 *
 *                0                 wt6=0      events                   b3 *
 *        .99957099  +-  .00051944  control wgt wctrl                   b4 *
 *                0                 marked photons           MarTot     a5 *
 *    .10000000E-02                 emin                                b6 *
 *    .10000000E-07                 delta                               b7 *
 *        .15689398                 raw ph. multipl.                    b8 *
 *       6.00000000                 Highest phot. mult.                 b9 *
 *                        YFSfin Finalize    END                           *
 ***************************************************************************
 ***************************************************************************
 *                            KarFin Finalize                              *
 *           180070                  generated events        nevgen     a2 *
 *        .99634044  +-  .00064027   general weight          wt         a1 *
 *        .15643916                  aver. ph. multi.        avmlt      a3 *
 ***************************************************************************
      -640          KK2f: Photon raw multiplicity                                                   
           nent            sum           bmin           bmax
           1000     .00000E+00     .00000E+00     .33000E+03
           undf           ovef           sumw           avex
     .00000E+00     .00000E+00     .21210E+04     .28322E+01
   .0000    .600000D+02 0XXXXXXXXXXXX                                                     I
  1.0000    .277000D+03 0XXXXXXXXXXXXXXXXXXXXXXXXXXXXXXXXXXXXXXXXXXXXXXXXXXXXXXX          I
  2.0000    .330000D+03 0XXXXXXXXXXXXXXXXXXXXXXXXXXXXXXXXXXXXXXXXXXXXXXXXXXXXXXXXXXXXXXXXXX
  3.0000    .196000D+03 0XXXXXXXXXXXXXXXXXXXXXXXXXXXXXXXXXXXXXXX                          I
  4.0000    .990000D+02 0XXXXXXXXXXXXXXXXXXX                                              I
  5.0000    .290000D+02 0XXXXX                                                            I
  6.0000    .800000D+01 0X                                                                I
  7.0000    .100000D+01 0                                                                 I
  8.0000    .000000D+00 0                                                                 I
  9.0000    .000000D+00 0                                                                 I
 ***************************************************************************
 *                       KK2f_Finalize  printouts                          *
 *     200.00000000                 cms energy total         cmsene     a0 *
 *             1000                 total no of events       nevgen     a1 *
 *               ** principal info on x-section **                         *
 *      47.39210315  +-  .51653089  xs_tot MC R-units        xsmc       a1 *
 *     102.90549867                 xs_tot    picob.         xSecPb     a3 *
 *       1.12157650                 error     picob.         xErrPb     a4 *
 *        .01089909                 relative error           erel       a5 *
 *       1.17765968                 WTsup, largest WT        WTsup     a10 *
 *                       ** some auxiliary info **                         *
 *      23.35792938                 xs_born   picobarns       xborn    a11 *
 *       2.12100000                 Raw phot. multipl.                 === *
 *       7.00000000                 Highest phot. mult.                === *
 *                         End of KK2f  Finalize                           *
 ***************************************************************************
........... skipping some lines ................
 ***************************************************************************
 *                MBrA: report on the main Weight                          *
 *            90875                 no of raw events         Ntot       b1 *
 *             1000                 accepted    events       Nacc       b2 *
 *                0                 wt<0        events       Nneg       b3 *
 *                1                 wt>WTmax    events       Nove       b4 *
 *       1.17765968                 WTsup, largest WT        WTsup      b5 *
 *        .00000195                 <Wt-WtMax>  Overfl.      AvOve      b6 *
 *        .00000000                 <Wt> for Wt<0            AvUnd      b7 *
 *        .00018028                 AvOve/<Wt>,WT>WtMax      ROverf     b8 *
 *        .00000000                 AvUnd/<Wt>,Wt<0          RUnder     b9 *
 ***************************************************************************
=====================================================================================================
            MBrA:    Detailed statistics for all branches    
=====================================================================================================
  KF     AveWt     ERela     WtSup      Wt<0   Wt>Wmax       Ntot       Nacc   Nneg   Nove   Nzer
   1   .017100   .024298 .7062       .000000   .000000      10024        181      0      0     96
   2   .005937   .026491 .6685       .000000   .000000      29851        184      0      0    343
   3   .024082   .024922 .7078       .000000   .000000       6738        157      0      0     57
   4   .034175   .026378 1.178       .000000   .001035       5022        149      0      1     53
   5   .046847   .022320 .7777       .000000   .000000       3332        168      0      0     38
  13   .002429   .036648 .8380       .000000   .000000      32573         91      0      0    474
  15   .020322   .038018 .6182       .000000   .000000       3335         70      0      0     40
All:   .010844   .010899 1.178       .000000   .000180      90875       1000      0      1   1101
=====================================================================================================
........... skipping some lines ................
      -630          Tau Pair: wt1, Spin Imprint weight                                              
           nent            sum           bmin           bmax
            281     .00000E+00     .00000E+00     .23000E+02
           undf           ovef           sumw           avex
     .00000E+00     .00000E+00     .28107E+03     .13707E+01
   .0000    .900000D+01 0XXXXXXXXXXXXXXXXXXXXXXXXX                                        I
   .1000    .120000D+02 0XXXXXXXXXXXXXXXXXXXXXXXXXXXXXXXXXX                               I
   .2000    .190000D+02 0XXXXXXXXXXXXXXXXXXXXXXXXXXXXXXXXXXXXXXXXXXXXXXXXXXXXXX           I
   .3000    .110000D+02 0XXXXXXXXXXXXXXXXXXXXXXXXXXXXXXX                                  I
   .4000    .900000D+01 0XXXXXXXXXXXXXXXXXXXXXXXXX                                        I
   .5000    .230000D+02 0XXXXXXXXXXXXXXXXXXXXXXXXXXXXXXXXXXXXXXXXXXXXXXXXXXXXXXXXXXXXXXXXXX
   .6000    .150000D+02 0XXXXXXXXXXXXXXXXXXXXXXXXXXXXXXXXXXXXXXXXXXX                      I
   .7000    .140000D+02 0XXXXXXXXXXXXXXXXXXXXXXXXXXXXXXXXXXXXXXXX                         I
   .8000    .160000D+02 0XXXXXXXXXXXXXXXXXXXXXXXXXXXXXXXXXXXXXXXXXXXXX                    I
   .9000    .230000D+02 0XXXXXXXXXXXXXXXXXXXXXXXXXXXXXXXXXXXXXXXXXXXXXXXXXXXXXXXXXXXXXXXXXX
  1.0000    .210000D+02 0XXXXXXXXXXXXXXXXXXXXXXXXXXXXXXXXXXXXXXXXXXXXXXXXXXXXXXXXXXXX     I
  1.1000    .150000D+02 0XXXXXXXXXXXXXXXXXXXXXXXXXXXXXXXXXXXXXXXXXXX                      I
  1.2000    .140000D+02 0XXXXXXXXXXXXXXXXXXXXXXXXXXXXXXXXXXXXXXXX                         I
  1.3000    .170000D+02 0XXXXXXXXXXXXXXXXXXXXXXXXXXXXXXXXXXXXXXXXXXXXXXXX                 I
  1.4000    .110000D+02 0XXXXXXXXXXXXXXXXXXXXXXXXXXXXXXX                                  I
  1.5000    .100000D+02 0XXXXXXXXXXXXXXXXXXXXXXXXXXXX                                     I
  1.6000    .600000D+01 0XXXXXXXXXXXXXXXXX                                                I
  1.7000    .100000D+02 0XXXXXXXXXXXXXXXXXXXXXXXXXXXX                                     I
  1.8000    .300000D+01 0XXXXXXXX                                                         I
  1.9000    .200000D+01 0XXXXX                                                            I
  2.0000    .500000D+01 0XXXXXXXXXXXXXX                                                   I
  2.1000    .700000D+01 0XXXXXXXXXXXXXXXXXXXX                                             I
  2.2000    .100000D+01 0XX                                                               I
  2.3000    .100000D+01 0XX                                                               I
  2.4000    .100000D+01 0XX                                                               I
  2.5000    .000000D+00 0                                                                 I
  2.6000    .100000D+01 0XX                                                               I
  2.7000    .300000D+01 0XXXXXXXX                                                         I
  2.8000    .000000D+00 0                                                                 I
  2.9000    .100000D+01 0XX                                                               I
  3.0000    .000000D+00 0                                                                 I
  3.1000    .100000D+01 0XX                                                               I
  3.2000    .000000D+00 0                                                                 I
  3.3000    .000000D+00 0                                                                 I
  3.4000    .000000D+00 0                                                                 I
  3.5000    .000000D+00 0                                                                 I
  3.6000    .000000D+00 0                                                                 I
  3.7000    .000000D+00 0                                                                 I
  3.8000    .000000D+00 0                                                                 I
  3.9000    .000000D+00 0                                                                 I
 ***************************************************************************
 *                      Tau Pair Finalize                                  *
 *       1.00026167  +-  .03630240  Spin Imprint <wt1>       wt1ave     a1 *
 *       3.16386575                 Maximum value wt1        wt1max     a2 *
 ***************************************************************************
\end{verbatim}}
\renewcommand{\baselinestretch}{1.0}

\newpage

\end{document}